\documentclass[onefignum,onetabnum]{siamonline250211}

\usepackage{enumitem}
\usepackage{lipsum}
\usepackage{amsfonts}
\usepackage{graphicx}
\usepackage{mathrsfs}
\usepackage{epstopdf}
\usepackage{subfiles} 
\usepackage{multirow} 
\usepackage{nicefrac}
\usepackage{algorithmic}
\usepackage{algorithm}
\ifpdf
  \DeclareGraphicsExtensions{.eps,.pdf,.png,.jpg}
\else
  \DeclareGraphicsExtensions{.eps}
\fi
\usepackage{amsmath, amssymb}

\usepackage{tikz}
\usepackage{hyperref}

\usetikzlibrary{mindmap}

\newcommand{\bA}{\mathbf{A}}

\newcommand{\bb}{\mathbf{b}}

\newcommand{\bd}{\mathbf{d}}

\newcommand{\bg}{\mathbf{g}}

\newcommand{\bQ}{\mathbf{Q}}

\newcommand{\bI}{\mathbf{I}}
\newcommand{\bL}{\mathbf{L}}
\newcommand{\bS}{\mathbf{S}}

\newcommand{\bK}{\mathbf{K}}
\newcommand{\bM}{\mathbf{M}}
\newcommand{\bm}{\mathbf{m}}
\newcommand{\bN}{\mathbf{N}}

\newcommand{\bP}{\mathbf{P}}

\newcommand{\bW}{\mathbf{W}}
\newcommand{\bw}{\mathbf{w}}
\newcommand{\bU}{\mathbf{U}}
\newcommand{\bu}{\mathbf{u}}
\newcommand{\bV}{\mathbf{V}}

\newcommand{\bZ}{\mathbf{Z}}

\newcommand{\bx}{\mathbf{x}}

\newcommand{\dual}{\bold{p}}
\newcommand{\dualp}{\bold{q}}
\newcommand{\sdual}{\boldsymbol{\lambda}}

\newcommand{\Wd}{\bW_{\!\texttt{D}}}
\newcommand{\Ws}{\bW_{\!\texttt{S}}}
\newcommand{\Wm}{\bW_{\!\texttt{M}}}
\newcommand{\phid}{w_{\texttt{D}}}
\newcommand{\phis}{w_{\texttt{S}}}
\newcommand{\phim}{w_{\texttt{M}}}
\newcommand{\bphid}{\bw_{\texttt{D}}}
\newcommand{\bphis}{\bw_{\texttt{S}}}
\newcommand{\bphim}{\bw_{\texttt{M}}}

\newcommand{\diag}[1]{\text{diag}(#1)}
\newcommand{\fref}[1]{Figure \ref{#1}}                  
\newcommand{\eref}[1]{\eqref{#1}}                  

\DeclareMathOperator*{\minimize}{minimize}

\DeclareMathOperator*{\argminimax}{argminimax}

\ifpdf
  \DeclareGraphicsExtensions{.eps,.pdf,.png,.jpg}
\else
  \DeclareGraphicsExtensions{.eps}
\fi

\newsiamremark{remark}{Remark}
\newsiamremark{hypothesis}{Hypothesis}
\crefname{hypothesis}{Hypothesis}{Hypotheses}
\newsiamthm{claim}{Claim}

\headers{Weighted Multiplier Waveform Inversion}{A. Gholami and K. Aghazade and A. Vishwakarma}

\title{Weighted Lagrange Multiplier Method for Robust Source-Independent Waveform Inversion 
\thanks{Submitted to the editors DATE
\funding{This research was financially supported by the SONATA BIS grant
(No. 2022/46/E/ST10/00266) of the National Science Center in
Poland. }}}

\author{Ali Gholami \thanks{Institute of Geophysics, Polish Academy of Sciences 
  (\email{agholami@igf.edu.pl}, \email{aghazade.kamal@igf.edu.pl}, \email{akshay@igf.edu.pl}). }
\and Kamal Aghazade \footnotemark[2] \and Akshay Vishwakarma \footnotemark[2]}

\usepackage{amsopn}

\usepackage{titlesec}
\titleformat{\paragraph}
  {\normalfont\normalsize\bfseries}{\theparagraph}{1em}{}
\titlespacing{\paragraph}{0pt}{3.25ex plus 1ex minus .2ex}{1.5ex plus .2ex}

\usepackage{soul}
\usepackage{optidef}

\renewcommand{\dual}{\boldsymbol{\lambda}}
\renewcommand{\dualp}{\boldsymbol{\lambda}_k}

\ifpdf
\hypersetup{
  pdftitle={Weighted Lagrange Multiplier Methods for Robust Waveform Inversion},
  pdfauthor={A. Gholami and K. Aghazade and A. Vishwakarma}
}
\fi

\begin{document}
\graphicspath{{./Figures/}}
\maketitle
%
\begin{abstract}
The Lagrange multiplier method has proven highly effective for mitigating the ill-conditioning of full waveform inversion (FWI), enabling robust and computationally efficient algorithms that converge to accurate velocity models even from poor initial estimates. Classical multiplier-based FWI methods optimize an augmented Lagrangian (AL) functional with a scalar penalty parameter that uniformly weights wave-equation constraint violations. While this balances data fit and wave-equation satisfaction, it applies uniform relaxation across the model, disregarding source locations and the natural decay of seismic energy. We propose a weighted proximal-point Lagrangian formulation that introduces spatially varying regularization, applying weaker enforcement near sources and progressively stronger enforcement with increasing distance. This compensates for the energy decay, promotes balanced wave-equation enforcement, and improves the convexity of the optimization landscape. The method also eliminates the need for explicit source signature estimation and relaxes the requirement for sources to lie on finite-difference grid points, increasing practical applicability. 
Enhanced computational efficiency is achieved through our dual-space ADMM implementation, which avoids repeated LU factorizations of the forward operator. Only a few LU factorizations are required, with all subsequent iterations solved via efficient forward–backward substitution, making the approach scalable to large-scale 2D and 3D problems. Numerical experiments on challenging synthetic benchmarks demonstrate that the proposed method broadens the basin of attraction of the AL objective, improves robustness to poor initial models and strong noise, and achieves faster, more stable convergence compared with standard multiplier-based methods.
\end{abstract}
%
\begin{keywords}
Full waveform inversion, Augmented Lagrangian, Proximal-point method, Multiplier methods, Extended source
\end{keywords}

\begin{MSCcodes}
86-08, 65F22, 86A22, 49M41, 35R30, 90C06  
\end{MSCcodes}

\section{Introduction.}\label{sect:intro}
Full waveform inversion (FWI) is an advanced PDE-constrained optimization method for high-resolution imaging of subsurface physical properties such as velocity, density, and anisotropy. These parameters appear as coefficients in the wave equation, which governs seismic wave propagation. In the frequency domain, the monochromatic forward wavefield $\bu(\bx)\in\mathbb{C}^n$, generated by a point source at $\bx_s$ in a medium with spatially varying squared slowness $\bm(\bx)\in\mathbb{R}^n$, satisfies the Helmholtz equation
\begin{equation} \label{wave_eq}
 \Delta \bu(\bold{x})+\omega^2\bm(\bold{x})\bu(\bold{x})=\delta(\bold{x-x}_s)f(\omega),
\end{equation}
where $\omega$ is the angular frequency, $\Delta$ is the Laplacian, $f(\omega)$ is the source signature, and $\delta$ is the Dirac delta. Sampling the solution at receiver locations via a restriction operator $\bP \in \mathbb{R}^{n_r\times n}$ yields the recorded data $\bd = \bP \bu$. FWI seeks model parameters $\bm$ that minimize the data misfit while satisfying the wave equation, leading to the PDE-constrained optimization problem
\begin{equation} \label{main_optim}
\minimize_{\bold{u,m}}~\frac12\|\bP\bu-\bd\|_{\Wd}^2 \quad\text{subject to}\quad \bold{A(m)u}=\bb,
\end{equation}
where $\|\bd\|_{\Wd}^2 = \bd^{\!\top}\Wd\bd$ denotes a weighted $\ell_2$-norm, and $(\cdot)^{\top}$ represents the Hermitian (conjugate) transpose. 
The matrix $\Wd$ is a positive-definite weighting matrix that scales the relative contribution of each data component in the misfit term. 
Here, $\bA(\bm)=\omega^2\diag{\bm}+\boldsymbol{\Delta}$ is the discretized Helmholtz operator, and $\bb(\bx)=\delta(\bx-\bx_s)f(\omega)$ is the source term. 
For clarity, the formulation is presented for a single source with fixed receivers, though the extension to multiple sources or nonstationary acquisitions is straightforward.

\subsection{Challenges of FWI.}
Despite its high resolution, FWI faces several well-known challenges \cite{Pratt_1998_GNF,Virieux_2009_OFW,vanLeeuwen_2016_PMP,Treister_2017_FWI_GTT,Huang_2018_SEW,Symes_2020_FWI, 
Rizzuti_2021_ADF,Buchatsky_2021_FWI_ESS,Gholami_2022_EFW,Operto_2023_FWI,Bao_2023_RFW}:
\begin{itemize}
    \item[(i)] Nonconvexity and cycle skipping: When the starting model is too far from the truth, the predicted data can be out of phase by more than half a cycle (cycle skipping), causing convergence to local minima.
 \item[(ii)] High computational cost: Each iteration requires solving large PDE systems, demanding multiple LU factorizations or iterative solutions, leading to significant computational and memory cost.
\item[(iii)]Limited data coverage: Low-frequency and long-offset data—critical for recovering the smooth background velocity—are often sparse or missing due to acquisition constraints.
\item[(iv)] Noise sensitivity: Field data is contaminated by unknown noise, which complicates inversion and causes instability.
\item[(v)] Unknown source signature: Accurate knowledge of the source wavelet $f(\omega)$ is crucial for waveform matching. Its misestimation can bias the solution and introduce artifacts.
\item[(vi)] Grid-alignment requirement: Standard finite-difference discretizations require that sources and receivers coincide with grid points, which restricts modeling flexibility and can introduce significant numerical errors if the true acquisition geometry does not align with the grid, which is the case in practical applications. This limitation can also necessitate impractically fine grids or interpolation methods to accurately implement acquisition geometry.
\end{itemize}

\subsection{Current Approaches and State of the Art.}

Various strategies have been proposed to mitigate these issues. Reduced-space (or reduced) formulations eliminate $\bu$ by solving the wave equation exactly, leading to an optimization problem solely over $\bm$ \cite{Pratt_1998_GNF,Metivier_2013_FWI_TNM}. While this reduces the dimensionality, it increases nonlinearity and exacerbates local minima issue. Penalty methods relax the wave equation as a soft constraint by adding a quadratic penalty term to the objective \cite{vanLeeuwen_2016_PMP,Buchatsky_2021_FWI_ESS}. Examples include Contrast Source Inversion (CSI) \cite{Abubakar_2008_FCS} and Wavefield Reconstruction Inversion (WRI) \cite{vanLeeuwen_2016_PMP}. These methods improve conditioning but require careful tuning of the penalty parameter, often using adaptive strategies \cite{Symes_2020_FWI}.
Lagrangian and augmented Lagrangian (AL) formulations retain wavefields as independent variables but also introduce Lagrange multipliers to enforce the wave equation more strictly \cite{Haber_2000_OTS,Powell_1969_NLC,Aghamiry_2019_IWR}. These multiplier-based methods combine the stability of penalty methods with the accuracy of constraint enforcement, improving robustness and reducing nonlinearity, but at the cost of additional variables and computational work.

Several efforts have focused on improving efficiency. Strategies include reusing LU factorizations over multiple iterations \cite{Abubakar_2008_FCS,Alkhalifah_2019_AEW} and reducing the number of factorizations per frequency via iterative refinement \cite{Aghamiry_2021_OEF}. Weighted norms for both data misfit and penalty terms have also been explored to improve convexity and convergence \cite{Rizzuti_2021_WaRIance,Lin_2023_FWR}.

Source signature estimation has been addressed by treating $f(\omega)$ as an additional unknown to be estimated during inversion \cite{Fang_2018_SEF,Aghamiry_2021_EES}. Alternatively, annihilator-based approaches \cite{Huang_2018_SEW,Symes_2020_FWI} eliminate explicit source estimation altogether and promote focused source extensions, enhancing stability and robustness to initial models. However, these methods are still based on penalty formulations, which may limit their convergence speed and final accuracy \cite{Gholami_2024_FWI}.

\subsection{Contributions.}
This work advances frequency-domain FWI within the framework of the multiplier method, a powerful approach for solving constrained optimization problems \cite{Gill_1981_PO,Bertsekas_1996_COL}. The multiplier method augments the classical Lagrangian with a quadratic penalty term, combining the stability of penalty methods with the rigorous constraint enforcement of Lagrange multipliers. When applied to FWI, the multiplier formulation can be expressed either in the source space \cite{Haber_2000_OTS,Aghamiry_2019_IWR} or the data space \cite{Rizzuti_2021_ADF,Gholami_2022_EFW,Gholami_2023_MWI}, depending on whether source or data multipliers are used.

We build on the source-space formulation and introduce a computationally efficient, robust, and practically applicable weighted multiplier waveform inversion approach that is suitable for both the frequency domain and the time domain implementations. Our contributions are threefold:

\begin{itemize}
\item[(i)] General weighted proximal-point Lagrangian formulation.
We propose a general weighted proximal-point Lagrangian objective function that introduces explicit weighting operators for the data misfit, model regularization, and multiplier terms. This formulation provides a unified and flexible framework for controlling the relative influence of each component of the objective function. Crucially, the classical Lagrangian, penalty, and augmented Lagrangian formulations arise as special cases of this general framework, allowing us to recover and compare existing methods within a single unifying theory. Properly designed weighting strategies enable us to improve both the convexity and conditioning of the problem, leading to faster and more stable convergence. 
   \item[(ii)] Spatially varying regularization for improved robustness and convexity.
   Classical multiplier-based FWI employs a scalar penalty parameter $\mu$ that uniformly weights wave-equation constraint violations \cite{Aghamiry_2019_IWR,Operto_2023_FWI,Li_2024_IAL}. We replace this scalar parameter with a spatially varying penalty function that increases with distance from the source. This weighting compensates for the natural decay of seismic energy, leading to more balanced enforcement of the wave equation. It brings the flexibility to focus the inversion to specific regions. This localization property allows to apply techniques like layer-stripping, where the model can be determined sequentially in small spatial subdomains. It result in improved convexity of the optimization landscape, a wider basin of attraction for the AL objective, and greater robustness to poor starting models and strong noise.
   \item[(iii)] Elimination of source-related practical constraints.
   By incorporating distance-dependent penalty weights, the proposed method removes the need for explicit source signature estimation and relaxes the requirement that sources be placed exactly on finite-difference grid points. This significantly improves modeling flexibility and makes the method more suitable for realistic acquisition geometries where source and receiver locations rarely coincide with grid nodes.
    \item[(iv)] Dual-space formulation for computational efficiency.
   We adopt a new dual formulation of the algorithm, where the focus is to estimate accurate Lagrange multipliers (dual variables) \cite{Aghazade_2025_FAF}. This eliminates explicit updates over model parameters, effectively framing the inversion as a source inverse problem. The resulting dual ADMM algorithms (e.g., Algorithms \ref{alg_dual} and \ref{alg_SI_dual}) requires only a few LU factorizations of the forward operator per frequency, while subsequent iterations rely solely on efficient forward–backward substitutions, making the method scalable for large-scale inversions.
\end{itemize}

Comprehensive numerical experiments on challenging synthetic benchmarks demonstrate that the proposed approach improves robustness, accelerates convergence, and enhances computational scalability compared with standard FWI methods.

\subsection{Outline of the paper.}
The remainder of the paper is organized as follows.
In Section~\ref{method}, we introduce the general weighted proximal-point Lagrangian objective function, establish its connection to classical formulations, and derive several numerical algorithms for its solution.
In Section~\ref{weightingmtx}, we design practical weighting functions, discuss strategies for selecting their free parameters, and demonstrate how they address key challenges in FWI.
Section~\ref{Interp} provides an interpretation of the proposed weighted multiplier-based formulation, offering insights into its improved robustness.
To further analyze its behavior, we present a series of numerical experiments and performance evaluations in Section~\ref{NumEx}. Finally, Section~\ref{Discussions} discusses current limitations and directions for future work, and Section~\ref{Conclusions} concludes the paper.

\section{Method.} \label{method}
The Lagrange multiplier methods, enhanced with proximal regularization \cite{Rockafellar_1976_PPA}, offers a robust algorithmic framework for constrained optimization problems such as \eqref{main_optim}. The proximal-point Lagrangian associated with \eqref{main_optim} is defined as follows:
\begin{equation} \label{PL}
\mathcal{L}_{\texttt{PL}}(\bu, \dual,\bm;\dual_k,\bm_k) = 
\frac{1}{2} \|\bP\bu-\bd\|_{\Wd}^2 
+ \langle \dual, \bold{A(m)u} - \bb \rangle 
+ \frac{1}{2}\|\bm-\bm_k\|_{\Wm}^2
- \frac{1}{2} \|\dual - \dual_k\|_{\Ws^{-1}}^2.
\end{equation} 
Here, the variable $\dual$ is the Lagrange multiplier (or dual variable), which simultaneously serves as the adjoint wavefield in the inversion process \cite{Plessix_2006_RAS} and as a source-extension term that drives the forward wavefield \cite{Symes_2020_WRI}. These two wavefields — forward and adjoint — constitute the fundamental building blocks of the gradient and Hessian information, making $\dual$ one of the most critical variables for updating the model and ultimately solving the inverse problem. $\bm_k$ and $\dualp$ are prior estimates of the model and multiplir, in the sense of proximal point methods \cite{Rockafellar_1976_PPA}. $\Wm$ and $\Ws$ are diagonal weighting matrices with positive weighting coefficients on the diagonal and zero elsewhere. The notation $\langle \cdot, \cdot \rangle$ denotes the inner product.

The quadratic regularization (proximal) terms in \eref{PL} play a crucial role in stabilizing and regularizing the inversion problem and has an important convexification effect. Removing these terms reduces the objective function to the standard Lagrangian formulation \cite{Haber_2000_OTS, Akcelik_2002_MNK}, which is often more difficult to optimize due to the non-smooth and potentially ill-conditioned nature of the Lagrangian term. The quadratic form of the proximal terms smooths the objective landscape, improves conditioning, and provides a natural mechanism for controlling step sizes. As a result, the proximal formulation generally exhibits better convergence properties and improved numerical stability compared with directly solving the original Lagrangian.

\subsection{Relationships between different methods.}
Several well-known objective functions can be derived from the proximal-point Lagrangian \eref{PL} by selecting appropriate values for the prior model $\bm_k$, prior multiplier $\dualp$, and weighting matrices $\Wm$ and $\Ws$. 

\textbf{The standard augmented Lagrangian function}. AL is obtained from $\mathcal{L}_{\texttt{PL}}$ by applying a variable projection \cite{Golub_2013_VPM,VanLeeuwen_2021_VPN} to eliminate $\dual$ from the optimization variables. Notably, $\mathcal{L}_{\texttt{PL}}$ is quadratic in $\dual$ and admits a closed-form solution:
\begin{equation} \label{lambda_opt}
    \dual(\bu,\bm;\dualp)=\dualp+\Ws[\bold{A(m)u} - \bb],
\end{equation}
which can be substituted back into \eqref{PL} to recover the classical form of the AL function, originally proposed for $\Wm=\bold{0}$ by Hestenes and Powell \cite{Hestenes_1969_MAG, Powell_1969_NLC}:
\begin{equation} \label{AL}
\mathcal{L}_{\texttt{AL}}(\bu,\bm;\dual_k,\bm_k) = 
\frac{1}{2} \|\bP\bu-\bd\|_{\Wd}^2 
+ \langle \dualp, \bold{A(m)u} - \bb \rangle 
+ \frac{1}{2} \|\bold{A(m)u} - \bb\|_{\Ws}^2+ \frac{1}{2}\|\bm-\bm_k\|_{\Wm}^2.
\end{equation}  
This function has been successfully applied to different forms of FWI \cite{Aghamiry_2019_IWR, Operto_2023_FWI,Gholami_2022_EFW} using homogeneous weighting matrices $\Wd, \Wm,$ and $\Ws$ as scale of the identity matrix.

\textbf{Partially reduced AL Function}. 
The AL function is also quadratic in $\bu$ and admits a closed-form solution:
\begin{equation}
    \bu(\bm;\dualp)=
        \left(\bP^{\!\top}\Wd\bP + \bA(\bm)^{\!\top}\Ws\bA(\bm)\right)^{-1}\!\!(\bP^{\!\top}\Wd\bd+ \bA(\bm)^{\!\top}\Ws\bb -\bA(\bm)^{\!\top}\dualp).\label{u}
\end{equation}
By substituting $\bu$ from \eqref{u} into \eref{AL} and simplifying, we obtain a reduced AL function \cite{Gholami_2022_EFW}:
\begin{equation} \label{RL}
\mathcal{L}_{\texttt{RL}}(\bm, \dualp;\bm_k) = 
\frac{1}{2} \Big\Vert \bd - \bS(\bm)[\bb -\Ws^{-1}\dualp]\Big\Vert^2_{\bQ(\bm)^{-1}}
+\frac{1}{2}\|\bm-\bm_k\|_{\Wm}^2-\|\dualp\|_{\Ws^{-1}}^2,
\end{equation} 
where $\bQ(\bm)=\bS(\bm)\Ws^{-1}\bS(\bm)^{\top}+\Wd^{-1}$.
This reduced formulation consists of three terms. The first is a weighted data misfit term, where the residual is defined as the classical data residual but evaluated with an extended source of the form $\bb - \Ws^{-1} \dualp$, with the multiplier $\dualp$ serving as the source extension. This term is defined by the model-dependent weighting matrix $\bQ(\bm)^{-1}$.
The second term is a regularization term which smooths the misfit term by forcing the model update to remaim close to the previous estimate. 
The third term is a regularization term that penalizes the multipliers via a weighted norm. This formulation, particularly in the case of zero multipliers and zero model weighting matrix $\Wm$, known as reduced penalty function, has been studied in \cite{Symes_2020_WRI, vanLeeuwen_2019_ANO}.


\textbf{The standard Lagrangian function} ($\Wm=\Ws^{-1}= \bold{0}$). In this case, the proximal terms vanish, and the objective reduces to the classical Lagrangian form:
    \begin{equation} \label{L}
\mathcal{L}(\bu, \dual,\bm) = 
\frac{1}{2} \|\bP\bu-\bd\|_{\Wd}^2 + \langle \dual, \bold{A(m)u} - \bb \rangle.
\end{equation}
This has been considered by \cite{Haber_2000_OTS,Akcelik_2002_MNK} with $\Wd$ as identity matrix.

    \textbf{The proximal-point penalty function} ($\dualp = \bold{0}$):
Setting the multiplier $\dualp$ to zero yields a regularized Lagrangian, forcing $\dual$ to be small \cite{Rizzuti_2021_ADF,Rizzuti_2021_WaRIance}:
    \begin{equation} \label{PP}
\mathcal{L}_{\texttt{PP}}(\bu, \dual,\bm;\bm_k) = 
\frac{1}{2} \|\bP\bu-\bd\|_{\Wd}^2 + \langle \dual, \bold{A(m)u} - \bb \rangle 
+\frac{1}{2}\|\bm-\bm_k\|_{\Wm}^2- \frac{1}{2} \|\dual\|_{\Ws^{-1}}^2.
\end{equation}
This has been considered in \cite{Rizzuti_2021_WaRIance} (for $\Wm=\bold{0}$) where the author suggest estimating $\Wd$ and $\Ws$ with low-rank stochastic approximations.

\textbf{The quadratic penalty function} ($\Wm=\bold{0}$, $\dualp = \bold{0}$, $\dual = \Ws[\bA(\bm)\bu - \bb]$):
This form results either by eliminating $\dual$ in \eqref{PP} through variable projection, or by substituting $\dualp = \bold{0}$ directly into \eref{AL}:
\begin{equation} \label{QP}
\mathcal{L}_{\texttt{P}}(\bu,\bm) = 
\frac{1}{2} \|\bP\bu-\bd\|_{\Wd}^2 + \frac{1}{2} \|\bold{A(m)u} - \bb\|_{\Ws}^2.
\end{equation}  
This approach has been considered in several studies \cite{Abubakar_2008_FCS, vanLeeuwen_2016_PMP,Operto_2023_FWI,Yang_2024_AAT}, typically assuming both $\Wd$ and $\Ws$ to be scaled identity matrices. In contrast, \cite{Huang_2018_SEW} uses a reformulation of \eref{QP} with $\Wd = \bold{I}$ but defines $\Ws$ as a diagonal matrix based on the distance from the source location, allowing to remove the source term $\bb$ from the objective function. \cite{Symes_2020_WRI,Bao_2023_RFW} further emphasizes that this action improves the conditioning of the inversion problem. 


\subsection{Optimization.} \label{subsec_optim}
The main computational challenge lies in solving the following min-max problem
\begin{subequations}
\label{main_AAO}
\begin{align} 
(\bu_{k+1}, \dual_{k+1},\bm_{k+1})&=  \arg \min_{\bm,\bu}\max_{\dual} \mathcal{L}_{\texttt{PL}}(\bu, \dual,\bm;\dual_k,\bm_k). \label{main_AAO_p}
\end{align}
\end{subequations}
Various approaches can be adopted for this step. A natural starting point for such nonlinear problems is Newton's method. This approach, which updates the model, wavefield, and multipliers simultaneously in an all-at-once manner, offers robust local convergence properties. This is done by performing an inner iterative loop, indexed by $l$, that updates $(\bu, \dual,\bm)$ simultaneously as:
\begin{equation}
    \bu^{l+1}=\bu^{l}+\delta\bu^{l}, \quad 
    \dual^{l+1}=\dual^{l}-\delta \dual^{l},\quad
        \bm^{l+1}=\bm^{l}+\delta\bm^{l}, 
\end{equation}
where the Newton direction $(\delta\bu^{l},\delta\dual^{l},\delta\bm^{l})$ satisfies the linear system
\begin{equation} 
\nabla^2 \mathcal{L}_{\texttt{PL}}(\bu^l, \dual^l,\bm^l;\dual_k,\bm_k)
    \begin{bmatrix}
        \delta\bu^{l}\\
        \delta\dual^{l} \\
        \delta\bm^{l}\\
    \end{bmatrix}=-
   \nabla \mathcal{L}_{\texttt{PL}}(\bu^l, \dual^l,\bm^l;\dual_k,\bm_k).
\end{equation}
The gradient and Hessian of $\mathcal{L}_{\texttt{PL}}$ are defined as
\begin{equation}\label{g_PL}
    \nabla \mathcal{L}_{\texttt{PL}}(\bu, \dual,\bm;\dual_k,\bm_k)
    =
        \begin{bmatrix}
        \bP^{\!\top}\Wd(\bP\bu-\bd)  + \bA(\bm)^{\!\top}\dual\\
        \bA(\bm)\bu-\bb -\Ws^{-1}(\dual-\dualp)\\
         \diag{\omega^2\bu}^{\top}\dual+\Wm(\bm-\bm_k)
    \end{bmatrix}
\end{equation}
and
\begin{equation} \label{H_PL}
    \nabla^2 \mathcal{L}_{\texttt{PL}}(\bu, \dual,\bm;\dual_k,\bm_k)
    =
     \begin{bmatrix}
        \bP^{\!\top}\Wd\bP  & \bA(\bm)^{\!\top} &  \diag{\omega^2\dual}\\
        \bA(\bm) &  -\Ws^{-1} &  \diag{\omega^2\bu} \\ 
         \diag{\omega^2\dual}^{\top} &  \diag{\omega^2\bu}^{\top} & \Wm
    \end{bmatrix}.
\end{equation}
This approach can be challenging for solving large-scale problems because the variables $\bu$ and $\dual$ are source dependent and must be kept in memory. The associated algorithm is summarized in Algorithm \ref{alg_full_Newton}.

\begin{algorithm}[H]
\caption{Basic Multiplier Method for finding a stationary point of \eref{PL} via the all-at-once Newton scheme.}
\label{alg_full_Newton}
\begin{algorithmic}[1]
\STATE \textbf{Input:} data $\bb$, $\bd$, frequency $\omega$, initial model $\bm_0$, initial multipliers $\dual_0=\bold{0}$
\FOR{$k = 0,1,2,\dots$ \textbf{until convergence}} 
    \STATE Initialize $(\bu^0,\bm^0,\dual^0) = (\bu_k,\bm_k,\dual_k)$
    \FOR{$l = 0,1,\dots,\texttt{maxit}-1$} 
        \STATE Compute gradient $\bg = \nabla \mathcal{L}_{\texttt{PL}}(\bu^l,\dual^l,\bm^l;\dual_k,\bm_k)$
        \STATE Compute Hessian $\bold{H} = \nabla^2 \mathcal{L}_{\texttt{PL}}(\bu^l,\dual^l,\bm^l;\dual_k,\bm_k)$
        \STATE Solve the linear system
        \[
        \bold{H}
        \begin{bmatrix}
            \delta \bu^l \\
            \delta \dual^l \\
            \delta \bm^l
        \end{bmatrix}
        = - \bg
        \]
        \STATE Update primal variables by descent:
        \[
        \bu^{l+1} = \bu^l + \delta \bu^l, 
        \quad
        \bm^{l+1} = \bm^l + \delta \bm^l
        \]
        \STATE Update dual variables by ascent:
        \[
        \dual^{l+1} = \dual^l - \delta \dual^l
        \]
        \IF{$\|\bg\| < \texttt{tol\_inner}$}
            \STATE \textbf{break}
        \ENDIF
    \ENDFOR
    \STATE Set $\bm_{k+1} = \bm^{\texttt{maxit}}$, \; $\dual_{k+1} = \dual^{\texttt{maxit}}$
    \IF{$\|\dual_{k+1}-\dual_k\| < \texttt{tol\_outer}$}
        \STATE \textbf{stop}
    \ENDIF
\ENDFOR
\end{algorithmic}
\end{algorithm}

\subsection{Reduction by eliminating $\dual$.} \label{subsec_red_dual}
To mitigate the memory and computational burden associated with the large $3\times 3$ Newton system of Algorithm \ref{alg_full_Newton}, particularly for large-scale FWI, a common strategy is to reduce the number of variables. 
The size of the Newton system above can be reduced to a $2 \times 2$ block system by applying variable projection to eliminate the primal multiplier $\dual$, using its analytical expression given in \eref{lambda_opt}. This reduction transforms the proximal Lagrangian $\mathcal{L}_{\texttt{PL}}$ into the AL function $\mathcal{L}_{\texttt{AL}}$, and the corresponding Newton direction is then computed for the remaining variables $\bu$ and $\bm$ as follows:
\begin{align}
     \nabla^2 \mathcal{L}_{\texttt{AL}}(\bu^{l}, \bm^{l};\dual_k,\bm_k)
     \begin{bmatrix}
        \delta\bu^{l}\\
        \delta\bm^{l}\\
    \end{bmatrix}
    =- \nabla \mathcal{L}_{\texttt{AL}}(\bu^{l},\bm^{l};\dual_k,\bm_k).
\end{align}
The gradient $\nabla \mathcal{L}_{\texttt{AL}}$ corresponds to the components of the full gradient in \eref{g_PL}, evaluated at the projected variables:
\begin{equation} \label{g_AL}
    \nabla \mathcal{L}_{\texttt{AL}}(\bu,\bm;\dual_k,\bm_k)
    =
    \begin{bmatrix}
        \bP^{\!\top}\Wd(\bP\bu-\bd)  + \bA(\bm)^{\!\top}\dual(\bm,\bu,\dual_k)\\
         \diag{\omega^2\bu}^{\top}\dual(\bm,\bu,\dual_k) + \Wm\bm_k
    \end{bmatrix}.
\end{equation}
The Hessian is equal to the Schur complement of the full Hessian in \eref{H_PL} with respect to the block (2, 2): %
\begin{equation} \label{H_AL}
    \nabla^2 \mathcal{L}_{\texttt{AL}}(\bu,\bm;\dual_k,\bm_k)
    =
    \begin{bmatrix}
        \bP^{\!\top}\Wd\bP+\bA(\bm)^{\!\top}\Ws\bA(\bm)  &
        \bold{K}(\bm,\bu,\dual_k)^{\top} \\
        \bold{K}(\bm,\bu,\dual_k)&
         \diag{\omega^2\bu}^{\top}\Ws\diag{\omega^2\bu}
    \end{bmatrix}
\end{equation}
where $\bold{K}(\bu,\bm;\dual_k)= \diag{\omega^2\dual(\bu,\bm;\dual_k)}^{\top} +  \diag{\omega^2\bu}^{\top}\Ws\bA(\bm)$. The associated algorithm is summarized in Algorithm \ref{alg_AL_Newton}

\begin{algorithm}[H]
\caption{Basic Multiplier Method for finding a stationary point of the augmented Lagrangian \eref{AL} via the all-at-once Newton scheme}
\label{alg_AL_Newton} 
\begin{algorithmic}[1]
\STATE \textbf{Input:} data $\bb$, $\bd$, frequency $\omega$, initial model $\bm_0$, initial multipliers $\dual_0=\bold{0}$
\FOR{$k = 0,1,2,\dots$ \textbf{until convergence}} 
    \STATE Initialize $(\bu^0,\bm^0) = (\bu_k,\bm_k)$
    \FOR{$l = 0,1,\dots,\texttt{maxit}-1$} 
        \STATE Compute gradient $\bg = \nabla \mathcal{L}_{\texttt{AL}}(\bu^l,\bm^l;\dual_k,\bm_k)$
        \STATE Compute Hessian $\bold{H} = \nabla^2 \mathcal{L}_{\texttt{AL}}(\bu^l,\bm^l;\dual_k,\bm_k)$
        \STATE Solve the linear system
        \[
        \bold{H}
        \begin{bmatrix}
            \delta \bu^l \\
            \delta \bm^l
        \end{bmatrix}
        = - \bg
        \]
        \STATE Update variables by descent:
        \[
        \bu^{l+1} = \bu^l + \delta \bu^l, 
        \quad
        \bm^{l+1} = \bm^l + \delta \bm^l
        \]
        \IF{$\|\bg\| < \texttt{tol\_inner}$}
            \STATE \textbf{break}
        \ENDIF
    \ENDFOR
    \STATE Update multipliers:
    \[
    \dual_{k+1} = \dual_k + \Ws \big[\bA(\bm^{\texttt{maxit}}) \bu^{\texttt{maxit}} - \bb \big]
    \]
    \STATE Update model for next iteration:
    \[
    \bm_{k+1} = \bm^{\texttt{maxit}}, \quad \bu_{k+1} = \bu^{\texttt{maxit}}
    \]
    \IF{$\|\dual_{k+1}-\dual_k\| < \texttt{tol\_outer}$}
        \STATE \textbf{stop}
    \ENDIF
\ENDFOR
\end{algorithmic}
\end{algorithm}

\subsection{Reduction by eliminating $\dual$ and $\bu$.} \label{subsec_red}
Building on the strategy of variable elimination, we may go one step further and reduce the AL function to an objective function that involves only the model parameter $\bm$, given by \eref{RL}. This ``reduced-space" approach is attractive for its conceptual simplicity, as it directly optimizes the primary unknown variable. In this case, the model update $\delta \bm^l$ is given by
$\nabla^2 \mathcal{L}_{\texttt{RL}}(\bm^l;\dual_k,\bm_k)\delta \bm^l=-\nabla \mathcal{L}_{\texttt{RL}}(\bm^l;\dual_k,\bm_k)$ where
\begin{equation} \label{g_RL}
    \nabla \mathcal{L}_{\texttt{RL}}(\bm;\dual_k,\bm_k)
    = \diag{\omega^2\bu(\bm;\dual_k)}^{\top}\dual(\bu(\bm;\dual_k),\bm;\dual_k)+\Wm\bm_k,
\end{equation}
and the corresponding Hessian is equal to the Schur complement of the AL Hessian in \eref{H_AL} with respect to the block (1, 1).
\begin{equation} \label{H_RL}
    \nabla^2 \mathcal{L}_{\texttt{RL}}(\bm;\dual_k,\bm_k)
    =
      \diag{\omega^2\bu}^{\top}\Ws\diag{\omega^2\bu}
    - \bK \left[\bP^{\!\top}\Wd\bP+\bA(\bm)^{\!\top}\Ws\bA(\bm)\right]^{-1} \bK^{\top}
\end{equation}
where $\bu\equiv \bu(\bm;\dualp)$ \eref{u}, $\bK\equiv \bK(\bu,\bm;\dualp)$.
Unlike the Hessians $\nabla^2 \mathcal{L}_{\texttt{PL}}$ and $\nabla^2 \mathcal{L}_{\texttt{AL}}$, which are sparse, the Hessian $\nabla^2 \mathcal{L}_{\texttt{RL}}$ is dense, making the model update significantly more computationally demanding \cite{vanLeeuwen_2016_PMP}.
The simple implementation of this approach is summarized in Algorithm \ref{alg_RL_Newton}.
\begin{algorithm}[H]
\caption{Basic Multiplier Method for finding a stationary point of the reduced proximal Lagrangian \eref{RL} via Newton's method}
\label{alg_RL_Newton}  
\begin{algorithmic}[1]
\STATE \textbf{Input:} data $\bb$, $\bd$, frequency $\omega$, initial multipliers $\dual_0$, initial model $\bm_0$, initial wavefield $\bu_0$
\FOR{$k = 0,1,2,\dots$ \textbf{until convergence}} 
    \STATE Initialize $\bm^0 = \bm_k$
    \FOR{$l = 0,1,\dots,\texttt{maxit}-1$} 
        \STATE Compute gradient $\bg = \nabla \mathcal{L}_{\texttt{RL}}(\bm^l;\dual_k,\bm_k)$
        \STATE Compute Hessian $\bold{H} = \nabla^2 \mathcal{L}_{\texttt{RL}}(\bm^l;\dual_k,\bm_k)$
        \STATE Solve the linear system
        \[
        \bold{H} \, \delta \bm^l = - \bg
        \]
        \STATE Update model:
        \[
        \bm^{l+1} = \bm^l + \delta \bm^l
        \]
        \IF{$\|\bg\| < \texttt{tol\_inner}$}
            \STATE \textbf{break}
        \ENDIF
    \ENDFOR
    \STATE Set updated model: $\bm_{k+1} = \bm^{\texttt{maxit}}$
    \STATE Solve for the wavefield $\bu_{k+1}$:
    \[
    \bu_{k+1} = 
    \left(\bP^{\!\top}\Wd\bP + \bA(\bm_{k+1})^{\!\top}\Ws\bA(\bm_{k+1})\right)^{-1} 
    \big(\bP^{\!\top}\Wd \bd + \bA(\bm_{k+1})^{\!\top} [\Ws \bb - \dual_k]\big)
    \]
    \STATE Update multipliers:
    \[
    \dual_{k+1} = \dual_k + \Ws \big[\bA(\bm_{k+1}) \bu_{k+1} - \bb \big]
    \]
    \IF{$\|\dual_{k+1}-\dual_k\| < \texttt{tol\_outer}$}
        \STATE \textbf{stop}
    \ENDIF
\ENDFOR
\end{algorithmic}
\end{algorithm}

\subsection{Alternating direction methods.}
Building on the fundamental Lagrangian formulations and their Newton-based solutions presented in Sections \ref{subsec_optim}–\ref{subsec_red}, we acknowledge that while these methods provide robust convergence in principle, their direct application to large-scale FWI problems is often hampered by significant computational costs. Specifically, the repeated calculation and factorization of large, model-dependent Hessian matrices in each inner iteration (as observed in Algorithms \ref{alg_full_Newton}, \ref{alg_AL_Newton}  and \ref{alg_RL_Newton}) can be prohibitively expensive due to high memory and processing demands. To overcome these challenges and achieve practical scalability for FWI, we now turn to Alternating Direction Methods (ADMs). These methods strategically decompose the complex, coupled optimization problem into a sequence of simpler subproblems. By judiciously fixing certain variables during inner iteration loops, ADMs allow for highly efficient updates that avoid the frequent re-factorizations characteristic of full Newton schemes, thereby significantly reducing computational demand and memory footprint. 
The optimization problem \eref{main_AAO_p} can be decomposed into coupled saddle-point subproblems solved in sequence. The first subproblem is a saddle-point system in $(\bu, \dual)$, with the model $\bm$ held fixed. Once $\bu$ and $\dual$ are updated, we switch to a second saddle-point system in $(\bm,\dual)$, keeping $\bu$ fixed. This approach effectively solves the $3\times 3$ block nonlinear system as a sequence of two coupled $2\times 2$ systems, leading to the following iteration:
\begin{subequations}
\label{saddle}
\begin{align} 
(\bu_{k+1}, \dual_{k+\frac{1}{2}})&=  \argminimax_{\bu,\dual} \mathcal{L}_{\texttt{PL}}(\bu, \dual,\bm_k;\dual_k,\bm_k) \label{saddle_up}\\
(\bm_{k+1},\dual_{k+1})&=  \argminimax_{\bm,\dual} \mathcal{L}_{\texttt{PL}}(\bu_{k+1}, \dual,\bm;\dual_k,\bm_k) \label{saddle_mp}
\end{align}
\end{subequations}
where $(\bu_{k+1}, \dual_{k+\frac{1}{2}})$ is the saddle point of $\mathcal{L}_{\texttt{PL}}$ for a fixed $\bm_k$ and $\dual_k$ and
 $(\bm_{k+1},\dual_{k+1})$ is the saddle point of $\mathcal{L}_{\texttt{PL}}$ for the fixed values of $\bu_{k+1}$ and $\dual_k$.
In what follows, we explain how these subproblems can be solved efficiently.

\subsubsection{Solving \eref{saddle_up} (joint update of $\bu$ and $\dual$).}
Computing the partial derivative of $\mathcal{L}_{\texttt{PL}}$ with respect to $\bu$ and $\dual$ and setting the resulting equations equal to zero leads to the following saddle point system:
\begin{equation}\label{saddle}
\underbrace{
    \begin{bmatrix}
        \bP^{\!\top}\Wd\bP & \bA(\bm)^{\!\top}\\
        \bA(\bm) & - \Ws^{-1}
    \end{bmatrix}}_{\bM}
        \begin{bmatrix}
        \bu\\
        \dual
    \end{bmatrix}
    =
        \begin{bmatrix}
        \bP^{\!\top}\Wd\bd \\
        \bb-\Ws^{-1}\dualp
    \end{bmatrix}.
\end{equation}
The subindex $k$ is removed for simplicity. Thanks to the proximal-point regularization, the nonzero (2, 2) block in the coefficient matrix stabilizes the saddle-point system. In this formulation, all blocks are nonsingular except for the (1, 1) block, which is singular. This block structure gives rise to three distinct sets of solution formulas, each suitable for different scenarios. \cite{Zhang_2006_SCA} provides a complete study of saddle point system solutions.
\begin{itemize}

\item[(i)] \textbf{Using the Schur complement of $\bA^{\!\top}$.}
The Schur complement of $(1,~2)$ block, $\bA^{\!\top}$, is defined as
   \begin{equation}
      \mathcal{S}_{12}^\bM= \left[\bA+ \Ws^{-1}\bA^{-\!\top}\bP^{\!\top}\Wd\bP\right]
      =\Ws^{-1}\bA^{-\!\top}\left[\bA^{\top}\Ws\bA+ \bP^{\!\top}\Wd\bP\right].
   \end{equation}
   Then, using $\mathcal{S}_{12}^\bM$, we obtain the following inverse formula:
   \begin{equation} \label{M12}
    \bM^{-1}
    =
    \begin{bmatrix}
        (\mathcal{S}_{12}^\bM)^{-1}\Ws^{-1}\bA^{-\!\top} & (\mathcal{S}_{12}^\bM)^{-1}\\
        \bA^{-\!\top}-\bA^{-\!\top}\bP^{\!\top}\Wd\bP(\mathcal{S}_{12}^\bM)^{-1}\Ws^{-1}\bA^{-\!\top} & -\bA^{-\!\top}\bP^{\!\top}\Wd\bP(\mathcal{S}_{12}^\bM)^{-1}
    \end{bmatrix}.
\end{equation}
If we multiply this inverse formula to both sides of \eref{saddle}, after simplifications, we obtain the following solution:
\begin{subequations}
\begin{align}
\bu_{k+1}&=
        \left[\bP^{\!\top}\Wd\bP + \bA^{\!\top}\Ws\bA\right]^{-1}[\bP^{\!\top}\Wd\bd+ \bA^{\!\top}\Ws\bb -\bA^{\!\top}\dualp].\label{u12}\\
\dual_{k+\frac{1}{2}}
        &=\bA^{-\!\top}\bP^{\!\top}\Wd[\bd-\bP\bu]. \label{nu12}
\end{align}
\end{subequations} 
In this formulation, we can compute the wavefield by \eref{u12} followed by computing the multiplier using \eref{nu12}.

\item[(ii)] \textbf{Using the Schur complement of $\bA$.}
 The Schur complement of $(2,~1)$ block, $\bA$, is given by
 \begin{equation}
\mathcal{S}_{21}^\bM=\left[\bA^{\!\top}+\bP^{\!\top}\Wd\bP\bA^{-1}\Ws^{-1}\right]
=\bA^{\!\top}\left[\bold{I}+ \bA^{\!-\top}\bP^{\!\top}\Wd\bP\bA^{-1}\Ws^{-1}\right].
 \end{equation}
This gives
\begin{equation} \label{M21}
    \bM^{-1}
    =
    \begin{bmatrix}
        \bA^{-1}\Ws^{-1} (\mathcal{S}_{21}^\bM)^{-1} &
        \bA^{-1}-\bA^{-1}\Ws^{-1}(\mathcal{S}_{21}^\bM)^{-1}\bP^{\!\top}\Wd\bP\bA^{-1}\\
        (\mathcal{S}_{21}^\bM)^{-1} & -(\mathcal{S}_{21}^\bM)^{-1}\bP^{\!\top}\Wd\bP\bA^{-1}
    \end{bmatrix},
\end{equation}
which leads to another solution form
\begin{subequations}
\begin{align}
 \dual_{k+\frac{1}{2}}
  &=\bA^{-\!\top}\bP^{\!\top}\left[\Wd^{-1}+ \bP\bA^{-1}\Ws^{-1}\bA^{\!-\top}\bP^{\!\top}\right]^{-1}\left[\bd - \bP\bA^{-1}[\bb-\Ws^{-1}\dualp]\right].\label{nu21}\\
    \bu_{k+1}
  &= \bA^{-1}\left[\bb+\Ws^{-1}[\dual_{k+\frac{1}{2}}-\dual_k]\right]. \label{u21}
\end{align}
\end{subequations}

\item[(iii)] \textbf{Using the Schur complement of $-\Ws^{-1}$.}
  The third option is to use the Schur complement of $(2,~2)$ block, $-\Ws^{-1}$, defined as $\mathcal{S}_{22}^\bM=\bP^{\!\top}\Wd\bP +  \bA^{\!\top}\Ws\bA$. Using $\mathcal{S}_{22}^\bM$ we get a different inverse matrix form of $\bM^{-1}$ as
\begin{equation} \label{M22}
    \bM^{-1}
    =
    \begin{bmatrix}
        (\mathcal{S}_{22}^\bM)^{-1} &  (\mathcal{S}_{22}^\bM)^{-1}\bA^{\!\top}\Ws\\
         \Ws\bA(\mathcal{S}_{22}^\bM)^{-1} & -\Ws +\Ws\bA(\mathcal{S}_{22}^\bM)^{-1}\bA^{\!\top}\Ws
    \end{bmatrix}.
\end{equation}
Multiplying this inverse formula to both sides of \eref{saddle}, after simplifications,  gives the following solution:
\begin{subequations}
\label{uv_22_w}
\begin{align}
\bu_{k+1}
&=
        \left[\bP^{\!\top}\Wd\bP + \bA^{\!\top}\Ws\bA\right]^{-1}[\bP^{\!\top}\Wd\bd+ \bA^{\!\top}\Ws\bb -\bA^{\!\top}\dualp].\label{u22}\\
\dual_{k+\frac{1}{2}}
        &= \Ws\bA\bu_{k+1}-\Ws\bb+\dualp. \label{nu22}
\end{align}
\end{subequations} 

\end{itemize}
The above analysis allows a flexible framework for estimation of the forward and adjoint wavefields (the half-step multiplier update). Dependent on which one of these wavefields plays the role of the primary variable, two different but equivalent algorithms are possible \cite{Gholami_2024_FWI}:

\begin{itemize}
     \item[(i)] \textbf{Multiplier-oriented approach.} 
     This formulation is based on the matrix inversion identity \eref{M21}, where the adjoint wavefield is treated as the primary variable. At each iteration, the $\dual$ is first computed using \eref{nu21}, followed by computation of the wavefield via \eref{u21}. 
\begin{subequations}
\label{MOAL}
\begin{align}
\dual_{k+\frac{1}{2}}&=
      \bA^{-\!\top}\bP^{\top}\!\!\left[\bP\bA^{-1}\Ws^{-1}\!\!\bA^{-\!\top}\bP^{\top}+\Wd^{-1}\right]^{-1}[\bd -\bP\bA^{-1}[\bb-\Ws^{-1}\dualp]]\\
\bu_{k+1}&=\bA^{-1}\left[\bb+\Ws^{-1}(\dual_{k+\frac{1}{2}}-\dualp)\right].
\end{align}
\end{subequations}
     Here, the main computational challenge lies in estimating the $\dual_{k+\frac{1}{2}}$, which involves inverting a dense data-space Hessian of the form $(\Wd^{-1}+ \bP\bA^{-1}\Ws^{-1}\bA^{-\!\top}\bP^{\top})\in \mathbb{C}^{N_r\times N_r}$. This approach is particularly well suited to acquisition scenarios involving source-dependent sampling matrices $\bP$—as encountered in streamer acquisitions—and source-dependent weighting matrices $\Wd$ and $\Ws$, both of which are considered in this paper. Moreover, since both $\dual_{k+\frac{1}{2}}$ and $\bu_{k+1}$ are governed by forward and adjoint wave equations, this strategy is also compatible with time-domain formulations that employ time-stepping solvers \cite{Gholami_2022_EFW,Rizzuti_2021_ADF}.
        \item[(ii)] \textbf{Wavefield-oriented approach.} 
    This formulation is based on the matrix inverse formulae \eref{M12} or \eref{M22}, where the forward wavefield is treated as the primary variable within the update algorithm. At each iteration, $\bu_{k+1}$ is first updated using \eref{u12} (identical to \eref{u22}), followed by the computation of the $\dual_{k+\frac{1}{2}}$ via \eref{nu12} or \eref{nu22}. By using \eref{nu22} which may be simpler to compute as long as the $\bu_{k+1}$ is available, we get 
\begin{subequations}
\label{WOAL}
\begin{align}
\bu_{k+1}&=
        \left[\bP^{\!\top}\Wd\bP +  \bA^{\!\top}\Ws\bA\right]^{-1}[\bP^{\!\top}\Wd\bd+ \bA^{\!\top}\Ws\bb -\bA^{\!\top}\dualp],\\
\dual_{k+\frac{1}{2}}&= \Ws[\bA\bu_{k+1}-\bb]+\dualp. 
\end{align}
\end{subequations} 
    Here, the dominant computational cost arises from solving for the forward wavefield, which requires inverting a sparse Hessian operator $\left(\bP^{\!\top}\Wd\bP +  \bA^{\!\top}\Ws\bA\right)\in \mathbb{C}^{N\times N}$ defined in the source space.  This approach is generally less practical for time-domain formulations \cite{Wang_2016_FIR, Rizzuti_2021_ADF}, and can also become computationally prohibitive in frequency-domain settings when $\bP$, $\Wd$, and $\Ws$ are source-dependent. In such cases, computing an LU factorization of the Hessian for each source may be infeasible due to excessive memory and computational demands.
\end{itemize}
Therefore, in the following, we focus on using the multiplier-oriented formulation to solve the saddle-point system and compute the wavefields $\dual_{k+\frac{1}{2}}$ and $\bu_{k+1}$. However, the presented formulations and results remain valid for both the multiplier- and wavefield-oriented approaches.

\subsubsection{Solving \eref{saddle_mp} (joint update of $\bm$ and $\dual$).}
Computing the partial derivative of $\mathcal{L}_{\texttt{PL}}$ with respect to $\bm$ and $\dual$ and setting the resulting equations equal to zero leads to the following saddle point system:
\begin{equation}\label{saddle_md}
\underbrace{
    \begin{bmatrix}
        -\Ws^{-1}& \bL_{k} \\
        \bL_{k}^{\top} & \Wm 
    \end{bmatrix}
    }_{\bN}
        \begin{bmatrix}
        \dual_{k+1}\\
        \bm_{k+1}
    \end{bmatrix}
    =
     \begin{bmatrix}
        \bb-\Ws^{-1}\dual_k-\bold{\Delta}\bu_{k+1}\\
        \Wm\bm_k 
    \end{bmatrix}
    =
        \begin{bmatrix}
        \bL_k\bm_k-\Ws^{-1}\dual_{k+\frac{1}{2}}\\
        \Wm\bm_k
    \end{bmatrix}.
\end{equation}
where $\bL_k=\diag{\omega^2 \bu_{k+1}}$.
The solution to this system may be computed in different equivalent forms, associated to the Schur complements of the diagonal blocks.
\begin{itemize}
    \item[(i)] \textbf{Using the Schur complement of $-\Ws^{-1}$.}
    The Schur complement of $(1,~1)$ block, $-\Ws^{-1}$, is given by $\mathcal{S}_{11}^\bN=\Wm + \bL_k^{\top}\Ws \bL_k$. Using $\mathcal{S}_{11}^\bN$ we get an inverse matrix form of $\bM^{-1}$ as
\begin{equation}
    \bN^{-1}=
    \begin{bmatrix}
    -\Ws+\Ws\bL_k(\mathcal{S}_{11}^\bN)^{-1} \bL_k^{\top}\Ws 
      & \Ws\bL_k (\mathcal{S}_{11}^\bN)^{-1} \\
        (\mathcal{S}_{11}^\bN)^{-1}\bL_k^{\top}\Ws  & (\mathcal{S}_{11}^\bN)^{-1}
        
    \end{bmatrix}
\end{equation}
This inverse formula gives the following solution in terms of the estimate $\dual_{k+\frac{1}{2}}$ from solving \eref{saddle_up}.
\begin{subequations}
\label{MP}
\begin{align} 
\bm_{k+1} &=\bm_k-\left[\bL_k^{\top}\Ws\bL_k+\Wm\right]^{-1} \bL_k^{\top}\dual_{k+\frac{1}{2}}\\
\dual_{k+1} &= \dual_k+ \Ws[\bA(\bm_{k+1})\bu_{k+1}-\bb].
\end{align}
\end{subequations}
\item[(ii)] \textbf{Using the Schur complement of $\Wm$.}
The Schur complement of $\Wm$ is given by $-\mathcal{S}_{22}^\bN=\Ws^{-1} + \bL_k^{\top}\Wm^{-1} \bL_k$. Using $\mathcal{S}_{22}^\bN$ we get
\begin{equation}
    \bN^{-1}=
    \begin{bmatrix}
    -(\mathcal{S}_{22}^\bN)^{-1} & (\mathcal{S}_{22}^\bN)^{-1}\bL_k\Wm^{-1}\\
    \Wm^{-1}\bL_k^{\top}(\mathcal{S}_{22}^\bN)^{-1}& 
    \Wm^{-1}-\Wm^{-1}\bL_k^{\top} (\mathcal{S}_{22}^\bN)^{-1} \bL_k \Wm^{-1} 
    \end{bmatrix}
\end{equation}
which leades to the following solution
\begin{subequations}
\label{PM}
\begin{align} 
    \dual_{k+1} &=\dual_{k+\frac{1}{2}}-\Ws \bL_k \left[\Wm + \bL_k^{\top}\Ws \bL_k\right]^{-1}\bL_k^{\top}\dual_{k+\frac{1}{2}}\\
   \bm_{k+1}&=\bm_k -\Wm^{-1}\bL_k^{\top}\dual_{k+1}
\end{align}
\end{subequations}
\end{itemize}



\subsubsection{Algorithms.}
Different algorithms can be derived by applying the alternating direction method to sequentially solve \eref{saddle_up} and \eref{saddle_mp}. The solution of \eref{saddle_up} can be expressed in either the multiplier-oriented form \eref{MOAL} or the wavefield-oriented form \eref{WOAL}. Likewise, for \eref{saddle_mp}, one may adopt either the \eref{MP} or the \eref{PM} formulation. Combining these choices yields four distinct but equivalent algorithms. In what follows, however, we adopt the multiplier-oriented form \eref{MOAL} together with the \eref{MP} formulation, which leads to the following iterations:
\begin{subequations}
\label{Scaled_MWI}
\begin{align}
\dual_{k+\frac{1}{2}}&=
      \bold{S}(\bm_k)^{\!\top}\left[\Wd^{-1}+ \bold{S}(\bm_k)\Ws^{-1}\bold{S}(\bm_k)^{\!\top}\right]^{-1}(\bd - \bold{S}(\bm_k)\bb +\bold{S}(\bm_k)\Ws^{-1}\dual_k) \label{Scaled_MWI_1}\\
\bu_{k+1}&=\bA(\bm_k)^{-1}\left[\bb+\Ws^{-1}(\dual_{k+\frac{1}{2}}-\dual_k)\right]\label{Scaled_MWI_2}\\
   \bm_{k+1}
   &=\bm_k- \left[\diag{\omega^2\bu_{k+1}}^{\top}\Ws\diag{\omega^2\bu_{k+1}}+\Wm\right]^{-1}\diag{\omega^2\bu_{k+1}}^{\top}\dual_{k+\frac{1}{2}}\label{Scaled_MWI_3}\\
   \dual_{k+1}&= \dual_k + \Ws [\bA(\bm_{k+1}) \bu_{k+1} - \bb].  \label{Scaled_MWI_4}
\end{align}
\end{subequations}
%
where $\bold{S}(\bm)=\bP\bA(\bm)^{-1}$ is the forward operator.

The iteration defined in \eref{Scaled_MWI} consists of four coupled equations for solving the proximal-point Lagrangian function in \eref{PL}. This iteration can be implemented in different algorithmic configurations within the framework of the Alternating Direction Method of Multipliers (ADMM) \cite{Gabay_1976_ADA,Boyd_2011_DOS}. Depending on how frequently the prior model $\bm_k$ and the multiplier 
$\dual_k$ are updated, three different ADMM variants can be distinguished to solve the optimization problem.

\begin{itemize}
    \item[(i)] \textbf{ADMM with frequent primal updates.} In this variant, the coupled saddle-point problems in \eref{saddle} are solved repeatedly to minimize the proximal-point Lagrangian while keeping the multiplier $\dual_k$ fixed. This results in a nested iteration structure (Algorithm~\ref{alg1}). At each outer iteration (indexed by $k$), the wavefield $\bu_l$, the half-step multiplier $\dual_{l+\tfrac{1}{2}}$, and the model $\bm_l$ are updated in an inner loop (indexed by $l$), with $\dual_k$ held fixed. The inner loop is performed for a prescribed number of iterations $\texttt{maxit}$ or until convergence. Once the inner loop terminates, the multiplier $\dual_{k+1}$ is updated using the final wavefield and model obtained from the inner loop.
    \item[(ii)] \textbf{ADMM.} 
    Here, only a single inner iteration is performed per outer step, so that the model, wavefield, and multiplier are updated once at each iteration $k$. In practice, this means that $\dual_{k+\frac12}$, $\bu_{k+1}$, and $\bm_{k+1}$ are computed sequentially from the current state $(\bm_k,\dual_k)$, after which the multiplier is updated to $\dual_{k+1}$. This synchronous update scheme avoids nested loops and typically is the most traditional form of ADMM, as summarized in Algorithm~\ref{alg2} \cite{Aghamiry_2019_IWR, Operto_2023_FWI,Gholami_2022_EFW}.
\item[(iii)] \textbf{ADMM with frequent dual updates.} In this variant, the model $\bm$ is held fixed during the inner iterations, while the wavefield $\bu$ and the multiplier $\dual$ are updated repeatedly. At each inner iteration $l$, a tentative model increment $\delta \bm_{\,l+1}$ is computed based on the current wavefield and half-step multiplier $\dual_{\,l+1/2}$, but the model itself is not immediately updated. The multiplier is then updated using the tentative model $\bm + \delta\bm_{\,l+1}$, allowing it to respond promptly to changes in the model direction. After completing the prescribed number of inner iterations $\texttt{maxit}$, the model is updated once using the final increment $\delta\bm_{\texttt{maxit}}$. The corresponding steps are summarized in Algorithm~\ref{alg_dual}. In this variant, rather than updating the model immediately, the model corrections are accumulated across the inner iterations. Once the inner loop converges or reaches a predefined iteration limit $\texttt{maxit}$, the model is updated in a single step using the accumulated correction.  This strategy enhances the accuracy of the forward-adjoint wavefields used in the model update, potentially improving the overall efficiency of the algorithm. Moreover, because the model remains fixed throughout the inner loop, the forward and adjoint wavefields can be refined using a single LU factorization of the forward operator associated with the current model, thereby maintaining computational efficiency.

\end{itemize}

\begin{algorithm}[H]
\caption{Primal Multiplier Waveform Inversion}
\label{alg1}
\begin{algorithmic}[1]
\STATE \textbf{Input:} $\bb, \bd, \bm_0, \omega, \dual_0 = \boldsymbol{0}$
\FOR{$k = 0,1,2,\dots$ until convergence} 
\STATE Set initial guess $\bm_0 = \bm_k$
   \FOR{$l = 0$ to $\texttt{maxit}-1$}
       \STATE $\displaystyle 
       \dual_{\,l+\frac{1}{2}} =
           \bold{S}(\bm_l)^{\!\top}
           \Big[\bold{S}(\bm_l)\Ws^{-1}\bold{S}(\bm_l)^{\!\top}+\Wd^{-1}\Big]^{-1}
           \Big(\bd - \bold{S}(\bm_l)\bb + \bold{S}(\bm_l)\Ws^{-1}\dual_k\Big)$
       \STATE $\displaystyle 
       \bu_{\,l+1} = \bA(\bm_l)^{-1} 
           \Big[ \bb + \Ws^{-1} (\dual_{\,l+\frac12} - \dual_k) \Big]$
       \STATE $\displaystyle 
       \bm_{\,l+1} = \bm_l - \left[\diag{\omega^2\bu_{\,l+1}}^{\top}\Ws\diag{\omega^2\bu_{\,l+1}}+\Wm\right]^{-1}\diag{\omega^2\bu_{\,l+1}}^{\top}\dual_{\,l+\frac12}$
   \ENDFOR
   \STATE $\displaystyle 
   \dual_{k+1} = \dual_k + \Ws \Big[ \bA(\bm_{\,\texttt{maxit}}) \bu_{\,\texttt{maxit}} - \bb \Big]$
   \STATE $\bm_{k+1} = \bm_{\texttt{maxit}}$
       \IF{$\|\dual_{k+1}-\dual_k\| < \texttt{tol\_outer}$}
        \STATE \textbf{stop}
    \ENDIF
\ENDFOR
\STATE \textbf{Output:} $\bm_{k+1}$
\end{algorithmic}
\end{algorithm}
\begin{algorithm}[H]
\caption{Synchronous ADMM Waveform Inversion}
\label{alg2}
\begin{algorithmic}[1]
\STATE \textbf{Input:} $\bb, \bd, \bm_0, \omega, \dual_0 = \boldsymbol{0}$
\FOR{$k = 0,1,2,\dots$ until convergence} 
   \STATE $\displaystyle 
       \dual_{\,k+\frac{1}{2}} =
           \bold{S}(\bm_k)^{\!\top}
           \Big[\bold{S}(\bm_k)\Ws^{-1}\bold{S}(\bm_k)^{\!\top}+\Wd^{-1}\Big]^{-1}
           \Big(\bd - \bold{S}(\bm_k)\bb + \bold{S}(\bm_k)\Ws^{-1}\dual_k\Big)$
   \STATE $\displaystyle 
       \bu_{\,k+1} = \bA(\bm_k)^{-1} 
           \Big[ \bb + \Ws^{-1} (\dual_{\,k+\frac12} - \dual_k) \Big]$
   \STATE $\displaystyle 
       \bm_{\,k+1} = \bm_k - \left[\diag{\omega^2\bu_{\,k+1}}^{\top}\Ws\diag{\omega^2\bu_{\,k+1}}+\Wm\right]^{-1}\diag{\omega^2\bu_{\,k+1}}^{\top}\dual_{\,k+\frac12}$
   \STATE $\displaystyle 
       \dual_{\,k+1} = \dual_k + \Ws \Big[ \bA(\bm_{\,k+1}) \bu_{\,k+1} - \bb \Big]$
    \IF{$\|\dual_{k+1}-\dual_k\| < \texttt{tol\_outer}$}
        \STATE \textbf{stop}
    \ENDIF
\ENDFOR
\STATE \textbf{Output:} $\bm_{k+1}$
\end{algorithmic}
\end{algorithm}

\begin{algorithm}[H]
\caption{Dual Multiplier Waveform Inversion}
\label{alg_dual}
\begin{algorithmic}[1]
\STATE \textbf{Input:} $\bb, \bd, \bm_0, \omega, \dual_0 = \boldsymbol{0}$
\FOR{$k = 0,1,2,\dots$ until convergence} 
   \STATE Set $\bm = \bm_k$, $\bS=\bP\bA(\bm)^{-1}$
   \FOR{$l = 0$ to $\texttt{maxit}-1$}
       \STATE $\displaystyle 
       \dual_{\,l+\frac{1}{2}} =
           \bold{S}^{\!\top}
           \Big[\bold{S}\Ws^{-1}\bold{S}^{\!\top}+\Wd^{-1}\Big]^{-1}
           \Big(\bd - \bold{S}\bb + \bold{S}\Ws^{-1}\dual_l\Big)$
       \STATE $\displaystyle 
       \bu_{\,l+1} = \bA(\bm)^{-1} 
           \Big[ \bb + \Ws^{-1} (\dual_{\,l+\frac12} - \dual_l) \Big]$
       \STATE $\displaystyle 
       \delta\bm_{\,l+1} = - \Big[\diag{\omega^2\bu_{\,l+1}} \Ws \diag{\omega^2\bu_{\,l+1}} + \Wm \Big]^{-1}\diag{\omega^2\bu_{\,l+1}} \, \dual_{\,l+\frac12}$
       \STATE $\displaystyle 
       \dual_{\,l+1} = \dual_l + \Ws \Big[ \bA(\bm+\delta\bm_{\,l+1})\bu_{\,l+1} - \bb \Big]$
   \ENDFOR
   \STATE $\displaystyle 
   \bm_{k+1} = \bm + \delta\bm_{\texttt{maxit}}$
       \IF{$\|\bm_{k+1}-\bm_k\| < \texttt{tol\_outer}$}
        \STATE \textbf{stop}
    \ENDIF
\ENDFOR
\STATE \textbf{Output:} $\bm_{k+1}$
\end{algorithmic}
\end{algorithm}


\section{Weighting Matrices.}\label{weightingmtx}
In the proximal Lagrangian formulation, three weighting matrices $\Wd$, $\Wm$, and $\Ws$ are introduced. The proper choice of these matrices is critical for improving the conditioning of the inversion problem. Without loss of generality, we define 
\begin{subequations}
\label{weights}
\begin{align}
\Wd &= \frac{1}{\mu}\diag{\phid(\bold{x}_1), \phid(\bold{x}_2), \dots, \phid(\bold{x}_{n_r})},\quad 0 < \phid(\bold{x}_j) \leq 1, \mu > 0, \label{Wd}\\
\Wm &= \diag{\phim(\bold{x}_1), \phim(\bold{x}_2), \dots, \phim(\bold{x}_{n})},\quad \phim(\bold{x}_i)>0,  \label{Wm} \\
\Ws &= \diag{\phis(\bold{x}_1), \phis(\bold{x}_2), \dots, \phis(\bold{x}_{n})},
\quad 0 < \phis(\bold{x}_i) \leq 1,   \label{Ws}
\end{align}
\end{subequations}
In this work, we focus on the selection of $\Ws$ and $\Wd$. For simplicity, $\phim(\bold{x})$ is not designed in detail and is instead chosen as $\phim(\bold{x})= \tau$, where $\tau>0$ is a small parameter that regularizes the model update. We primarily examine the choice of the weighting function $\phis(\bold{x})$, since it directly controls the scaling of the wave equation constraints. The role of $\mu$ is addressed separately in Section~\ref{Res_est}, while $\phid(\bold{x})$ is fixed to unity throughout the experiments.


\subsection{Design of $\Ws$.} \label{subsec_Ws}
The proximal regularization is controlled by the spatially varying weights $\phis(\bold{x})$ (encoded in $\Ws$) and the global parameter $\mu$, both of which strongly affect conditioning and convergence. The primary role of $\phis(\bold{x})$ is to balance the amplitudes of the wave-equation residuals. Without such weighting, the standard $\ell_2$-norm tends to underemphasize small residuals (e.g., from weak late-arrival waves), even though they may carry important information. 

Table~\ref{Table:reg} summarizes the effect of the spatial weights $\phis(\bold{x})$ (for fixed $\mu$ and $\tau$) on the strength of proximal regularization, the data misfit, and the wave-equation residual, highlighting the trade-offs between enforcing the wave equation and fitting the data.
Small values of $\phis(\bold{x})$ imply stronger regularization, leading to smaller multipliers, larger residuals $\|\bA(\bm)\bu - \bb\|_2$, and generally a better fit to the data. Larger weights relax the regularization, producing larger multipliers that enforce the wave equation more strictly, but often at the expense of ill-conditioning and increased data misfit. In practice, stronger regularization is typically applied in early iterations and gradually reduced to tighten wave-equation enforcement. Because residual amplitudes naturally decay with distance from the source, constant weights tend to overemphasize accuracy near the source and underemphasize it farther away. This imbalance can be mitigated by distance-dependent weights, which rescale the residuals and promote balanced enforcement of the wave equation across the model.

\begin{table}[h!]
\centering
\caption{Effect of spatial weights $\phis(\bold{x})$ (for fixed $\mu$ and $\tau$) on the proximal regularization, data misfit $\|\bP\bu - \bd\|_2$, and wave equation error  $\|\bold{A(m)}\bu - \bb\|_2$.}
\label{Table:reg}
\begin{tabular}{|c|c|c|c|}
\hline
\textbf{$\phis(\bold{x})$} & \textbf{Regularization level} & \textbf{$\|\bP\bu - \bd\|_2$} & \textbf{ $\|\bA(\bm)\bu - \bb\|_2$} \\
\hline
Small ($\phis \ll 1$) & Strong (small multipliers) & Smaller (better data fit) & Larger (looser enforcement) \\
\hline
Large ($\phis \approx 1$) & Weak (large multipliers) & Larger (poorer data fit) & Smaller (strict enforcement) \\
\hline
\end{tabular}
\end{table}

In Table~\ref{Table:w} we give four examples of weighting functions. The $\phis^\texttt{L}$, proposed by \cite{Huang_2018_SEW}, is a linearly increasing function of distance from the source that compensates for geometrical spreading. This strategy has been shown to improve the convexity of the optimization landscape \cite{Symes_2020_WRI, Symes_2022_EBE}. However, this definition introduces a singularity at the source location, and thus is applicable only when the source point lies outside the imaging domain. To mitigate this issue, \cite{Rizzuti_2021_ADF} suggested the hyperbolic function $\phis^{\texttt{H}} $ as a smoothed version, where $h$ is a tuning parameter. Alternatively, \cite{Symes_2022_EBE} proposed a capped linear weighting, $\phis^{\texttt{CL}}$ which increases linearly with distance up to a cutoff $R$ and remains constant beyond that. These weighting functions can also be normalized to map coefficients between 0 and 1 for consistent scaling.
In this paper, we introduce an alternative complementary Gaussian weighting function,
\begin{equation}\label{w}
\phis(\bx) = \left[ 1 - (1-\epsilon)\exp\!\!\left(-\frac{\|\bx - \bx_{\texttt{s}}\|_2^2}{2\sigma^{2}}\right) \right]^{2}.
\end{equation}
where $0<\epsilon \ll 1$ and $\sigma>0$ are tuning parameters, further discussed in Section ~\ref{Sigma_est}. This weighting function avoids singularities at the source,  is smooth, and provides a flexible transition between near-source and far-field regions.

\begin{table}[H]
\renewcommand{\arraystretch}{1.5} 
\centering
\caption{Comparison of different definitions of the spatial weighting function $w(\bx)$.}
\label{Table:w}
\begin{tabular}{|c|c|c|}
\hline
\textbf{Weighting function } &\textbf{Expression of $\phis(\bx)$} & \textbf{Ref.} \\ \hline
$\phis^\texttt{L}(\bx)$ & $\|\bx - \bx_{\texttt{s}}\|_2^2$ &  \cite{Huang_2018_SEW} \\ \hline
$\phis^\texttt{H}(\bx)$ & $ \tfrac{1}{h^2}(\|\bx - \bx_{\texttt{s}}\|_2^2 + h^2)$, $h>0$ &  \cite{Rizzuti_2021_ADF} \\ \hline
$\phis^\texttt{CL}(\bx)$ & $\min(\|\bx - \bx_{\texttt{s}}\|_2,R)^2$, $R>0$ &  \cite{Symes_2022_EBE}  \\ \hline
\end{tabular}
\end{table}

Notably, $\phis(\bold{x})$ assigns minimal weight at the source location. 
Under the point source assumption, the source term $\bb$ is nonzero only at the source location $\bold{x}_s$, where the associated weight is $\phis(\bold{x}_s) =\epsilon^2$. This leads to the important approximation:
\begin{equation} \label{Wb}
\Ws \bb \approx \bold{0}.
\end{equation} 
Consequently, since the dual variable $\dual_0$ is initialized as zero, it remain approximately zero at the source location, i.e., $\dual_k(\bold{x}_s) \approx 0$. 
This behavior can be derived from the optimality condition for $\dual$.
Using the approximation $\Ws \bb \approx 0$ from equation~\eqref{Wb} in \eref{lambda_opt}, we get
\begin{equation} \label{lambda_opt_w}
    \dual(\bu, \bm, \dualp)=\Ws \bA(\bm) \bu - \Ws \bb - \dualp 
    \approx \Ws \bA(\bm) \bu.
\end{equation}
Additionally, since $\dual(\bold{x}_s) \approx 0$, under the point-source assumption, we have $\langle \dual, \bb \rangle \approx 0$. Therefore,
\begin{equation}
\langle \dual, \bA(\bm) \bu - \bb \rangle = \langle \dual, \bA(\bm) \bu \rangle - \langle \dual, \bb \rangle 
\approx \langle \dual, \bA(\bm) \bu \rangle.
\end{equation}
This elegant approximation renders the objective function \eref{PL} independent of the source signature.
\begin{equation}
\mathcal{L}_{\texttt{PL}}(\bm,\bu, \dual,\bm_k,\dual_k) = \frac{1}{2} \|\bP \bu - \bd\|_{\Wd}^2 
+ \langle \dual, \bA(\bm) \bu \rangle 
+ \frac{1}{2} \|\bm - \bm_k\|_{\Wm}^2
- \frac{1}{2} \|\dual - \dual_k\|_{\Ws^{-1}}^2.
\end{equation}
This approximation yields the source-independent equivalent of \eref{Scaled_MWI} as 
\begin{subequations}
\label{Scaled_MWI_SI}
\begin{align}
\dual_{k+\frac12}&=
     \bold{S}(\bm_k)^{\!\top}\left[ \Wd^{-1}+\bold{S}(\bm_k)\Ws^{-1}\bold{S}(\bm_k)^{\!\top}\right]^{-1}(\bd +\bold{S}(\bm_k)\Ws^{-1}\dual_k) \label{Scaled_MWI_SI_lam}\\
\bu_{k+1}&= \bA(\bm_k)^{-1}\Ws^{-1}(\dual_{k+\frac12}-\dual_k) \label{Scaled_MWI_SI_u}\\
   \bm_{k+1}
   &=\bm_k-\left[\diag{\omega^2\bu_{k+1}}^{\top}\Ws\diag{\omega^2\bu_{k+1}}+\Wm\right]^{-1}\diag{\omega^2\bu_{k+1}}^{\top}\dual_{k+\frac12} \label{Scaled_MWI_SI_m}\\
   \dual_{k+1}&= \dual_k + \Ws\bA(\bm_{k+1}) \bu_{k+1}.  \label{Scaled_MWI_SI_lam2}
\end{align}
\end{subequations}
In this equations, we use $``="$ instead of $``\approx"$.
It is important to note that, this formulation completely eliminates the dependence on the source signature in the resulting algorithms. 
It can be used to develop the source-independent counterpart of the algorithms presented above. The source-independent dual algorithm which will be implemented in this paper is summarized in Algorithm \ref{alg_SI_dual}.
This independence from the source signature is a key advantage of the proposed weighting approach. It should be noted that the algorithm presented by equations \eref{Scaled_MWI_SI} is based on the specific weighting function and the point source assumption.\\
Below, we summarize the key properties of the Algorithm \eref{alg_SI_dual}:
\begin{enumerate}
\item The source signature is fully eliminated from the iteration process. However, knowledge of an approximate source location is still required to construct the weighting function \cite{Huang_2018_SEW}.
\item 
Traditional numerical solutions of the wave equation rely on grid-based discretization, where sources positioned between grid points often necessitate interpolation, potentially leading to inaccuracies \cite{Hicks_2002_ASR}. In contrast, the multipliers in the proposed method are represented as spatially distributed quantities, a defining characteristic of extended source functions \cite{Symes_2020_FWI}. This spatial distribution reduces the sensitivity to the exact alignment of the source with the grid, enhancing the method's robustness.

\item All matrix operators, including $\bA$ and thus $\bold{S}=\bold{PA}^{-1}$, are derived from the model that remain fixed during the inner loop. These operators can be precomputed and inverted (e.g., via LU factorization) once before starting the iterations, significantly improving computational efficiency \cite{Abubakar_2008_FCS,Aghazade_2025_FAF}.
\item The data-space Hessian matrix, $\left(\bold{S} \Ws^{-1}\bold{S}^{\top}+\Wd^{-1}\right)$, is sized according to the number of receivers, making its computation and inversion efficient once the forward operator  $\bold{S}$ is determined.
\end{enumerate}

%
%

\begin{algorithm}[H]
\caption{Source Independent Dual Multiplier Waveform Inversion}
\label{alg_SI_dual}
\begin{algorithmic}[1]
\STATE \textbf{Input:} $\bd, \bm_0, \omega$
\FOR{$k = 0,1,2,\dots$ until convergence} 
   \STATE Set $\bm = \bm_k$, $\bS=\bP\bA(\bm)^{-1}$, $\dual_0 = \bold{0}$ \label{alg_0}
   \FOR{$l = 0$ to $\texttt{maxit}-1$}
       \STATE $\displaystyle 
       \dual_{\,l+\frac{1}{2}} =
           \bold{S}^{\!\top}
           \Big[\bold{S}\Ws^{-1}\bold{S}^{\!\top}+\Wd^{-1}\Big]^{-1}
           \Big(\bd + \bold{S}\Ws^{-1}\dual_l\Big)$ \label{alg_lam}
       \STATE $\displaystyle 
       \bu_{\,l+1} = \bA(\bm)^{-1} \Ws^{-1} (\dual_{\,l+\frac12} - \dual_l) $ \label{alg_u}
       \STATE $\displaystyle 
       \delta\bm_{\,l+1} = -\left[\diag{\omega^2\bu_{\,l+1}}^{\top}\Ws\diag{\omega^2\bu_{\,l+1}}+\Wm\right]^{-1}\diag{\omega^2\bu_{\,l+1}}^{\top}\dual_{\,l+\frac12}$ \label{alg_dm}
       \STATE $\displaystyle 
       \dual_{\,l+1} = \dual_l + \Ws\bA(\bm+\delta\bm_{\,l+1})\bu_{\,l+1}$ \label{alg_Lam}
   \ENDFOR
   \STATE $\displaystyle 
   \bm_{k+1} = \bm + \delta\bm_{\texttt{maxit}}$  \label{alg_m}
    \IF{$\|\bm_{k+1}-\bm_k\| < \texttt{tol\_outer}$}
        \STATE \textbf{stop}
    \ENDIF
\ENDFOR
\STATE \textbf{Output:} $\bm_{k+1}$
\end{algorithmic}
\end{algorithm}


\subsubsection{On the parameters $\sigma$ and $\epsilon$.} \label{Sigma_est}
These parameters are crucial for achieving robust inversion. In the following, we present practical strategies for selecting appropriate values.

The parameter $\epsilon$ controls the behavior of the distance-weight function in the vicinity of the source location $\bx_{\texttt{s}}$.
When $\epsilon \ll 1$, its effect on the shape of the weighting function $\phis$ is negligible, but it has a strong influence on the inverse weighting function $\phis(\bx)^{-1}$, which directly regularizes the multipliers. Since these multipliers act as extended sources in the inversion, $\epsilon$ provides the flexibility for concentrating their energy in specific regions of the space.

Our strategy for choosing $\epsilon$ is motivated by the fact that propagating waves interfere constructively as long as the phase shift is smaller than $\frac12\lambda_{\texttt{w}}$, where $\lambda_{\texttt{w}}$ is the wavelength of the propagating wave \cite{Sheriff_1980_NFC}. Consequently, we aim to impose stronger regularization within a neighborhood of radius $\frac{1}{4}\lambda_{\texttt{w}}$ around the source. This penalization encourages the estimated extended source to remain concentrated in that region.
At the source location and at a distance of $\frac{1}{4}\lambda_{\texttt{w}}$, the regularization levels are given by
\begin{equation}
   \phis(\bx_{\texttt{s}})^{-1}=\epsilon^{-2} \quad \text{and} \quad
    \phis\!\!\left(\bx_{\texttt{s}}\pm\frac{1}{4}\lambda_{\texttt{w}}\right)^{-1} = \left[1 - (1-\epsilon)\exp\!\!\left(-\frac{\lambda_{\texttt{w}}^2}{32\sigma^2}\right)\right]^{-2}.
\end{equation}
We define the amplitude ratio $\gamma > 1$ as
\begin{equation}
 \gamma =\frac{\phis(\bx_{\texttt{s}}+\frac{1}{4}\lambda_{\texttt{w}})}{\phis(\bx_{\texttt{s}})},
\end{equation}
and solving for $\epsilon$ yields
\begin{equation} \label{eps_alpha}
\epsilon
= \frac{\sinh{\!(\frac{\lambda_{\texttt{w}}^2}{64\sigma^{2}})}}{\gamma^{\nicefrac{1}{4}}\,\sinh{\!(\frac{\lambda_{\texttt{w}}^2}{64\sigma^{2}}+\frac{1}{4}\ln{\gamma})}}.
\end{equation}

\begin{figure}[h]
    \centering
    \includegraphics[width=0.75\linewidth,trim={0 0cm 0 0},clip]{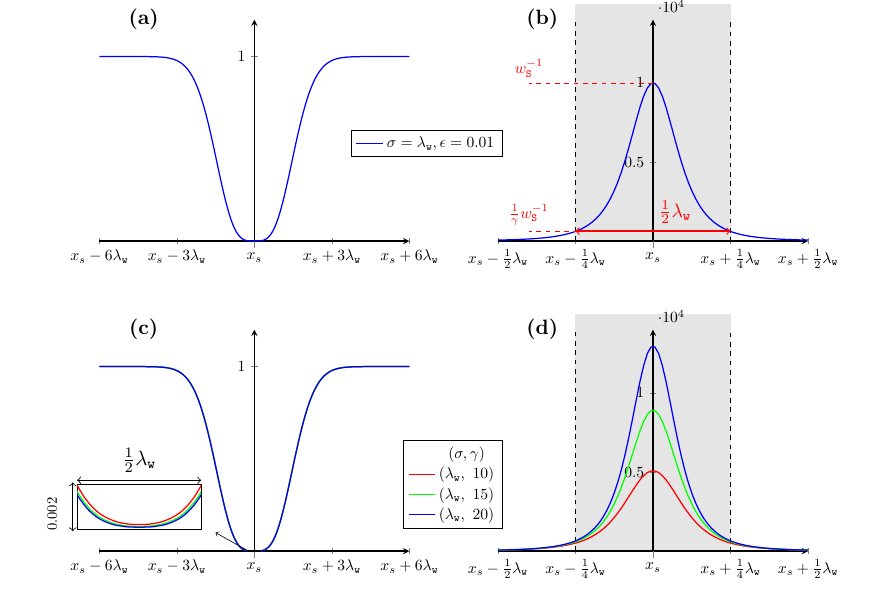}
    \caption{(a) Weight function for $\sigma=\lambda_{\texttt{w}}$ and $\epsilon=0.01$. The horizontal axis shows distance in units of the dominant wavelength $\lambda_{\texttt{w}}$. (b) Inverse weighting function, $\phis(x)^{-1}$. The vertical dashed lines mark the distances $\pm\tfrac{\lambda_{\texttt{w}}}{4}$ from the source location $x_s$, where the weight strength drops to the constant fraction $\tfrac{1}{\gamma}$. (c–d) The weighting function for $\sigma=\lambda_{\texttt{w}}$ and different parameters $\gamma=10,~15,~20$.}
    \label{wfunc}
\end{figure}

\fref{wfunc}a shows the weighting function for $\sigma=\lambda_{\texttt{w}}$ and $\epsilon=0.01$. The horizontal axis is expressed in terms of the propagating wavelength $\lambda_{\texttt{w}}$ measured from the source location. \fref{wfunc}b displays the corresponding inverse weighting functions. Since $\epsilon$ is fixed, the maximum value at the source is $\phis(\bx_{\texttt{s}})^{-1}=1/\epsilon^2=10^4$. 
In \fref{wfunc}c, we show the weightig function for $\sigma=\lambda_{\texttt{w}}$ and vary $\epsilon=[0.014,0.01,0.008]$, which correspond to $\gamma=[10,15,20]$. The overall shape of $\phis(\bx)$ remains nearly unchanged and visually indistinguishable, except for a subtle difference near the source, highlighted in the zoomed inset. However, this seemingly minor change produces a substantial impact on $\phis(\bx)^{-1}$ within a region of width $\tfrac{1}{2}\lambda_{\texttt{w}}$ around the source, as shown in \fref{wfunc}d. The parameter $\gamma$ provides effective flexibility for regularization of the multipliers.
For $\gamma=10$, $\tfrac{1}{2}\lambda_{\texttt{w}}$ is equal to the full width at tenth of maximum of the resulting weighting function $\phis^{-1}$.

\subsection{Design of $\Wd$.} \label{subsec_Wd}
This operator acts in the data space and rescales the contribution of each data component in the misfit term of the objective function. In a Bayesian framework, such weighting operators can be interpreted as the inverses of the data covariance operators \cite{Rizzuti_2021_WaRIance}

\subsubsection{On selection of $\phid$.} \label{subsec_Wd_w}
A common strategy is to define $\phid(\bx)$ as a function of the source–receiver offset, thereby emphasizing specific portions of the data. For example, to strengthen the contribution of large-offset data in crustal-scale imaging, one may use \cite{Operto_2006_CIM} 
\begin{equation}
    \phid(\bx_{\texttt{r}}) = \exp(\beta\ln\|\bx_{\texttt{r}}-\bx_{\texttt{s}}\|_2)=\|\bx_{\texttt{r}}-\bx_{\texttt{s}}\|_2^\beta
\end{equation}
where $\beta$ is a tuning parameter. This weighting reduces the relative influence of short-offset data, which typically contain limited information about deep structures, and enhances the sensitivity to long-offset information.

\subsubsection{Selection of regularization parameter $\mu$.} \label{Res_est}
The regularization parameter $\mu$, defining the data weighting matrix $\Wd^{-1}(\mu)$ \eref{Wd}, must be carefully chosen to balance data fitting and wave-equation enforcement. As $\mu \to \infty$, the algorithm strictly enforces the wave equation, potentially at the cost of poor data fitting. In this limit, the wave equation is satisfied exactly, but the regularization effect vanishes. In this case, the algorithm reduces to conventional reduced-space FWI \cite{Pratt_1998_GNF}. Conversely, as $\mu \to 0$, the algorithm emphasizes data fitting while effectively ignoring the wave equation.

In the weighted formulation of the algorithm, the data weighting matrix appears only in the first subproblem associated with estimating the adjoint wavefield, solving \eref{adjoint_eq}. Here we show how to properly determine an appropriate value for $\mu$.
\begin{itemize}
    \item[(i)] \textbf{Discrepancy Principle.} 
    It can be determined adaptively using the discrepancy principle, which selects $\mu$ to satisfy the condition $ \|\bold{r}(\mu)\|_2^2 = \delta^2$, where
    \begin{equation}
       \bold{r}(\mu)=\delta\bd_l-\bold{S}\Ws^{-1}\dual_{l+1/2}^\mu
    \end{equation}
 is the residual and $\delta$ is an upper bound on the norm of the noise contaminating the data. However, the discrepancy principle requires prior knowledge of the noise level, which may not be available in practical scenarios.
    \item[(ii)] \textbf{Generalized cross-validation \cite{Wahba_1990_SMO}.} 
    GCV is a widely used method for selecting the regularization parameter in discrete ill-posed problems without requiring prior knowledge of the noise level. Based on GCV, an optimal regularization parameter should provide good predictions of missing or unseen data values. 
For a given forward operator $\bS$ and data residual $\delta\bd_l$, the optimum parameter is defined as
\begin{equation} \label{mu_gcv}
 \mu =  \arg\min_{\mu} \phi_{\texttt{GCV}}(\mu)=\frac{\|\bold{r}(\mu)\|_2^2}{\left[\texttt{Trace}\left(\bI-\bS\bS^{\dag}\right)\right]^2}.
\end{equation}
where $\texttt{Trace}$ denotes the trace of a matrix and $\bS^{\dag}=\Ws^{-1}\bold{S}^{\top} \left[\bold{S}\Ws^{-1}\bold{S}^{\top}+\Wd^{-1}(\mu)\right]^{-1}$ is the regularized inverse operator. At each iteration, the GCV function can be computed efficiently by using Generalized Singular Value Decomposition (GSVD).
The GSVD of the matrix pair $\Ws^{-{\nicefrac{1}{2}}}\bold{S}^{\top}$ and $\diag{\bw_{\texttt{D}}^{-1}}$ is given by:
\begin{equation}
    \Ws^{-{\nicefrac{1}{2}}}\bold{S}^{\top} = \bU\bold{\Gamma}_{\texttt{S}}\bold{Z}^{\top},
    \quad
    \diag{\bw_{\texttt{D}}^{-1}} = \bold{V}\bold{\Gamma}_{\texttt{D}}\bold{Z}^{\top},
\end{equation}
where $\bU\in \mathbb{C}^{n\times n}$ and $\bold{V}\in \mathbb{C}^{n_r \times n_r}$ are unitary matrices, $\bold{Z}\in \mathbb{C}^{n_r\times n_r}$ is nonsingular, and $\bold{\Gamma}_{\texttt{S}}\in \mathbb{C}^{n\times n_r}$ and $\bold{\Gamma}_{\texttt{D}}\in \mathbb{C}^{n_r\times n_r}$ are diagonal  matrices satisfying $\bold{\Gamma}_{\texttt{S}}^{\top}\bold{\Gamma}_{\texttt{S}}+\bold{\Gamma}_{\texttt{D}}^{\top}\bold{\Gamma}_{\texttt{D}}=\bold{I}$. Accordingly,
\begin{align}
    \bold{r}(\mu)&=  (\bI-\bS\bS^{\dag})\delta\bd_l\\
    &=\Wd^{-1} \left[\bS\Ws^{-1}\bS^{\top}+ \Wd^{-1}\right]^{-1}\delta\bd_l\\
     &=\mu\bZ\bold{\Gamma}_{\texttt{D}}^{\top}\bold{\Gamma}_{\texttt{D}}\bold{Z}^{\top} \left[\bold{Z}\bold{\Gamma}_{\texttt{S}}^{\top}\bU^{\top}\bU\bold{\Gamma}_{\texttt{S}}\bold{Z}^{\top}+\mu \bold{Z}\bold{\Gamma}_{\texttt{D}}^{\top}\bold{V}^{\top}\bold{V}\bold{\Gamma}_{\texttt{D}}\bold{Z}^{\top}\right]^{-1}\delta\bd_l\\
     &=\mu\bold{Z} \bold{\Gamma}_{\texttt{D}}^{\top}\bold{\Gamma}_{\texttt{D}}\left[\bold{\Gamma}_{\texttt{S}}^{\top}\bold{\Gamma}_{\texttt{S}}+\mu \bold{\Gamma}_{\texttt{D}}^{\top}\bold{\Gamma}_{\texttt{D}}\right]^{-1}
     \bold{Z}^{-1}\delta\bd_l.
\end{align}
Then, the GCV function can be computed as
\begin{equation} \label{mu_gcv}
 \phi_{\texttt{GCV}}(\mu)=\frac{\|\bZ\diag{\bold{w}(\mu)}\bZ^{-1}\delta\bd\|_2^2}{\|\bold{w}(\mu)\|_1^2}, \quad [\bold{w}(\mu)]_i=\frac{\mu[\bold{\Gamma}_{\texttt{D}}]_{ii}^2}{\mu[\bold{\Gamma}_{\texttt{D}}]_{ii}^2+[\bold{\Gamma}_{\texttt{S}}]_{ii}^2}.
\end{equation}
where $\circ$ denotes the Hadamard product.
\item[(iii)] \textbf{Residual Whiteness Principle (\cite{Almeida_2013_PEB,Lanza_2013_VID}).}
The RWP is an automatic, parameter-free strategy for selecting regularization parameters in variational inverse problems, particularly in image processing tasks such as restoration, deblurring, and super-resolution \cite{Almeida_2013_PEB,Lanza_2013_VID}.
The core idea is that the residual $\bold{r}(\mu)$ should behave like a realization of white noise if the regularization parameter $\mu$ is chosen correctly. Conversely, a poor choice of $\mu$ typically produces coherent artifacts in the residual, resulting in non-white behavior. To quantify this, a non-negative whiteness measure ($\phi_{\texttt{RWP}}(\mu)$) is defined. A common choice is the kurtosis of the Fourier spectrum of the residual, computed as the ratio between the $\ell_4$-norm to the $\ell_2$-norm raised to the fourth power:
\begin{equation}
    \phi_{\texttt{RWP}}(\mu)=\frac{\|\texttt{fft}[\bold{r}(\mu)]\|_4^4}{\|\texttt{fft}[\bold{r}(\mu)]\|_2^4}.
\end{equation}
The RWP then selects the optimal parameter by minimizing this whiteness measure:
\begin{equation} \label{mu_rwp}
\mu = \arg\min_{\mu} \phi_{\texttt{RWP}}(\mu)
= \frac{\|\texttt{fft}\left[\bZ\diag{\bold{w}(\mu)}\bZ^{-1}\delta\bd_l\right]\|_4^4}
{\|\texttt{fft}\left[\bZ\diag{\bold{w}(\mu)}\bZ^{-1}\delta\bd_l\right]\|_2^4}.
\end{equation}

\end{itemize}
\subsection{Design of $\Wm$.} \label{subsec_Wm}
The matrix $\Wm$ is a model-space weighting operator that appears only in the model-update stage of the algorithm. Its design can serve multiple purposes. The simplest choice is a diagonal weighting matrix. When $\Wm$ is chosen as a scaled identity matrix, the update corresponds to zero-order Tikhonov regularization \cite{Tikhonov_1977_SIP}, which forces the updated model $\bm_{k+1}$ to remain close to the previous estimate $\bm_k$ and thereby stabilizes the inversion.

More generally, $\Wm$ can be designed to perform localized updates, similar in spirit to layer-stripping techniques. In this case, the diagonal weights are constructed to focus the update near specific regions—such as the shallow part of the model in surface-acquisition geometries—and are progressively relaxed to allow deeper updates as the iterations proceed.

Alternatively, $\Wm$ can be defined as a finite-difference operator to penalize model roughness, effectively imposing smoothness constraints on the solution.

\section{Interpretation of the weighted dual algorithm.}\label{Interp}
 The algorithm iteratively estimates a set of Lagrange multipliers that both predict the recorded data and satisfy the wave equation. We provide an implementation details of the proposed algorithm: 
\begin{itemize}
\item[(i)] Estimate the global regularization parameter $\mu$ for building the data weighting matrix $\Wd^{-1}=\mu\bI$. Methods based on the Generalized Cross-validation (GCV) and Residual Whiteness Principal are presented in Subsection \ref{Res_est} for estimation of $\mu$ .
\item[(ii)] For a given source multiplier $\dual_l$, the predicted data are computed in the background model $\bm$ as $\bd^{\texttt{pre}} =-\bS(\bm)\Ws^{-1} \dual_l$ which are used to compute the data residuals,  $\delta\bd=\bd - \bd^{\texttt{pre}}$. The matrix $\bS(\bm)$ is precomputed. The weighting matrix  $\Ws^{-1}$ gains multiplier energy at and near the physical source location by $\phis(\bx)^{-1}$ (\fref{wfunc}). This weighting process serves as a regularization mechanism, guiding the inversion toward a more focused source representation \cite{Huang_2018_SEW}. The data residual is then back-propagated by the adjoint operator $\bold{S}^{\top}$  after deblurring by the data-space Hessian matrix $\left( \bold{S} \Ws^{-1} \bold{S}^{\top} + \Wd^{-1} \right)$, giving the adjoint wavefield $\sdual_{l+\frac12}$ (line \ref{alg_lam}). The preconditioning role of matrix $\Ws$ for estimating $\sdual_l$ and will be explored more in the next section.

The inversion process begins with an initial guess of zero multipliers, $\dual_0=\bold{0}$ (line \ref{alg_0}). Thus, the adjoint wavefield in the first iteration is computed simply by a back-propagation of the recorded data. Unlike first-order Born algorithms like reverse time migration (RTM) which propagate only the first order reflections, in this algorithm all the events in the data including first arrivals, diving waves, and multiples will be back-propagated. 
\item[(iii)]  The inversion of the data-space Hessian ensures that, for sufficiently small values of $\mu$, the computed $\sdual_{l+\frac12}$, when interpreted as an equivalent source term, accurately reproduces the observed data when propagated through the model $\bm$. This property reduces the risk of cycle-skipping by aligning the predicted and observed wavefields despite potential kinematic mismatches in $\bm$. The forward wavefield $\bu_{l+1}$ is computed in line \ref{alg_u} by propagating the total extended source, $\Ws^{-1}(\dual_{l+\frac12}-\dual_l)$.
\item[(iv)] The model perturbation $\delta\bm_{l+1}$ is then estimated through the zero-lag cross-correlation of the forward and adjoint wavefields, akin to the standard imaging condition in RTM (line \ref{alg_dm}).  
\item[(v)] The multipliers are updated (line \ref{alg_Lam}) based on the estimated $\delta\bm_{l+1}$ and $\bu_{l+1}$ and the process is repeated for $\texttt{maxit}$ iterations (or until the residual between successive multiplier estimates falls below a predefined threshold, indicating convergence). 
\item[(vi)] Finally, the model is updated with the final perturbation $\delta\bm_{\texttt{maxit}}$ (line \ref{alg_m}). If necessary, the updated model can used to repeat the inner loop.
\end{itemize}

We demonstrate how distance-weight function precondition each subproblem in FWI using multiplier methods.

\subsection{Estimation of  adjoint wavefield.} 
Regarding the adjoint wavefield (line \ref{alg_lam}), we note that for a computed data $\delta\bd_l=\bd + \bold{S}(\bm)\Ws^{-1} \dual_l$, the adjoint wavefield $\dual_{l+\frac12}$ is a regularized solution to inverse source problem
\begin{equation} \label{adjoint_eq}
\bold{S}(\bm)\Ws^{-1}\sdual_{l+\frac12} = \delta\bd_l,
\end{equation}
given by the following weighted $\ell_2$-norm regularization \cite{Tarantola_2005_IPT}
\begin{equation} \label{lambda_reg}
  \sdual_{l+\frac12}=  \arg \min_{\sdual} \frac{1}{2}\|\bold{S}(\bm)\Ws^{-1}\dual-\delta\bd_l\|_{\Wd}^2+\frac{1}{2}\|\dual\|_{\Ws^{-1}}^2.
\end{equation}
The regularized solution is obtained by computing the gradient and setting the result equal to zero:
\begin{subequations}\label{lambda_regg}
\begin{align}
  \sdual_{l+\frac12}&=\left[\bold{S}(\bm)^{\top} \Wd \bold{S}(\bm)\Ws^{-1}+ \bI\right]^{-1}\bold{S}(\bm)^{\top} \Wd\delta\bd_l \label{a11}\\
    &=\bold{S}(\bm)^{\top} \Wd\left[\bold{S}(\bm)\Ws^{-1}\bold{S}(\bm)^{\top} \Wd+ \bold{I}\right]^{-1}\delta\bd_l \label{a22} \\
    &=\bold{S}(\bm)^{\top} \left[\bold{S}(\bm)\Ws^{-1}\bold{S}(\bm)^{\top}+ \Wd^{-1}\right]^{-1}\delta\bd_l. 
\end{align}
\end{subequations}
Note that we derived \eqref{a22} from \eqref{a11} by using the identity $(\bU\bV+\bI)^{-1}\bV = \bV(\bV\bU+\bI)^{-1}$ which is a special case of the Sherman–Morrison–Woodbury formula  \cite{Guttman_1946_EMC}.
 To demonstrate the role of this regularization, we notice that there are two main challenges in solving the inverse source problem defined by $\bold{S}(\bm)\sdual = \delta\bd_l$:
 (i) Underdetermined nature:  
   The matrix $\bold{S}$ is of size $n_r \times n$, where $n_r \ll n$, leading to an underdetermined system with infinitely many possible solutions. A straightforward minimum-norm solution, $\bold{S}^{\top}(\bold{S} \bold{S}^{\top})^{-1} \delta\bd_l$, does not account for the physics of wave propagation or the localized nature of the true source \cite{Symes_2020_FWI}.   
(ii) Energy decay of seismic waves:  
   The seismic wave energy naturally decay with distance from the source.
Each row of the matrix $\bold{S}$ corresponds to a Green’s function associated with a receiver, with larger amplitudes near the receiver and smaller amplitudes further away. Consequently, the columns of $\bold{S}$ exhibit non-uniform energy distribution, with higher energy concentrated near receiver locations \cite{Lee_2020_SFW}. This uneven energy distribution contributes to the ill-conditioning of the system. Therefore, the unweighted regularization tends to favor solutions where the energy tend to concentrate near the receivers, reflecting the higher energy of $\bold{S}$'s columns in these regions. 

The weighting matrices adjust for these drawbacks, giving more weights to the columns of $\bold{S}$ near the source. This adjustment mitigates the bias introduced by the unweighted solution, allowing for more physically meaningful and spatially focused estimates of $\sdual_{l+\frac12}$. This weighting approach not only improves the conditioning of the system, but also ensures that the estimated solution align more closely with the expected physics of seismic wave propagation.

\begin{figure}[h]
    \centering
    \includegraphics[width=0.7\linewidth,trim={0 0cm 0 0},clip]{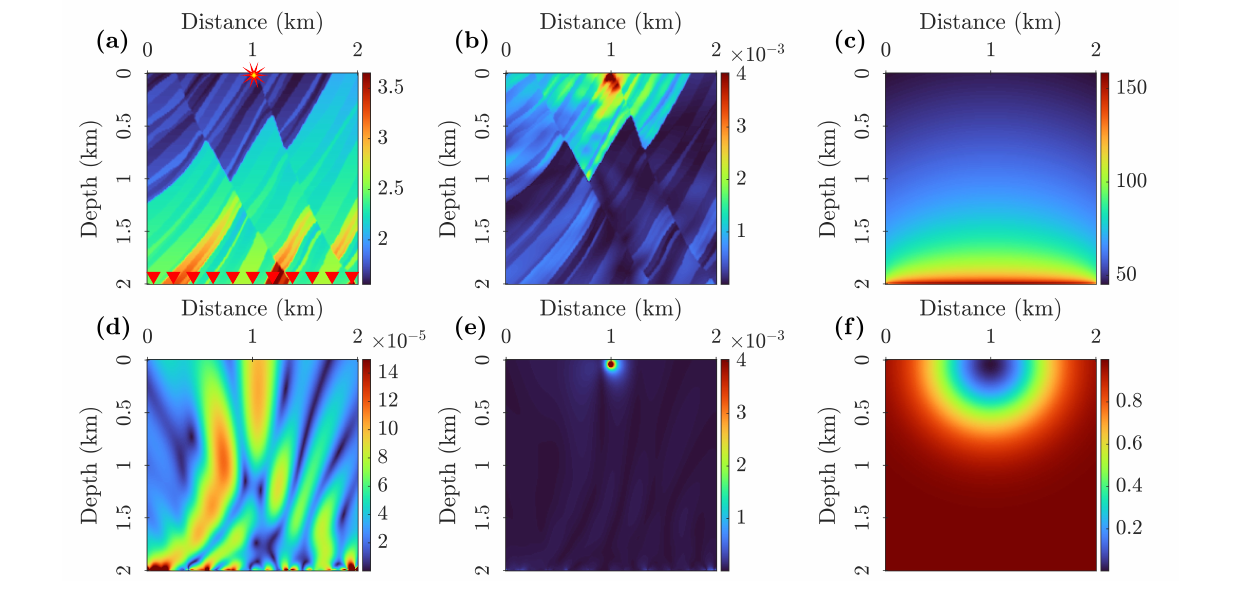}
    \caption{An example showing the use of a distance-weight function in source reconstruction to compensate for the natural decay of wave amplitude with distance. (a) A homogeneous velocity model with the positions of the source and receivers.  
(b) The ground-truth source extension, defined as the product of the wavefield and a homogeneous reflectivity of 1. This source was used to generate the data \(\bd\) at a frequency of 6 Hz.  
(c) The distribution of the norms of the columns of the forward operator $\bold{S}$.  
(d) The source reconstructed using the original, unpreconditioned system.  
(e) The source reconstructed using the preconditioned system, which employs the distance-weight function shown in (f).}
    \label{Decay}
    \vspace{-.4cm}
\end{figure}

We demonstrate the preconditioning role of the distance-weight function using a transmission problem. The model consists of a homogeneous background velocity of 2 km/s (\fref{Decay}a), spanning 2 km in both depth and lateral extent, discretized on a 10 m grid. A single source is placed near the top of the model at \(\bold{x}_s = (1000 \, \text{m}, 50 \, \text{m})\), with 200 receivers equally spaced every 10 m along the bottom boundary at a depth of 1.99 km. 
We computed a 6 Hz wavefield, $\bu^*$, in the true model and then construct an $\dual^*=\bA(\bm)\bu^*-\bb$ using a homogeneous model $\bm$. The result, shown in \fref{Decay}b, serves as the ground truth for $\sdual^*$. The figure clearly illustrates the decay of amplitudes with distance from the source.
In \fref{Decay}c, we display the $\ell_2$-norm of the columns of the forward operator $\bold{S}$, reshaped into a matrix form, showing that the larger amplitudes are concentrated near the receivers. 
We generated a synthetic dataset $\delta\bd = \bold{S}\sdual^*$, where $\sdual^*$ is shown in \fref{Decay}b. For the first attempt, we inverted the data using a damped least-squares $\bold{S}^{\top} \left( \bold{S} \bold{S}^{\top} + \mu \bold{I} \right)^{-1} \delta\bd$ with a small damping parameter. The reconstructed solution is shown in \fref{Decay}d. The energy of the solution is smeared, extending from the true source location at the top of the model toward the receiver positions at the bottom.
In the second attempt, we computed the solution using the weighted approach \eref{lambda_reg} with $\Wd^{-1}=\mu\bold{I}$ and $\Ws$ built from the weights $\phis(\bold{x})$ in \eref{w} with $\sigma=200$ and $\epsilon=0.001$, shown in \fref{Decay}f. The solution, shown in \fref{Decay}e, demonstrates that the preconditioning effectively concentrates the energy at the true source location, mitigating the bias observed in the first reconstruction. 

\subsection{Estimation of  forward wavefield.}
After computing $\dual_{l+\frac12}$, the term $\Ws^{-1}(\dual_{l+\frac12}-\dual_l)$ acts as the total source \cite{Gholami_2024_FWI}. The weight $\Ws^{-1}$ emphasizes the energy around the physical source location. The resulting source term is then used to solve the wave equation for the forward wavefield, $\bu_{l+1} = \bA(\bm)^{-1}\Ws^{-1}(\dual_{l+\frac12}-\dual_l)$, effectively injecting a spatially focused extended source \cite{Symes_2020_FWI,Symes_2020_WRI}. It is important to note that the source signature is implicitly estimated during the first subproblem by fitting the data, and this estimate is progressively refined through iterations.

\subsection{Estimation of model update.}
The other effect of distance-dependent regularization is their role in left preconditioning the linear system associated with the model update. 
Simple computations show that the model perturbation $\delta\bm_{l+1}$ is a regularized solution to
\begin{equation} \label{lam_pre}
\text{diag}(-\omega^2\bu_{l+1})\delta\bm= \Ws^{-1}\sdual_{l+\frac12},
\end{equation}
given by the following weighted Tikhonov regularization
\begin{equation} \label{dm_reg}
  \delta \bm_{l+1}=  \arg \min_{\delta \bm} \frac{1}{2}\|\text{diag}(-\omega^2\bu_{l+1})\delta\bm- \Ws^{-1}\sdual_{l+\frac12}\|_{\Ws}^2+\frac{1}{2}\|\delta\bm\|_{\Wm}^2.
\end{equation}
Here, the matrix $\Ws$ modulates the relative contributions of the forward and adjoint wavefields in the model reconstruction process.
Its effect is to downweight the wavefields near the source positions while assigning greater weights to regions farther away.
As seen from \eref{lam_pre}, the extended-source term $\Ws^{-1}\sdual_{l+\frac12}$ becomes highly concentrated near the sources, enforcing the wave equation more strictly at greater distances. This causes the model updates to be largest near the sources and to gradually decay with distance.
An intuitive interpretation of this mechanism, and its benefit for convexity, can be given in the context of a surface-acquisition scenario, where sources and receivers are located on the surface of the model.
In such cases, the weighting encourages model updates to focus first near the surface — where the relationship between data and model is approximately linear — and then progressively penetrate deeper into the subsurface, effectively reconstructing the model layer by layer.
This behavior resembles the classical layer-stripping approach \cite{Virieux_2009_OFW}, which improves inversion stability by sequentially building up the model from shallow to deep regions.

\subsection{Scale of the multipliers.}
Finally, the distance-dependent regularization plays a crucial role by effectively scaling the multiplier $\dual$ within the algorithm.
The multiplier update equation \eqref{Scaled_MWI_SI_lam2} can be interpreted as a steepest-ascent iteration for maximizing the dual function \cite{Bertsekas_1996_COL}. Under this interpretation, the weighting matrix $\Ws$ acts as a spatially varying step-size operator, controlling how the multipliers are updated across the domain. In particular, larger weights lead to faster updates at locations farther from the source, whereas smaller weights slow the updates near the source. This mechanism enforces the wave-equation constraint more strictly in regions distant from the source, which helps to promote a more focused and spatially balanced source extension.



\subsection{Acceleration.}
Since the model parameters $\bm$ remain fixed during the inner loop of the dual algorithm, Algorithm \ref{alg_SI_dual}, basically it can be considered as a fixed-point iteration 
\begin{equation}\label{FP}
\dual_{l+1} = g(\dual_l),
\end{equation}
where $g(\dual)$ is the fixed-point mapping describing a single iteration of the inner loop of Algorithm \ref{alg_SI_dual}.
Equation \eref{FP} is a fixed-point problem for the multipliers associated to the given model $\bm$.
This allows one to apply acceleration methods such as the Anderson acceleration \cite{Anderson_1965_IPN,Aghazade_2022_AAA} to increase convergence of the inner loop.
Anderson acceleration aims at speeding up convergence of the iteration to a fixed point where the residual $f(\dual)=\dual - g(\dual)$ vanishes.
Its main idea is to use the residuals of the latest $h+1$ estimates $\dual_{l},\dual_{l-1},...,\dual_{l-h}$ to find a new $\dual_{l+1}$ giving a small residual. This is achieved via a linear combination of these previous estimates under the fixed-point mapping:
\begin{equation}
    \dual_{l+1} = \sum_{j=0}^h \theta_j g(\dual_{l-h+j})
\end{equation}
where the weights $\theta_j$ are determined by solving a least-squares problem:
\begin{equation}
 \theta= \arg\min \Big\Vert \sum_{j=0}^h \theta_j f(\dual_{l-h+j})\Big\Vert_2^2\quad 
 \text{subject to}  \quad \sum_{j=0}^h \theta_j=1.
\end{equation}

\section{Numerical examples.} \label{NumEx}
We evaluate the proposed method on several 2D single-parameter synthetic examples. Forward modeling employs a 9-point finite-difference stencil with anti-lumped mass and 10-point PMLs on all boundaries \cite{Chen_2013_OFD}.
For the efficiency, we compare the performance of two algorithms: the standard dual MWI (DMWI) (Algorithm \ref{alg_dual}) and the weighted (source-independent) DMWI (Algorithm \ref{alg_SI_dual}).
The algorithm parameters are chosen as follows:
\begin{itemize}
    \item[(i)] The PDE weighting function $\bphis$ in \eref{w} is controlled by $\sigma$ and $\gamma$; we set $\gamma = 10$ and determine $\sigma$ empirically by trial and error, using the rule $\sigma = \alpha \lambda_{\texttt{w}}$ with $\alpha \in [1,3]$, where $\lambda_{\texttt{w}}$ is the dominant wavelength.
    \item[(ii)] The data weighting function is set to $\bphid(\bx) = 1$, and the global regularization parameter $\mu$ is automatically updated at each iteration using the RWP method (Section~\ref{Res_est}).
    \item[(iii)] The model weighting function is chosen as $\bphim(\bx)=\epsilon$, where $\epsilon$ is a small positive constant to stabilize the inversion.
    \item[(iv)] Anderson acceleration with a history size $h = 6$ is applied to accelerate convergence of the inner iterations.
    \item[(v)] The number of inner iterations, $\texttt{maxit}$, is kept the same for both algorithms and is selected empirically to balance computational cost and convergence quality.
    \item[(vi)] For the DMWI case, the source signature is assumed to be known. This assumption is not required for the weighted DMWI formulation. Moreover, in the weighted DMWI we do not require sources and receivers to coincide with the finite-difference grid points, and we exploit this flexibility in the numerical experiments unless otherwise stated. In contrast, for DMWI we assume that sources and receivers are located on grid points.
\end{itemize}


\begin{figure}[h]
    \centering
    \includegraphics[width=0.6\linewidth,trim={0 0cm 0 0},clip]{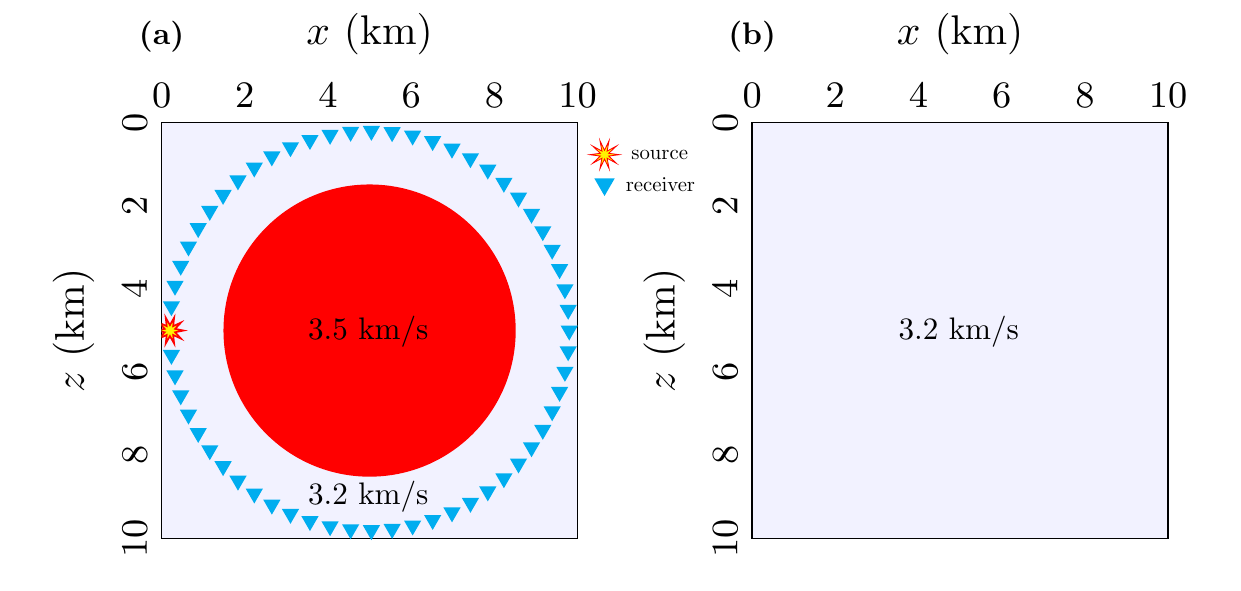}
    \caption{(a) True velocity model for the Camembert experiment.  The model consists of a large circular velocity anomaly with velocity of 3.5 km/s and a radius of 3.5 km, embedded in a homogeneous background with a velocity of 3.2 km/s.
    The acquisition setup consists of a circular array of 60 sources and receivers, evenly spaced every 6 degrees along a circle with a 4.8 km radius surrounding the anomaly. Blue triangles mark receiver positions, and the star indicates a source location. (b) Initial velocity model.}
    \label{Camembert}
    \vspace{-.4cm}
\end{figure}

\begin{figure}[h]
    \centering
    \includegraphics[width=0.6\linewidth,trim={0 0cm 0 0},clip]{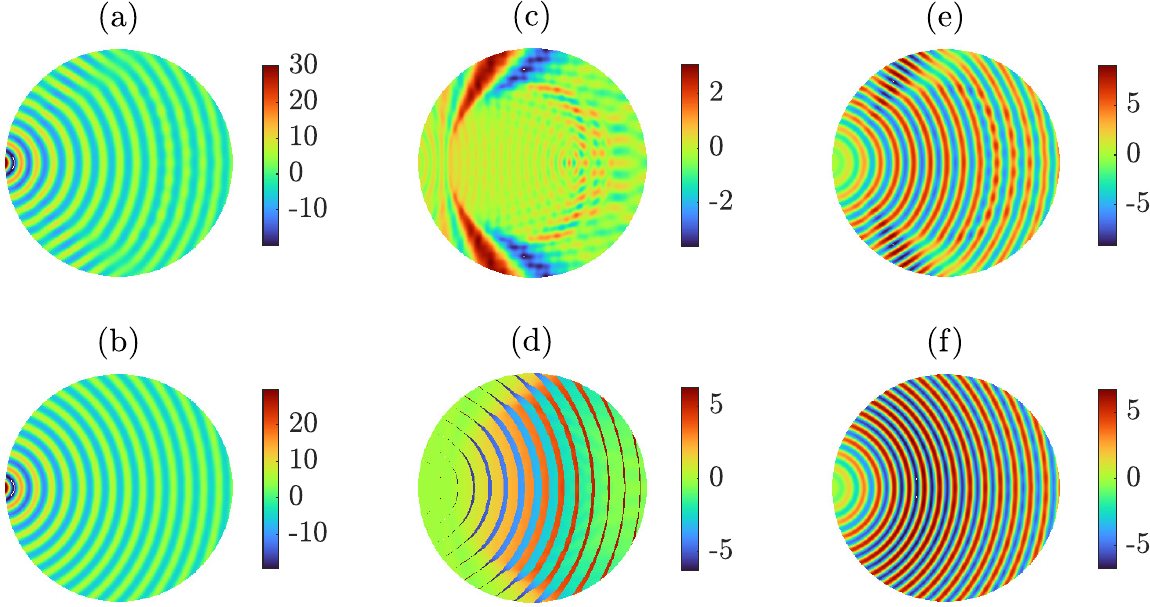}
    \caption{(a-b) Wavefields simulated in the (a) true model (\fref{Camembert}a) and (b) homogeneous background model (\fref{Camembert}b) at $f = 5$ Hz. 
    (c-d) (c) Magnitude and (d) phase differences between the true and background wavefields.
    (e-f) Weighted versions of the wavefields in (a-b) using $\bphis$ with $\gamma=10$ and $\sigma=1.5$ km.
 }
    \label{Camembert_wave}
\end{figure}
\begin{figure}[h]
    \centering
    \includegraphics[width=0.5\linewidth,trim={0 0cm 0 0},clip]{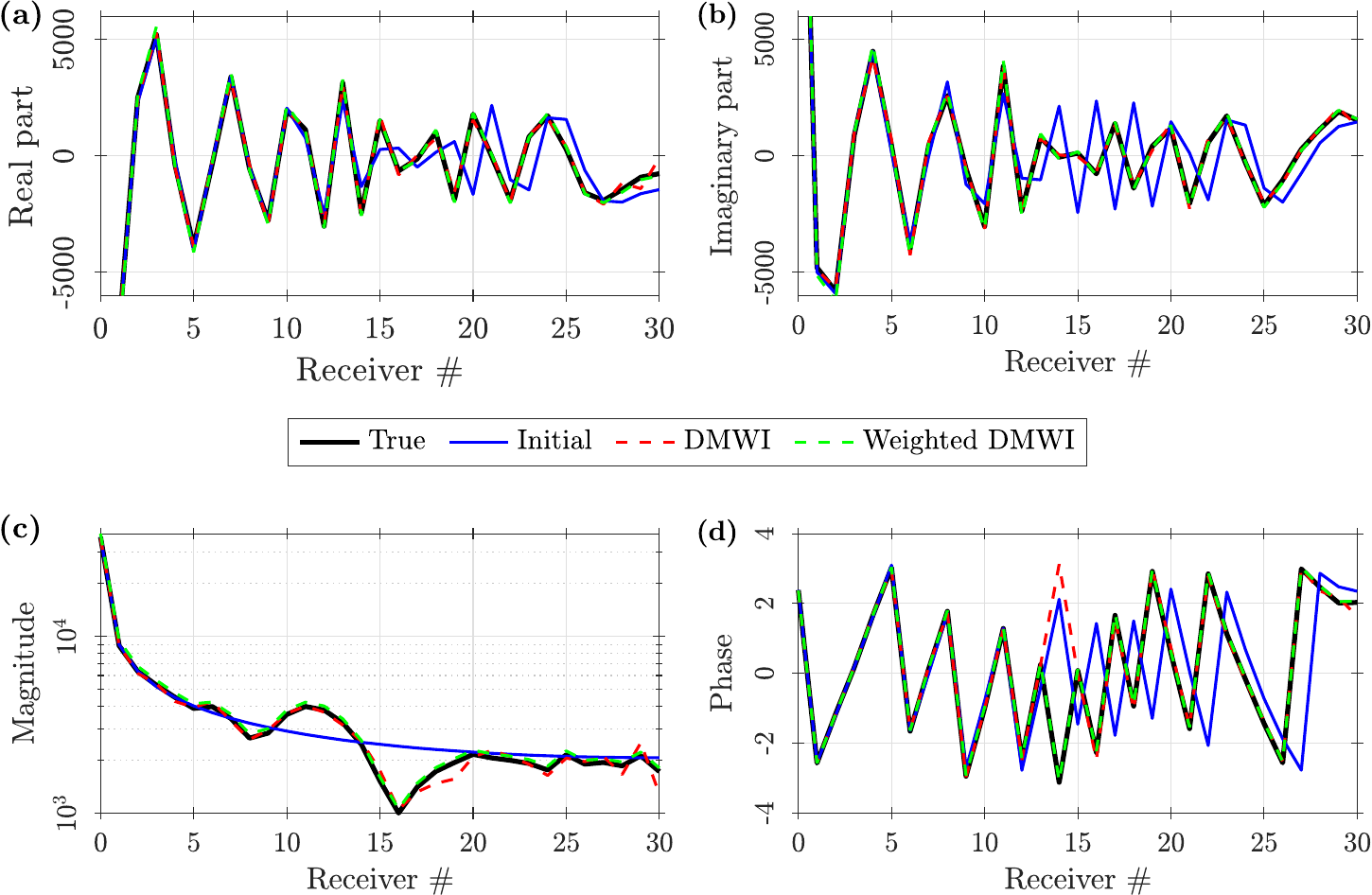}
    \caption{Observed data for the Camembert model example in \fref{Camembert}. (a) Real and (b) imaginary components of the recorded data for half of the receivers. (c) Magnitude and (d) phase of the data. Cycle skipping is evident in the phase plot (d), particularly for receivers 14-30.}
    \label{Camembert_data}
\end{figure}
\begin{figure}[h]
    \centering
    \includegraphics[width=0.5\linewidth,trim={0 0cm 0 0},clip]{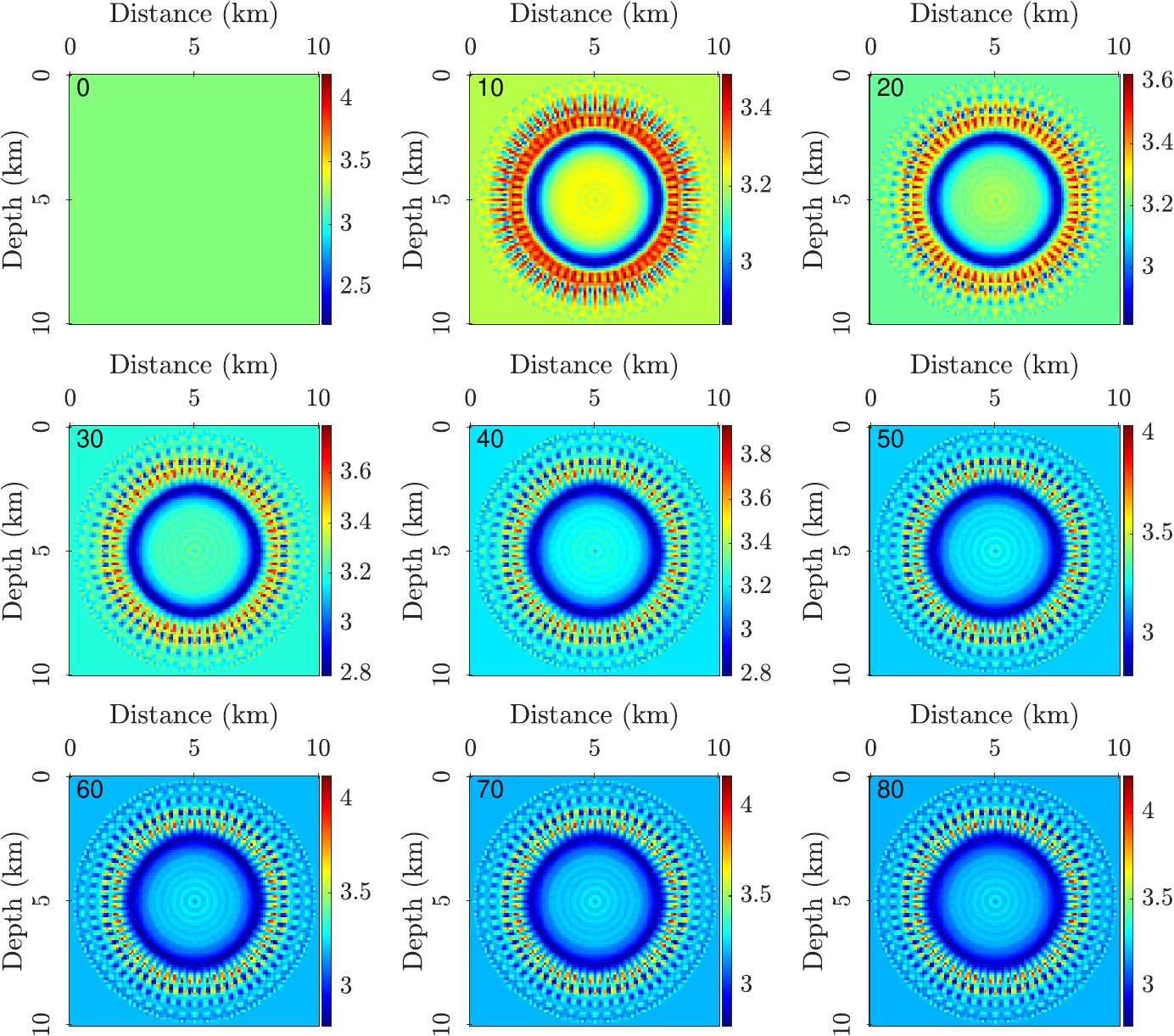}
    \caption{Inverted velocity models using the standard DMWI for the mono-frequency 5 Hz inversion. Results are shown after every 10 iterations, starting from the homogeneous initial model.}
    \label{Camembert_AL}
\end{figure}
\begin{figure}[h]
    \centering
    \includegraphics[width=0.5\linewidth,trim={0 0cm 0 0},clip]{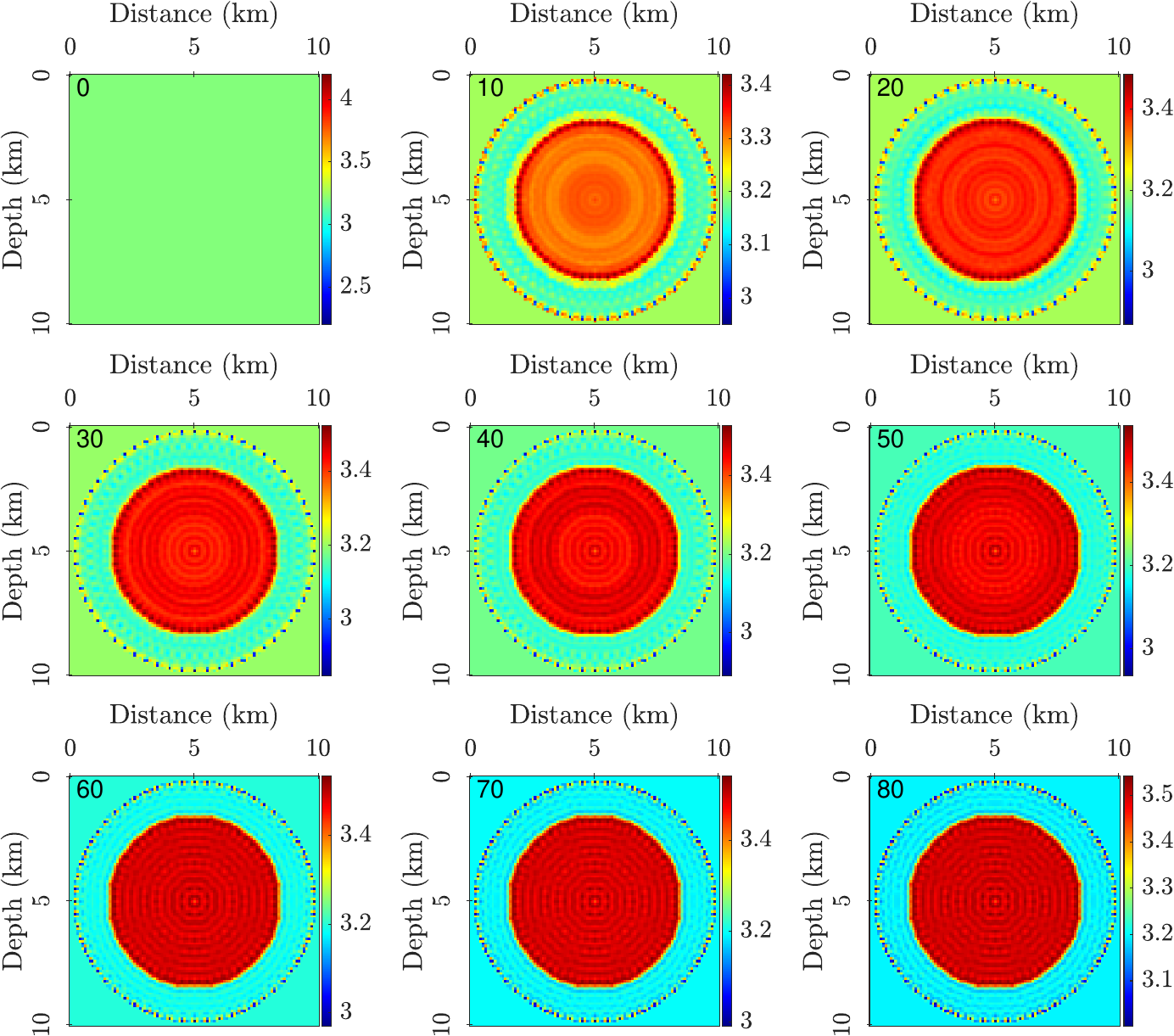}
    \caption{Same as Figure \ref{Camembert_AL} but for the weighted DMWI with $\bphis$.}
    \label{Camembert_WAL}
\end{figure}

\begin{figure}[h]
    \centering
    \includegraphics[width=1\linewidth,trim={0 0cm 0 0},clip]{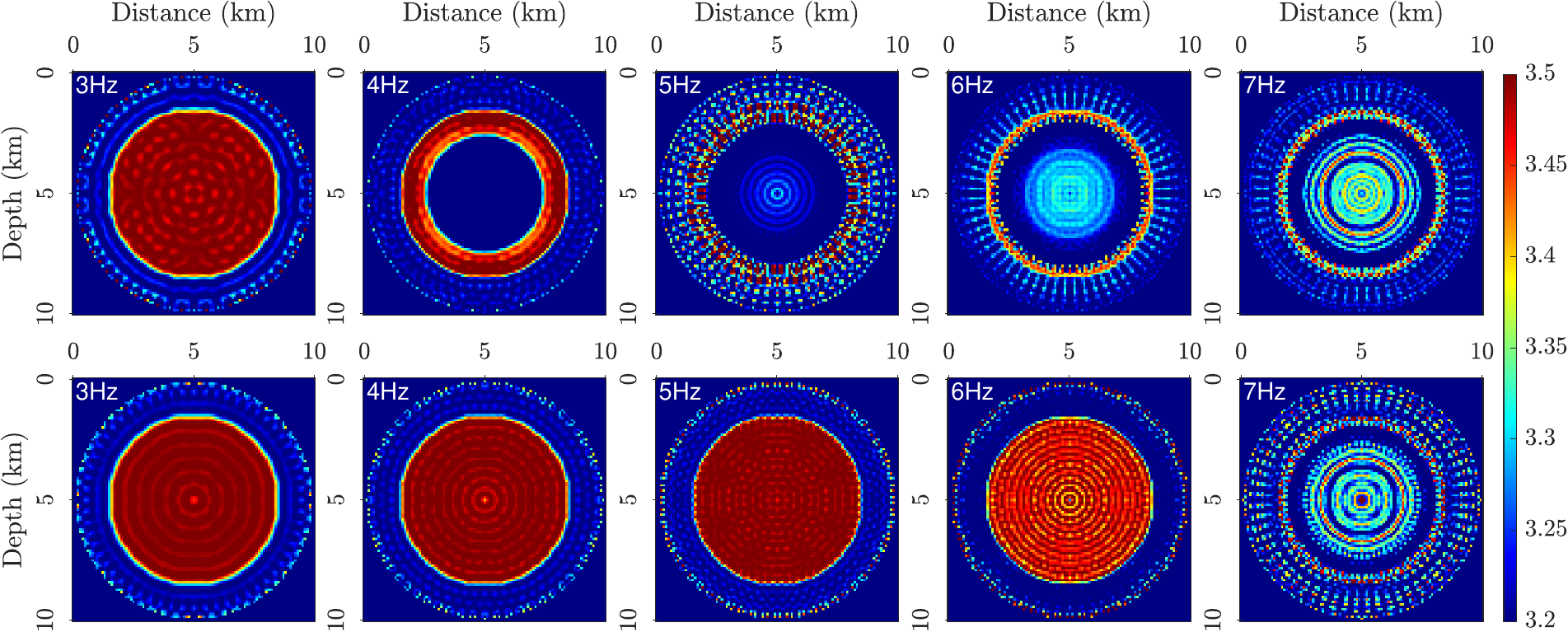}
    \caption{Inverted velocity models for different mono-frequencies (3 Hz to 7 Hz). (Top row) results from the traditional multiplier method with uniform penalty weights. (Bottom row) results from the weighted algorithm with $\bphis$.}
    \label{Camembert_freqs}
\end{figure}

\subsection{Transmission problem: Camembert anomaly.}
As the first example, we consider the transmission problem using the Camembert model, as studied by \cite{Gauthier_1986_TNI,Operto_2023_FWI}. The true model, shown in Figure~\ref{Camembert}a, spans 10 km in both depth and horizontal extent and is discretized on a 100 m grid. The source and receiver locations are marked by star and triangle, respectively. The model features a large circular velocity anomaly with a velocity of 3.5 km/s and a radius of 3.5 km, embedded in a homogeneous background with a velocity of 3.2 km/s. The acquisition setup consists of a circular array of 60 sources and receivers, evenly spaced every 6 degrees along a circle with a 4.8 km radius surrounding the anomaly. We conduct a single-frequency inversion at 5 Hz, a scenario known to induce cycle skipping when the initial model is a homogeneous velocity of 3.2 km/s (Figure~\ref{Camembert}b). 

Figures \ref{Camembert_wave}(a) and (b) compare the simulated wavefields in the true and initial models at $f = 5$ Hz, respectively. In order to highlight the challenge with the inversion process, we display the magnitude and phase differences between the true and background wavefields in \fref{Camembert_wave}(c) and (d). 
Figure~\ref{Camembert_data} presents the recorded data for half of the receivers: the real and imaginary parts are shown in Figures~\ref{Camembert_data}a-b, while Figures~\ref{Camembert_data}c-d display the magnitude and phase. The phase plot (Figure~\ref{Camembert_data}d) clearly reveals cycle skipping particularly in receivers 14-30. This challenge makes it extremely difficult for traditional gradient-based optimization algorithms to recover a model that correctly predicts the phase information. \cite{Operto_2023_FWI} tackled this issue by implementing total-variation (TV) regularization in the MWI with a stationary penalty parameter. However, we omit any model regularization here to focus exclusively on the impact of $\bphis$.
A weighting function with $\gamma = 10$ and $\sigma = 1.5$ km, was used to build $\Ws$. Applying the weighting function to the wavefields in \fref{Camembert_wave}(a-b) produces the weighted results in \fref{Camembert_wave}(e-f). We see that the amplitude of the wavefields get to be well balanced after a distance from the source. 

We performed 80 iterations of the inversion using both the standard DMWI (uniform weighting) and the proposed weighted DMWI. Notably, the standard DMWI required the true source signature, whereas the weighted DMWI did not rely on this information.  Starting from the homogeneous background model, Figure~\ref{Camembert_AL} presents the inverted velocity models obtained every 10 iterations using the DMWI. The results clearly indicate that the algorithm converged to a local minimum, yielding an incorrect velocity model. While the final estimate deviates significantly from the true model, this discrepancy is not immediately apparent from the observed data. In Figure~\ref{Camembert_data}a-b, we see that although the velocity model is incorrect, it still predicts the data reasonably well. The real and imaginary components of the predicted data (red line) align closely with the observed data (black line). Even the amplitude match is quite satisfactory, particularly up to receiver 14 (Figure~\ref{Camembert_data}c). This agreement is unsurprising, given that the algorithm started with the correct background model.  
However, a major issue with the estimated model becomes evident in the phase plot (Figure~\ref{Camembert_data}d). While the phase is accurately fitted across most receivers, a clear discrepancy appears at receiver 14. At this location, the predicted phase follows the incorrect phase of the initial wavefield, indicating that the algorithm failed to recover the correct phase information at this receiver.  
In contrast, Figure~\ref{Camembert_WAL} presents the results obtained using the weighted DMWI. Two key observations emerge:  
(i) The algorithm successfully converged to an accurate solution, demonstrating that the $\bphis$ effectively enhances the convexity of the objective function.  
(ii) A reasonable approximation of the velocity anomaly was recovered within just 10 iterations, showing the strong preconditioning effect of the $\bphis$.  
The data predicted by the final model obtained with the weighted algorithm is shown in Figure~\ref{Camembert_data} (green line). Unlike the unweighted case, both the phase and amplitude information are accurately reconstructed, confirming the advantage of the proposed approach.

We further tested the inversion using data at different frequencies, keeping the setup identical to the previous example but varying the input frequency. Figure~\ref{Camembert_freqs} shows the inverted velocity models after 80 iterations for frequencies ranging from 3 to 7 Hz. Both the DMWI and weighted DMWI algorithms successfully recovered the velocity model when inverting the 3 Hz data. However, the DMWI algorithm failed to converge to the true model for higher frequencies, as shown in the top row of Figure~\ref{Camembert_freqs}.  
In contrast, the weighted algorithm remained effective up to 6 Hz, although the quality of the inversion degraded as the frequency increased (bottom row of Figure~\ref{Camembert_freqs}). At 7 Hz, however, both algorithms failed to converge to the correct solution. This result indicates that while the proposed weighting approach significantly enhances the convexity of the objective function and extends the frequency range over which accurate inversion is possible, it cannot fully eliminate non-convexity at higher frequencies.



\begin{figure}[h]
    \centering
    \includegraphics[width=0.5\linewidth,trim={0 0cm 0 0},clip]{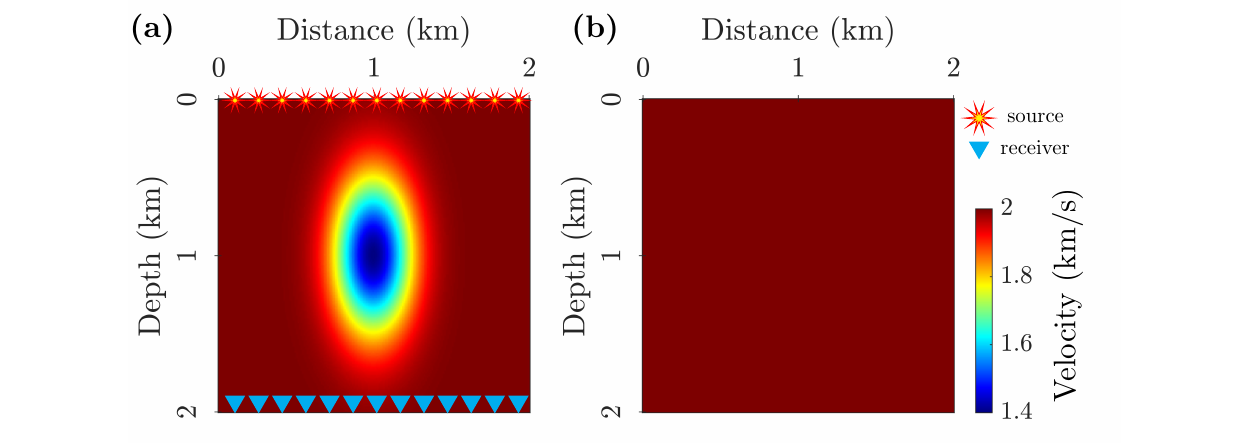}
    \caption{(a) True velocity model for the Gaussian anomaly. The positions of the sources and receivers are shown by stars and triangles. (b) Initial velocity model.}
    \label{Gaussian_model}
\end{figure}
\begin{figure}[h]
    \centering
    \includegraphics[width=0.5\linewidth,trim={0 0cm 8.1cm 0},clip]{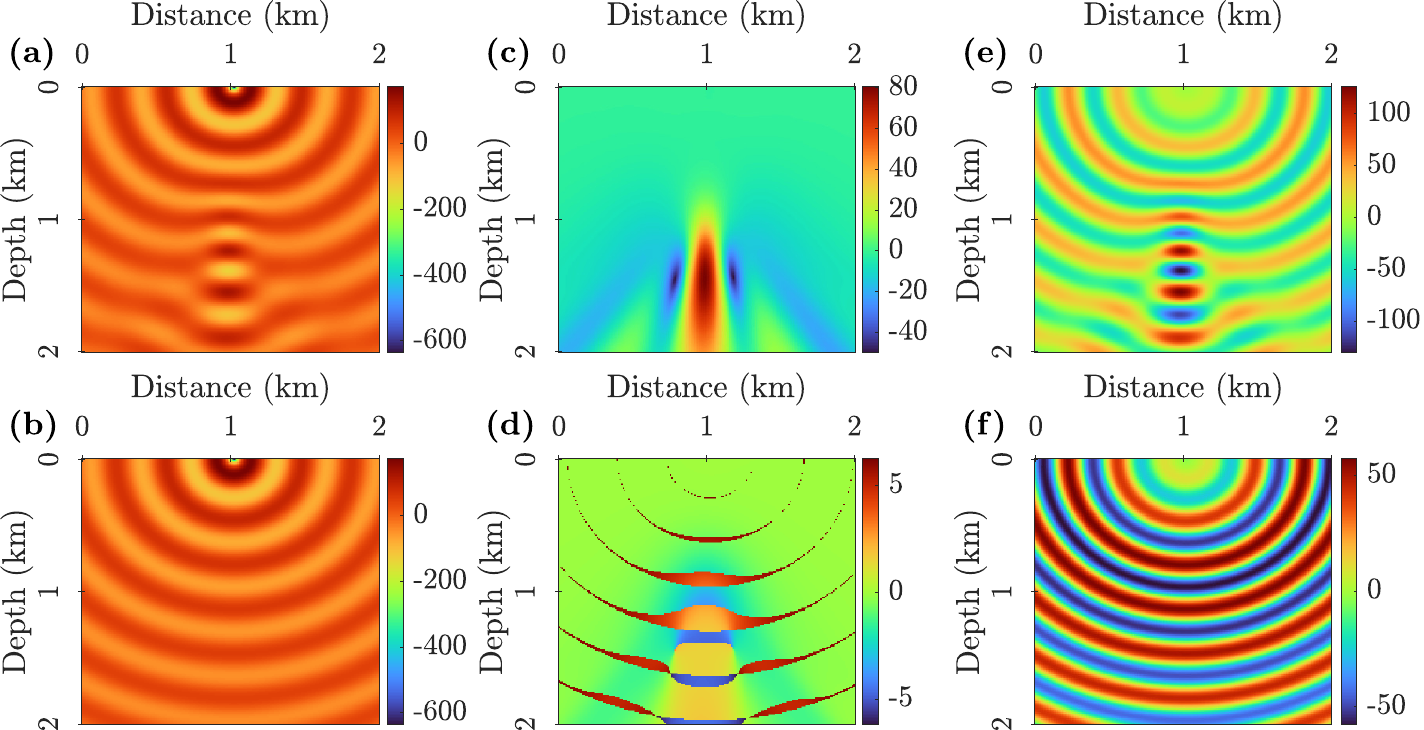}
    \caption{(a-b) Wavefields simulated in the (a) true model and (b) homogeneous background models  at $f = 6$ Hz for the Gaussian model in \fref{Gaussian_model}. 
    (c-d) (c) Magnitude and (d) phase differences between the true and background wavefields for a source at 1 km distance.
    (e-f) Weighted versions of the wavefields in (a-b).
    }
    \label{Gaussian_waves}
\end{figure}
\begin{figure}[h]
    \centering
    \includegraphics[width=0.5\linewidth,trim={0 0cm 0 0},clip]{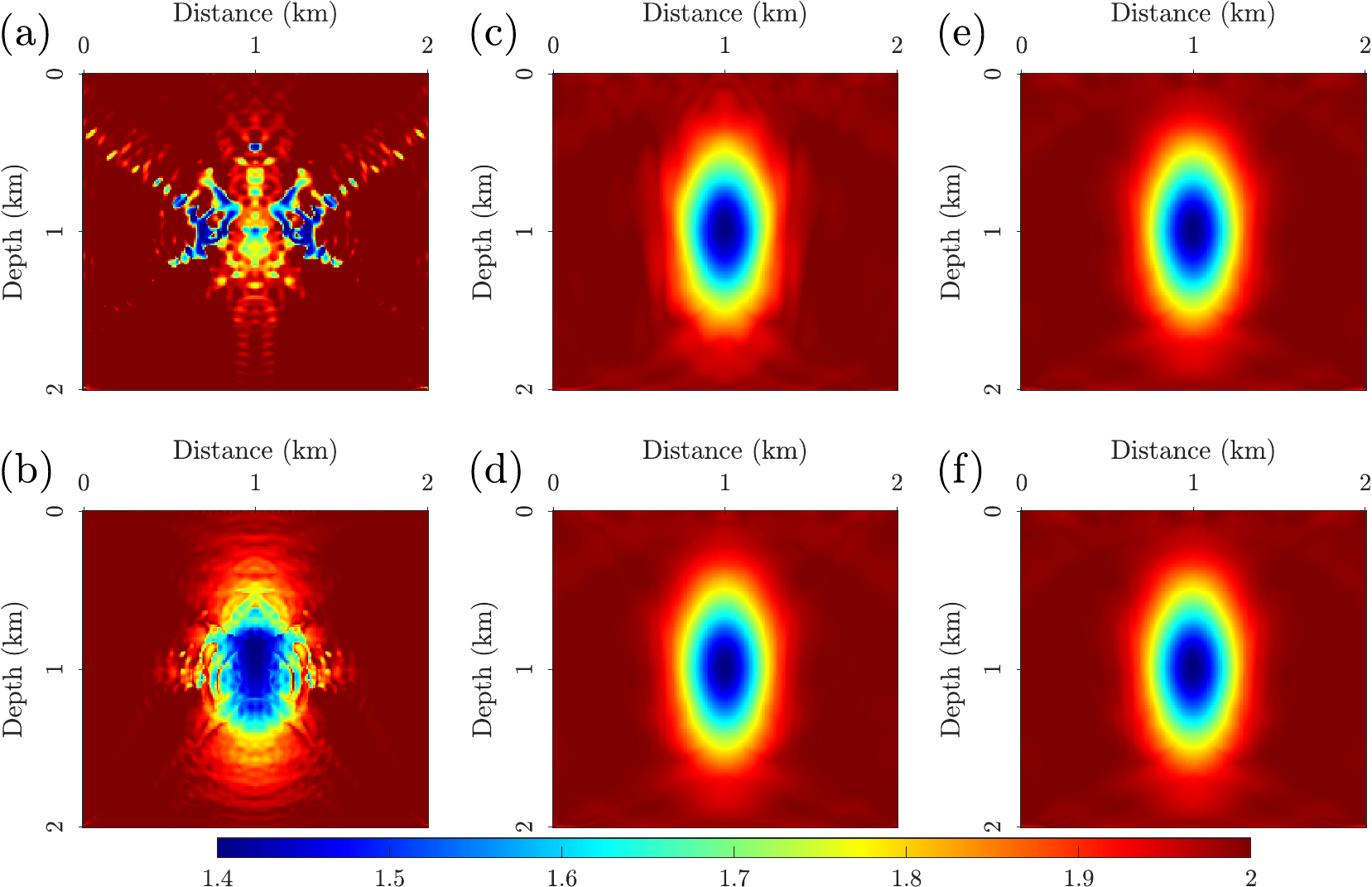}
    \caption{(a)  Reconstructed velocity model after inverting the 6 Hz data by DMWI. 
    (b)  Final DMWI result by sequentially inverting the 6, 10, 14, and 18 Hz data.
    (c) Reconstructed velocity model after inverting the 6 Hz data by weighted DMWI.
    (d)  Final weighted DMWI result by sequentially inverting the 6, 10, 14, and 18 Hz data.
    (e) Same as (d), but with receivers aligned to the grid while sources are off-grid. (f) Same as (d), but with both sources and receivers positioned off-grid.}
    \label{Gaussian_results}
\end{figure}

\begin{figure}[h]
    \centering
    \includegraphics[width=0.5\linewidth,trim={0 0cm 0 0},clip]{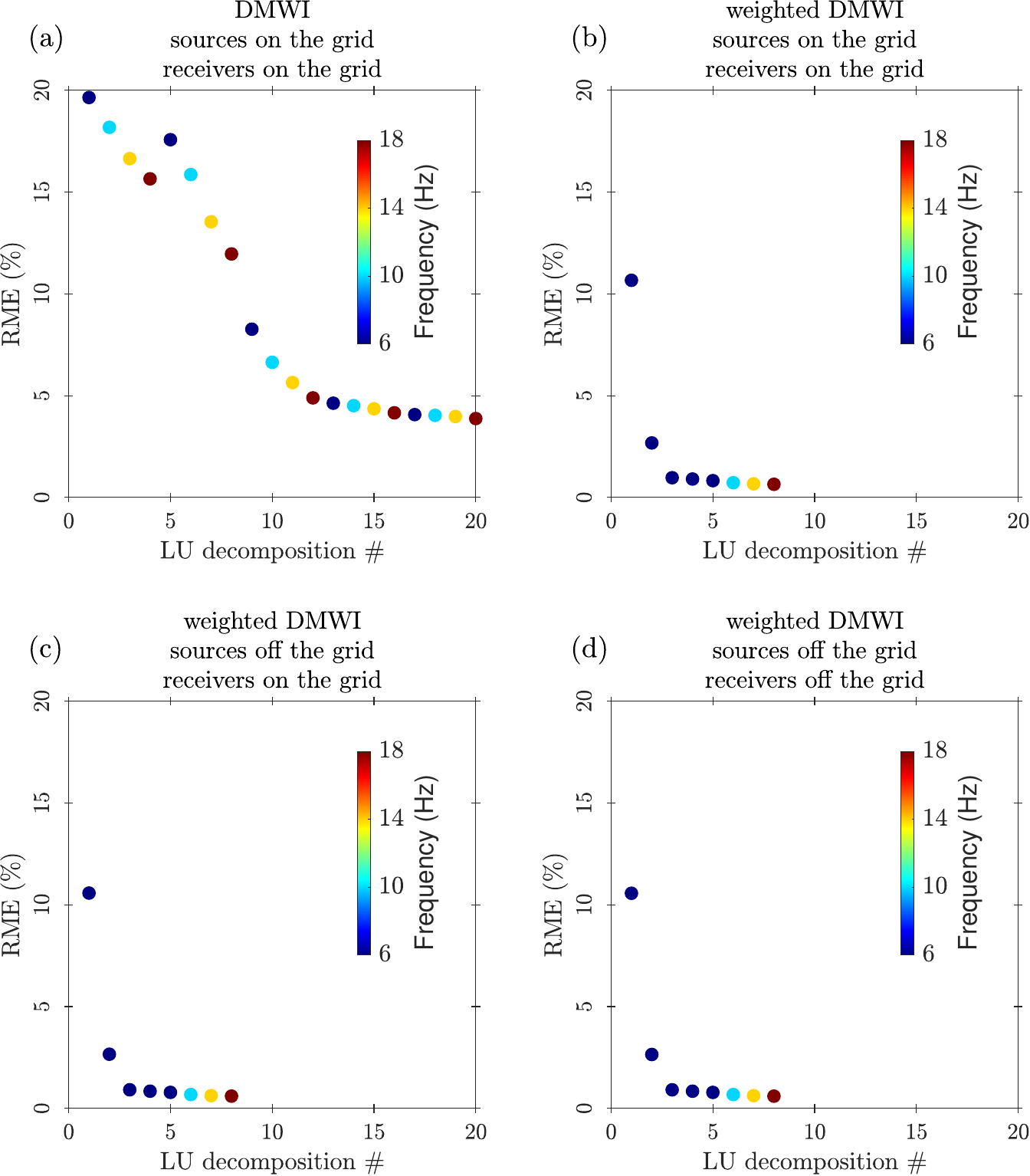}
    \caption{(a-d) Relative model error versus the number of LU factorizations for the results shown in \fref{Gaussian_results}(c-f). The color coding indicates the frequency of the input data used for each inversion step.}
    \label{Gaussian_errors}
\end{figure}

\subsection{Transmission problem: Gaussian anomaly.}

As the second example we use the transmission problem described in \cite{Huang_2018_SEW}. The model, spanning 2 km in both width and depth, poses significant challenges for FWI algorithms due to multipathing caused by a strong low-velocity anomaly (Figure~\ref{Gaussian_model}a).  The acquisition setup includes 200 equally spaced receivers at the bottom and 50 sources equally distributed along the top, simulating a transmission problem. FWI is conducted using single-frequency data at 6, 10, 14, and 18 Hz, starting from a homogeneous background velocity (Figure~\ref{Gaussian_model}b). 

Figures \ref{Gaussian_waves}(a) and (b) show the simulated wavefields in the true and initial models at $f = 6$ Hz, respectively. To emphasize the inversion challenge, the magnitude and phase differences between the true and background wavefields are displayed in \fref{Gaussian_waves}(c) and (d). The phase plot reveals strong phase wrapping in the central receiver region around 1 km distance, coinciding with large amplitude differences. This situation makes the inversion susceptible to local minima, since the $\ell_2$-norm misfit tends to prioritize fitting the large amplitudes at these receivers, even if it means matching the wrong phase.

The inversion is carried out sequentially in frequency, starting from the lowest frequency (6 Hz). For the first frequency, we construct the $\bphis$ using $\sigma=0.5$ km, and for all subsequent higher frequencies we use $\sigma=0.25$ km. Applying the weighting function to the wavefields in \fref{Gaussian_waves}(a–b) yields the weighted results shown in \fref{Gaussian_waves}(e–f). 

To perform the standard DMWI, we first carried out 100 iterations for the initial 6 Hz data, updating the model every 20 iterations ($\texttt{maxit}=20$ in Algorithm~\ref{alg_dual}). The result, shown in Figure~\ref{Gaussian_results}a, indicates that the algorithm became trapped in a local minimum. Without an appropriate weighting function, escaping such local minima and reaching the global minimum is extremely challenging. To improve the performance of DMWI, we next performed multi-frequency inversion using 20 iterations for each frequency (6, 10, 14, and 18 Hz) in sequence, repeating the full cycle five times. This procedure required a total of 20 LU factorizations and 400 iterations. The final result is presented in Figure~\ref{Gaussian_results}(b), with the corresponding error-convergence curve (measured against the number of LU factorizations) shown in Figure~\ref{Gaussian_errors}(a).

We then applied the proposed weighted DMWI. Using 100 iterations for the initial 6 Hz inversion with model updates every 20 iterations ($\texttt{maxit}=20$ in Algorithm~\ref{alg_SI_dual}), the method produced a satisfactory reconstruction even at the lowest frequency alone (Figure~\ref{Gaussian_results}c). Inversion of higher frequencies subsequently refined the structural details (\fref{Gaussian_results}d). The associated convergence curve versus the number of LU factorizations is shown in Figure~\ref{Gaussian_errors}b. Overall, the weighted approach not only removed the need for explicit source signature estimation, but also significantly improved problem conditioning and robustness against cycle skipping—a challenge emphasized in previous studies (e.g., \cite{Symes_2020_FWI}).

\subsubsection{Sources and Receivers Off the Grid.}
In the previous tests, we assumed that the sources and receivers were positioned on grid points. However, in practical applications, sources and receivers are often located off the grid, requiring interpolation, which can introduce inaccuracies. This limitation is mitigated in the proposed algorithm because it is based on extended sources, which inherently distribute the source energy over a volume or surface. Furthermore, the $\bphis$ depends on the continuous source location, $\bold{x}_s$, without requiring it to align with the grid. 

We first considered the case where sources were positioned off the grid while receivers remained on the grid. Source locations were perturbed by $\pm \frac{\Delta x}{2}$, and data were simulated using the method described in \cite{Hicks_2002_ASR}. During the inversion, the exact source locations were used to construct the $\bphis(\bx)$. The inverted velocity model obtained within a single inversion cycle is shown in Figure~\ref{Gaussian_results}e. The result is visually indistinguishable from the case where sources were aligned with the grid (Figure~\ref{Gaussian_results}d and \ref{Gaussian_errors}). This example demonstrates the robustness of the proposed method in handling off-the-grid source locations, as we see from the relative error curve in \fref{Gaussian_errors}c.

Next, we extended our analysis to the case where both sources and receivers were positioned off the grid. Source locations were perturbed by \(\pm \frac{\Delta x}{2}\), while receiver positions were similarly shifted away from grid points. The data were modeled using the approach of \cite{Hicks_2002_ASR}. During the inversion, we employed a simple linear interpolation scheme to compute the predicted data at each receiver location, based on the values of the wavefield at the four surrounding grid points. Although this interpolation introduces some errors, these errors are effectively managed by the error correction action of the multipliers in the algorithm. Remarkably, the inverted velocity model obtained within a single inversion cycle (Figure~\ref{Gaussian_results}f) is nearly identical to the case where receivers were positioned on the grid (Figure~\ref{Gaussian_results}e). Moreover, the relative error curve remains unchanged (\fref{Gaussian_errors}d), demonstrating that the proposed algorithm retains its accuracy and robustness even under off-grid conditions for both sources and receivers. This highlights the flexibility of the method in handling practical acquisition setups with minimal impact on performance.



\begin{figure}[!th]
    \centering
    \includegraphics[width=1\linewidth,trim={0 0cm 0 0},clip]{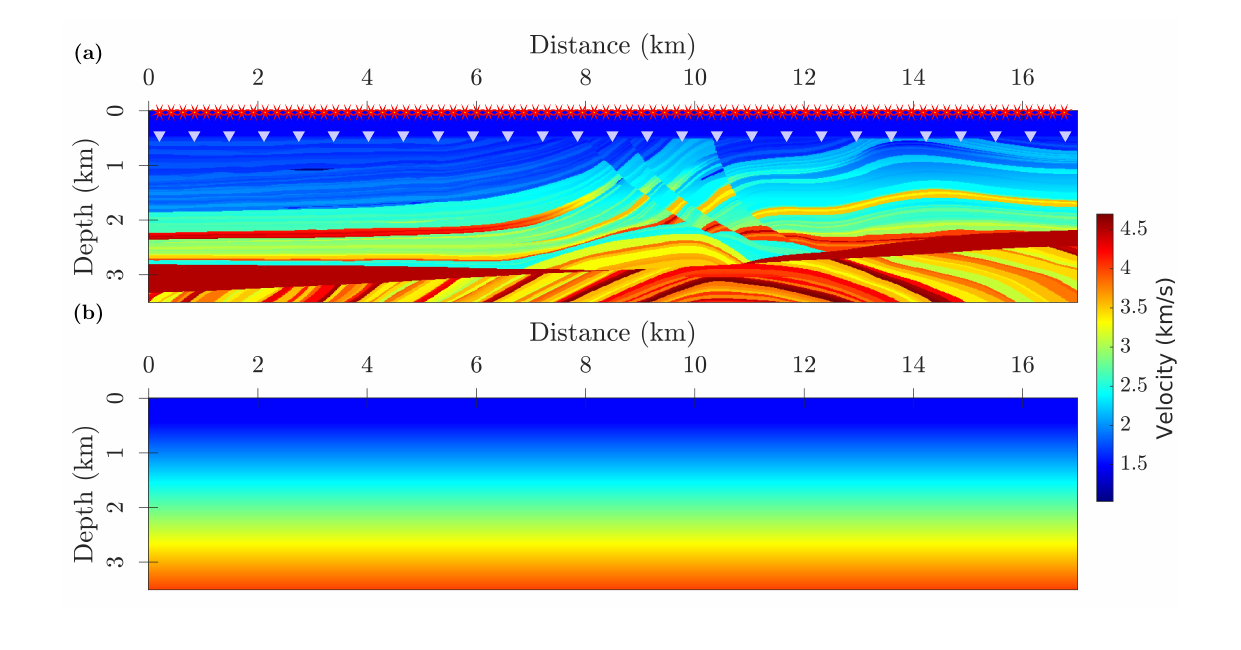}
    \caption{(a) The Marmousi II velocity model.  The positions of the sources and receivers are shown by stars and triangles. (b) The 1D initial model used for inversion.}
    \label{Marmousi_model}
\end{figure}
\begin{figure}[!th]
    \centering
    \includegraphics[width=1\linewidth,trim={0 0cm 0 0},clip]{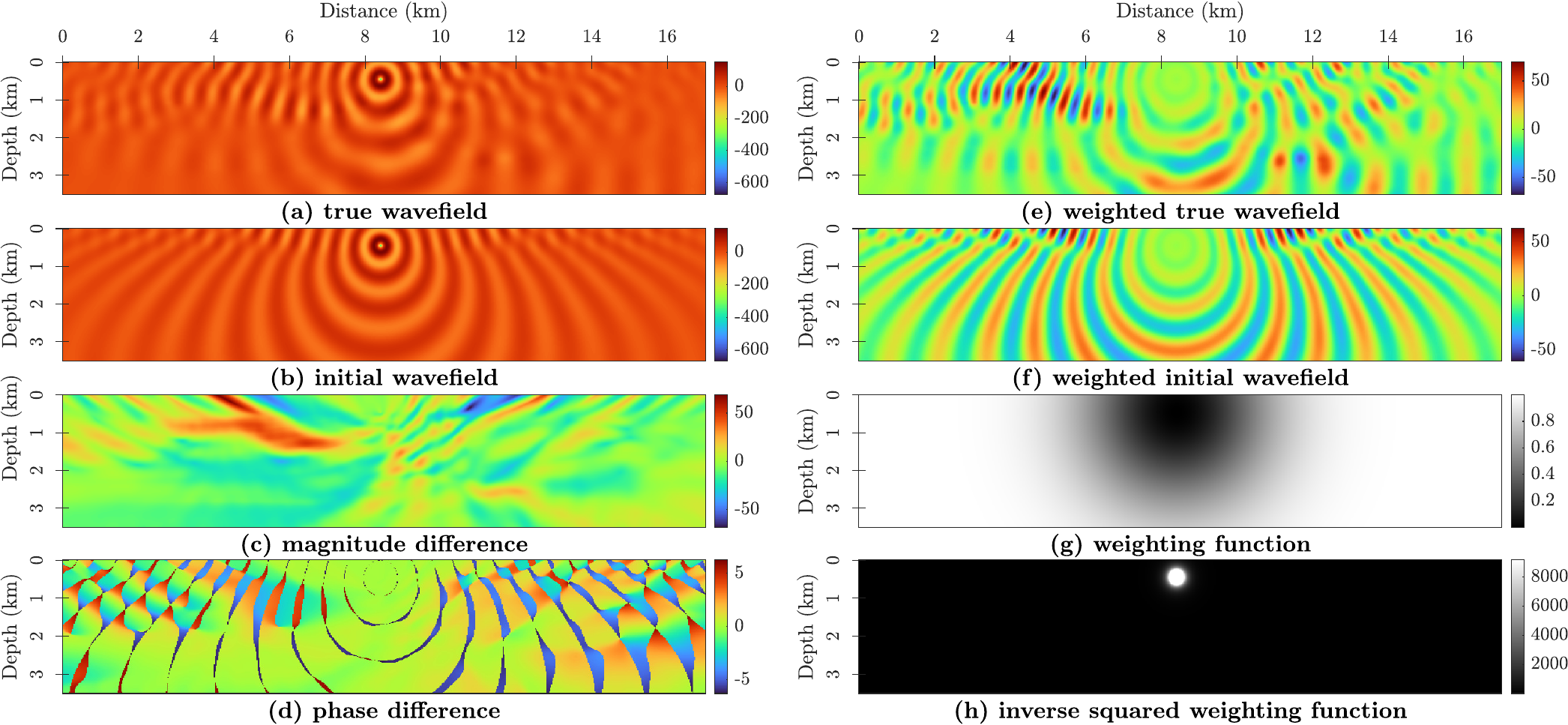}
    \caption{(a-b) Wavefields simulated in the (a) true model and (b) initial model at $f = 3$ Hz for the Marmousi model in \fref{Marmousi_model}. 
    (c-d) (c) Magnitude and (d) phase differences between the true and background wavefields.
    (e-f) Weighted versions of the wavefields in (a-b) using a $\bphis(\bx)$ with $\gamma=10$ and $\sigma=1$ km.
    (g) The used weighting function. (h) Inverse squared weighting function.
    }
    \label{Marmousi_wavefields}
\end{figure}

\begin{figure}[!th]
    \centering
    \includegraphics[width=0.8\linewidth,trim={0 0cm 0 0},clip]{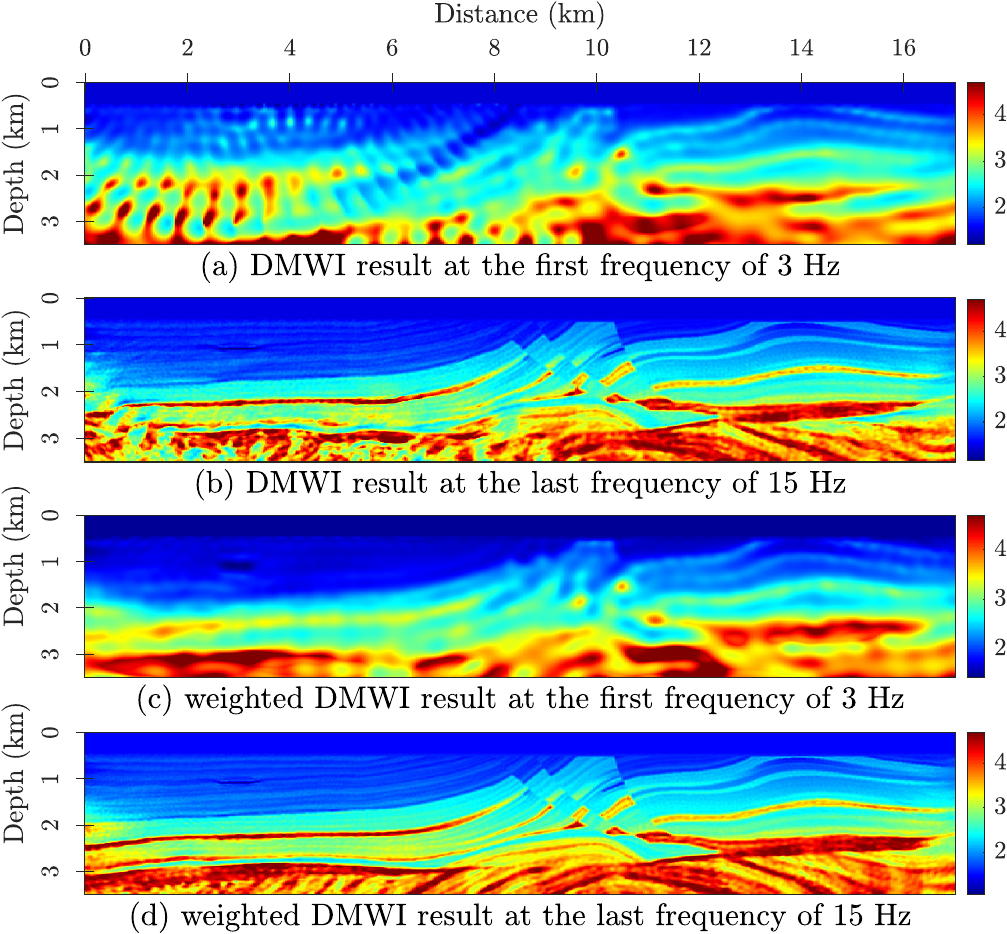}
    \caption{
    (a)  Reconstructed velocity model after inverting the 3 Hz data by DMWI. 
    (b)  Final DMWI result by sequentially inverting the 3-15 Hz data.
    (c) Reconstructed velocity model after inverting the 3 Hz data by weighted DMWI.
    (d)  Final weighted DMWI result by sequentially inverting the 3-15 Hz data.
    }
    \label{Marmousi_results}
\end{figure}
\begin{figure}[!th]
    \centering
    \includegraphics[width=0.8\linewidth,trim={0 0cm 0 0},clip]{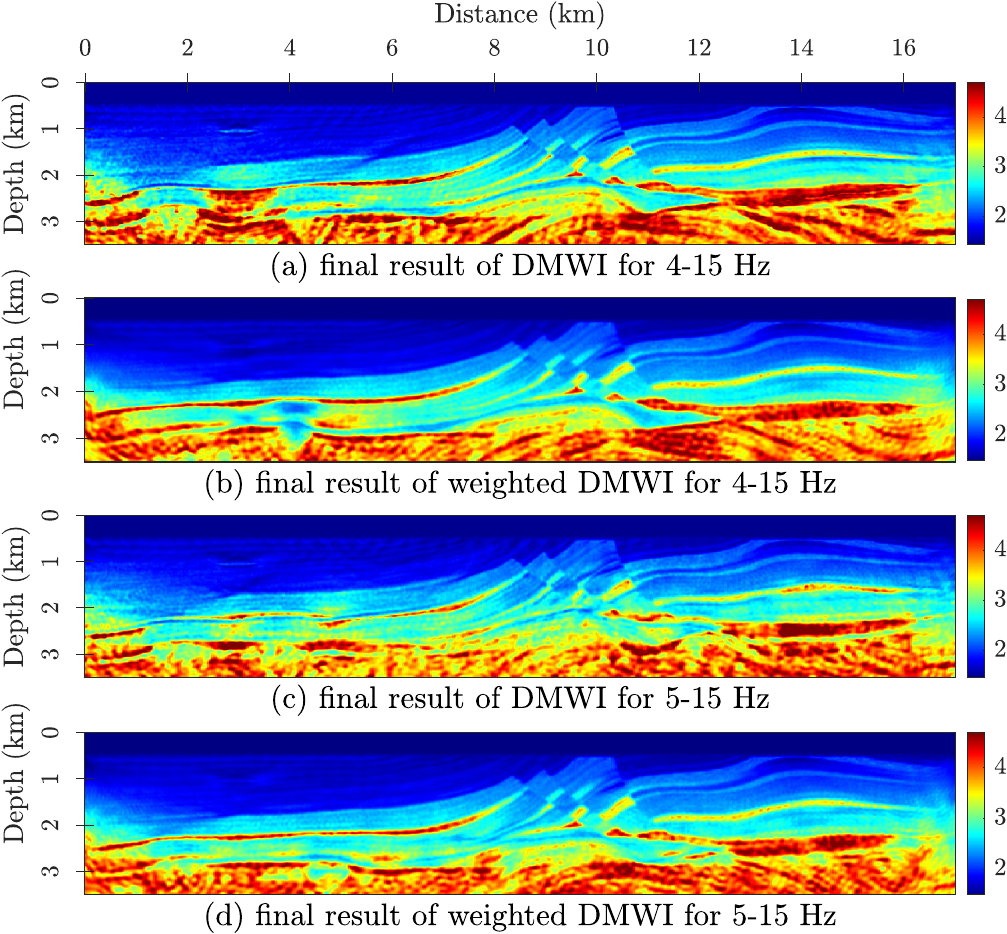}
    \caption{
    (a) Inverted velocity model from (a) 4-15 Hz data by DMWI, (b) 4-15 Hz data by weighted DMWI, (c) 5-15 Hz data by weighted DMWI, and (d) 5-15 Hz data weighted DMWI.
    }
    \label{Marmousi_4Hz_5Hz}
\end{figure}
\begin{figure}[!h]
    \centering
    \includegraphics[width=0.5\linewidth,trim={0 0cm 0 0},clip]{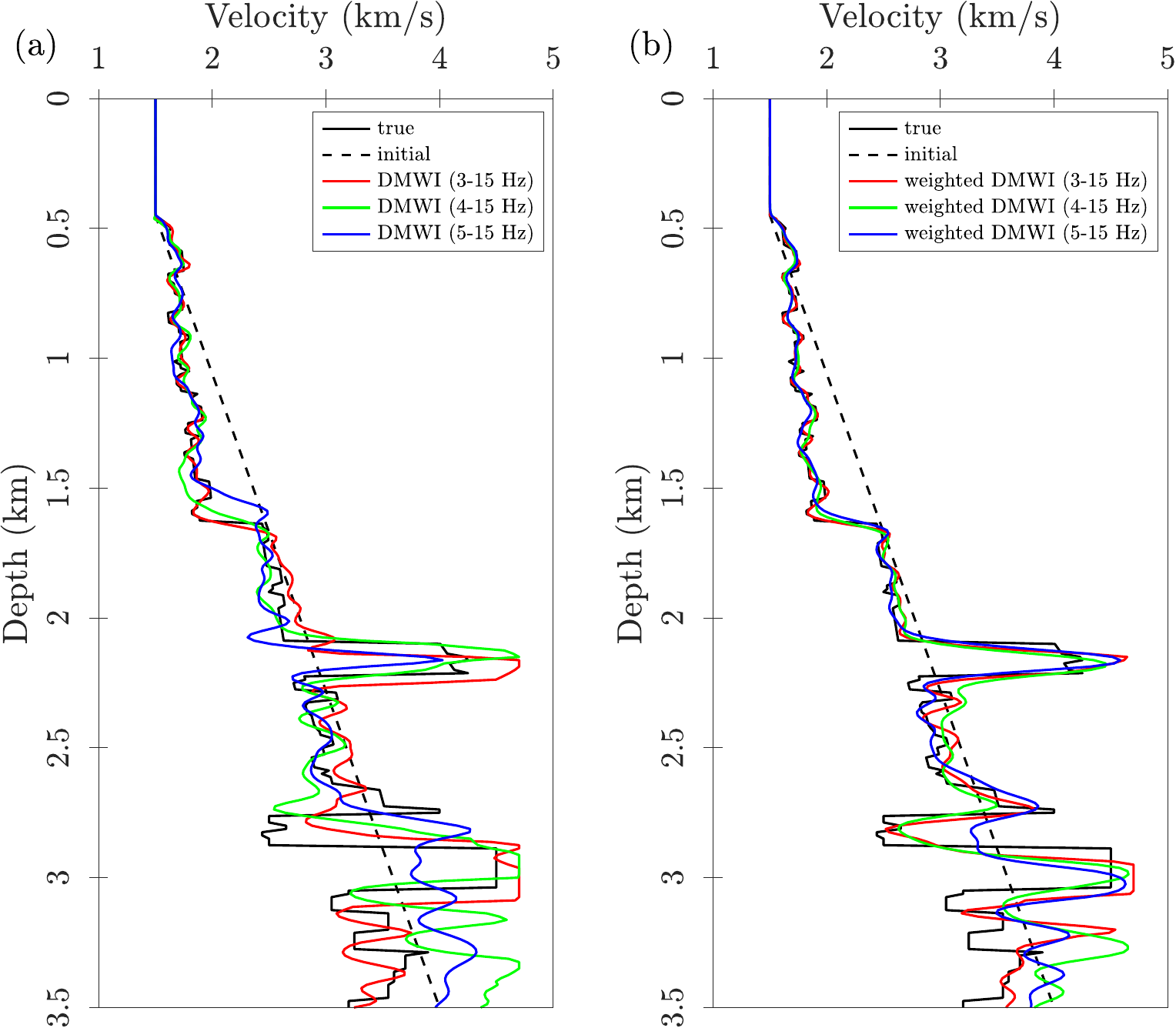}
    \caption{Direct comparison, at horizontal distance 6 km, between the true (black), initial (dashed black), and inverted velocity models from 3-15 Hz data (red), 4-15 Hz data (green),  5-15 Hz data (blue), obtained by (a) DMWI and (b) weighted DMWI.
    }
    \label{Marmousi-logs}
\end{figure}
\begin{figure}[!h]
   \centering
   \includegraphics[width=0.5\linewidth,trim={0 0cm 0 0},clip]{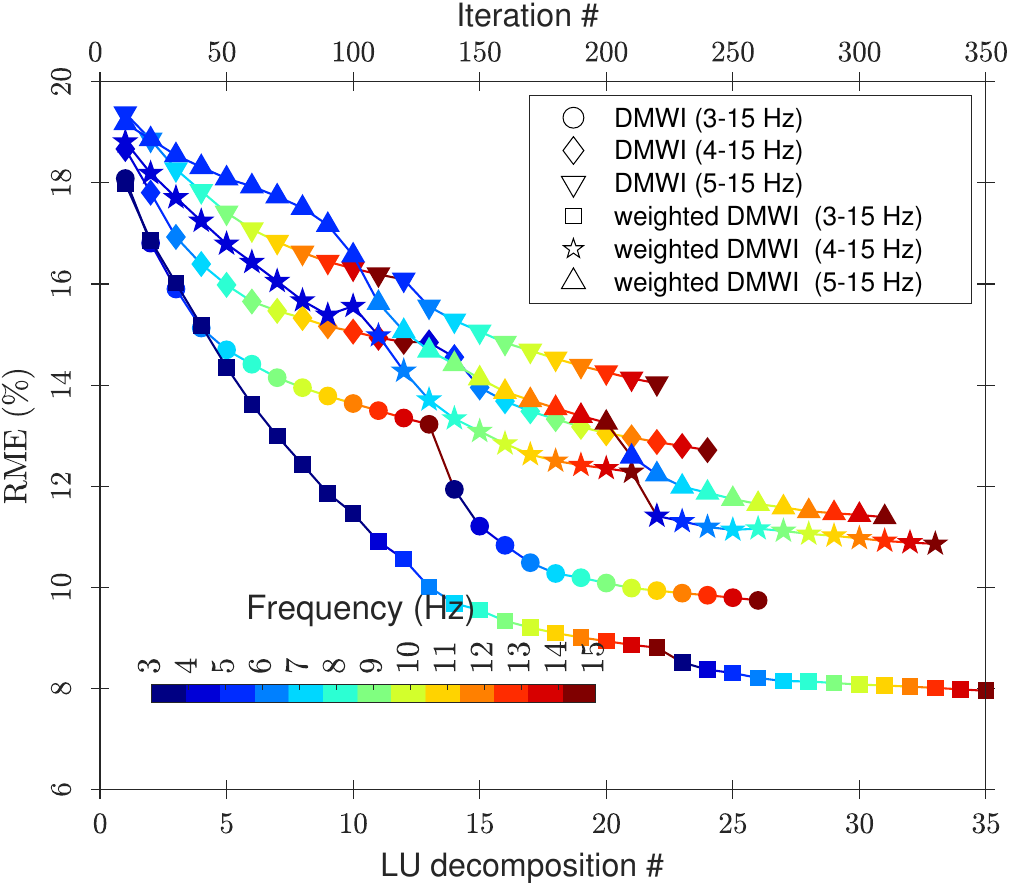}
   \caption{Relative model error as a function of the iteration and the number of LU factorizations during the inversion of the Marmousi model. The color coding indicates the frequency of the input data used for each inversion step.    
   }
   \label{Marmousi-error}
\end{figure}

\subsection{Marmousi II model.}
We further evaluated the weighted DMWI using the Marmousi II velocity model (Figure~\ref{Marmousi_model}a), which spans 17 km in width and 3.5 km in depth. The subsurface model was discretized with a 12.5 m grid interval, resulting in a mesh of 281 × 1361 points. The long-offset stationary acquisition consisted of 49 hydrophones spaced approximately 350 m apart on the seabed and 137 pressure sources positioned about 125 m apart at a depth of 50 m. Notably, for the weighted DMWI, neither the sources nor the receivers were aligned with grid points. Single-frequency data from 3 Hz to 15 Hz were used for inversion, starting from a simple 1D initial velocity model (Figure~\ref{Marmousi_model}b).  

Figures \ref{Marmousi_wavefields}(a) and (b) display the simulated wavefields in the true and initial models at $f = 3$ Hz, respectively. The corresponding magnitude and phase differences are shown in \fref{Marmousi_wavefields}(c) and (d). The phase plot reveals strong phase wrapping on both sides of the source, particularly on the left, where it coincides with large amplitude differences. This combination makes the inversion highly susceptible to local minima, particularly challenging to reconstruct the left part of the model. 
To address this, we construct the $\bphis(\bx)$ using $\sigma=1$ km. Applying the weighting function to the wavefields in \fref{Marmousi_wavefields}(a–b) produces the weighted results shown in \fref{Marmousi_wavefields}(e–f). Finally, Figures \ref{Marmousi_wavefields}(g) and (h) show the weighting function and its inverse square, respectively. In the latter case, the color axis is clipped at the value corresponding to $\tfrac{1}{4}\lambda_{\mathrm{w}}$ for better visualization.

The inversion was carried out sequentially, starting from the lowest frequency (3 Hz). Both the standard and weighted DMWI methods were applied. For each method, we performed 10 iterations to invert each frequency data, updating the LU factorizations every 10 iterations (\texttt{maxit} = 10 in Algorithms \ref{alg_dual} and \ref{alg_SI_dual}). Then, we progressively incorporated higher frequencies up to 15 Hz in 1 Hz increments. After completing this cycle, the final model was used as the starting point for a second cycle of inversion, again covering frequencies from 3–15 Hz with 10 iterations per frequency. 
The inversion result obtained from the 3 Hz data using standard DMWI is shown in \fref{Marmousi_results}(a), while the final result after two full cycles is displayed in \fref{Marmousi_results}(b). As expected, the resolution improves significantly, particularly in the right part of the model. Although the result from the 3 Hz inversion alone is of limited quality, it is greatly enhanced after incorporating the higher frequencies. For the weighted DMWI, the weighting function was constructed with $\sigma = 1$ km for the first frequency and $\sigma = 0.5$ km for all subsequent higher frequencies. The corresponding inverted velocity models at 3 Hz and after the final cycle are shown in Figures \ref{Marmousi_results}(c) and (d). These results demonstrate a clear improvement in reconstruction quality compared to the DMWI. 

To examine the robustness of the algorithm in the absence of low-frequency data, we repeated the inversion using both methods under the assumption that the lowest frequency of 3 Hz was unavailable. The final results obtained from inverting the 4–15 Hz data with DMWI and weighted DMWI are shown in \fref{Marmousi_4Hz_5Hz}(a) and (b), respectively. While the right part of the model is recovered reasonably well by both methods, DMWI fails to reconstruct the left part of the model. Next, we further increased the lowest available frequency to 5 Hz and repeated the inversion. The corresponding results obtained from the 5–15 Hz data for DMWI and weighted DMWI are shown in \fref{Marmousi_4Hz_5Hz}(c) and (d), respectively. These results demonstrate that the weighted DMWI is more robust to the lack of low-frequency data. This observation is further supported by the vertical log plots in Figure~\ref{Marmousi-logs}, which compare the true model, the initial model, and the estimates from DMWI and weighted DMWI along a vertical profile at 6 km offset. The weighted DMWI consistently produces velocity profiles that remain closer to the true model across depth for all cases considered.
Finally, the relative model error plotted in Figure~\ref{Marmousi-error} confirms the improved stability and convergence of the weighted DMWI compared to the standard formulation.



\begin{figure}[th]
    \centering
    \includegraphics[width=1\linewidth,trim={0 0cm 0 0},clip]{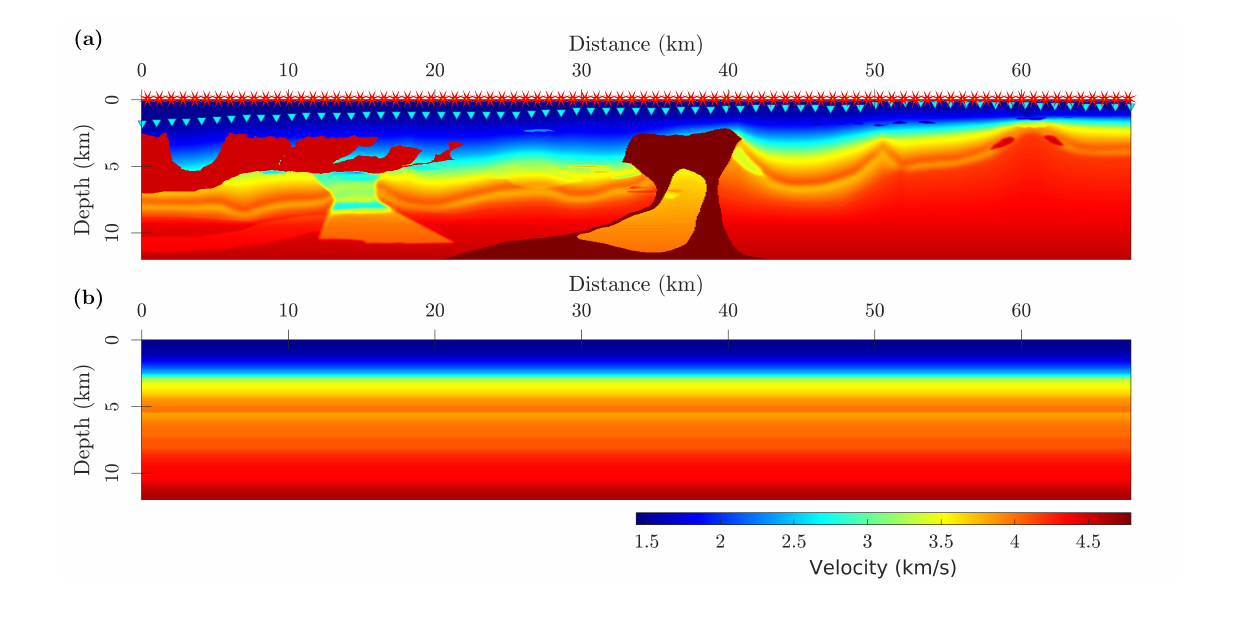}
    \caption{(a) The 2004 BP salt model. (b) The 1-D initial model used for the inversion.}
    \label{BP_model}
\end{figure}
\begin{figure}[th]
    \centering
    \includegraphics[width=1\linewidth,trim={0 0cm 0 0},clip]{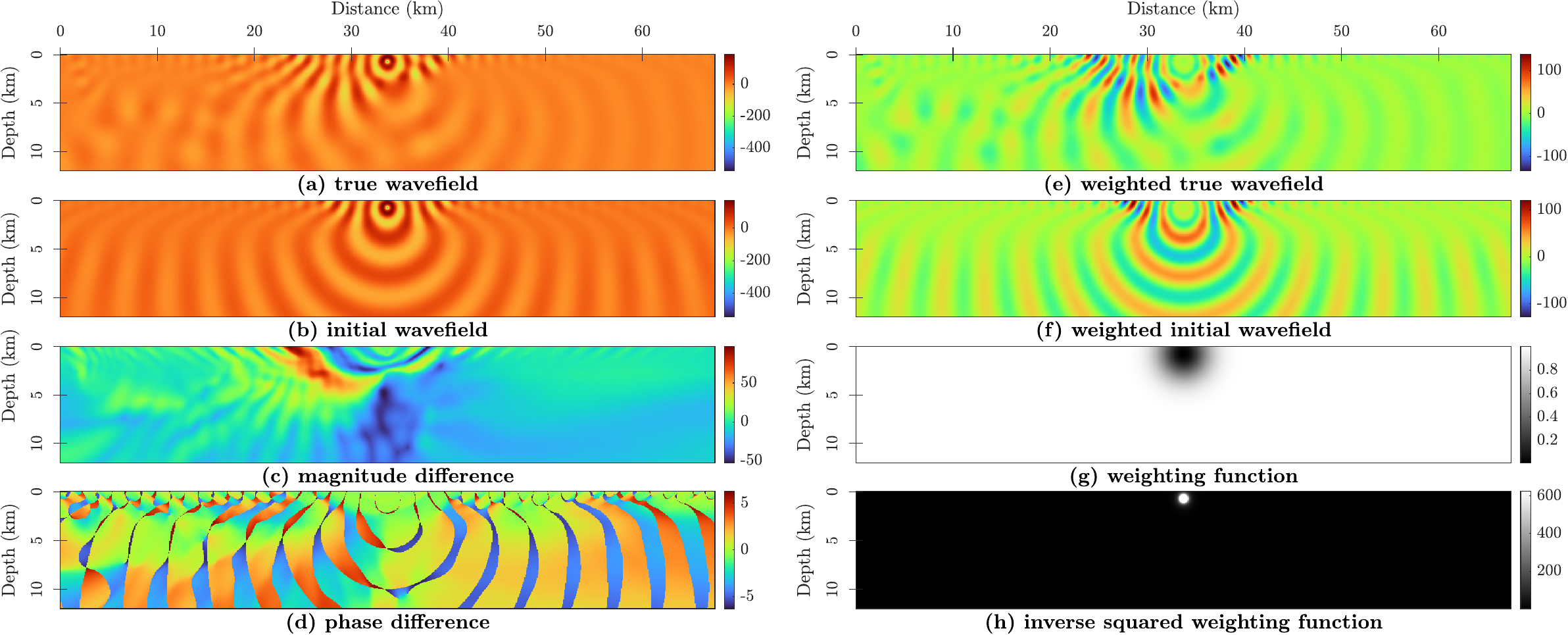}
    \caption{(a-b) Wavefields simulated in the (a) true model and (b) initial model at $f = 1$ Hz for the BP model in \fref{BP_model}. 
    (c-d) (c) Magnitude and (d) phase differences between the true and background wavefields.
    (e-f) Weighted versions of the wavefields in (a-b) using $\bphis(\bx)$ with $\gamma=10$ and $\sigma=1.5$ km.
    (g) The used weighting function. (h) Inverse squared weighting function.
    }
    \label{BP_wavefields}
\end{figure}
\begin{figure}[th]
    \centering
    \includegraphics[width=0.8\linewidth,trim={0 0cm 0 0},clip]{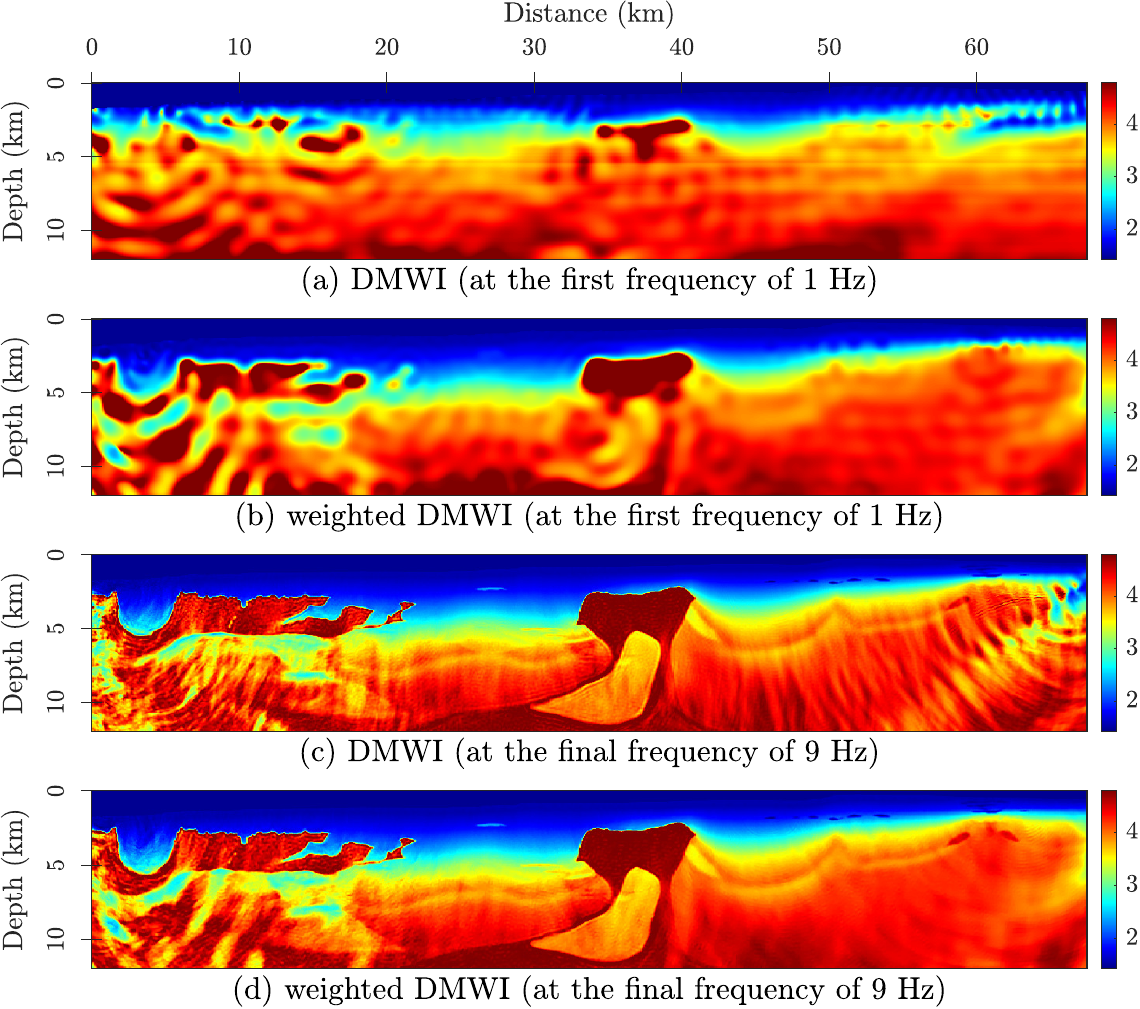}
    \caption{
     (a-b)  Reconstructed velocity model after inverting the 1 Hz data by (a) DMWI and (b) weighted DMWI. 
    (c-d)  Final DMWI result by sequentially inverting the 1-9 Hz data by (c) DMWI and (d) weighted DMWI. 
    }
    \label{BP_results}
\end{figure}
\begin{figure}[h]
   \centering
   \includegraphics[width=0.75\linewidth,trim={0cm 0cm 0 0},clip]{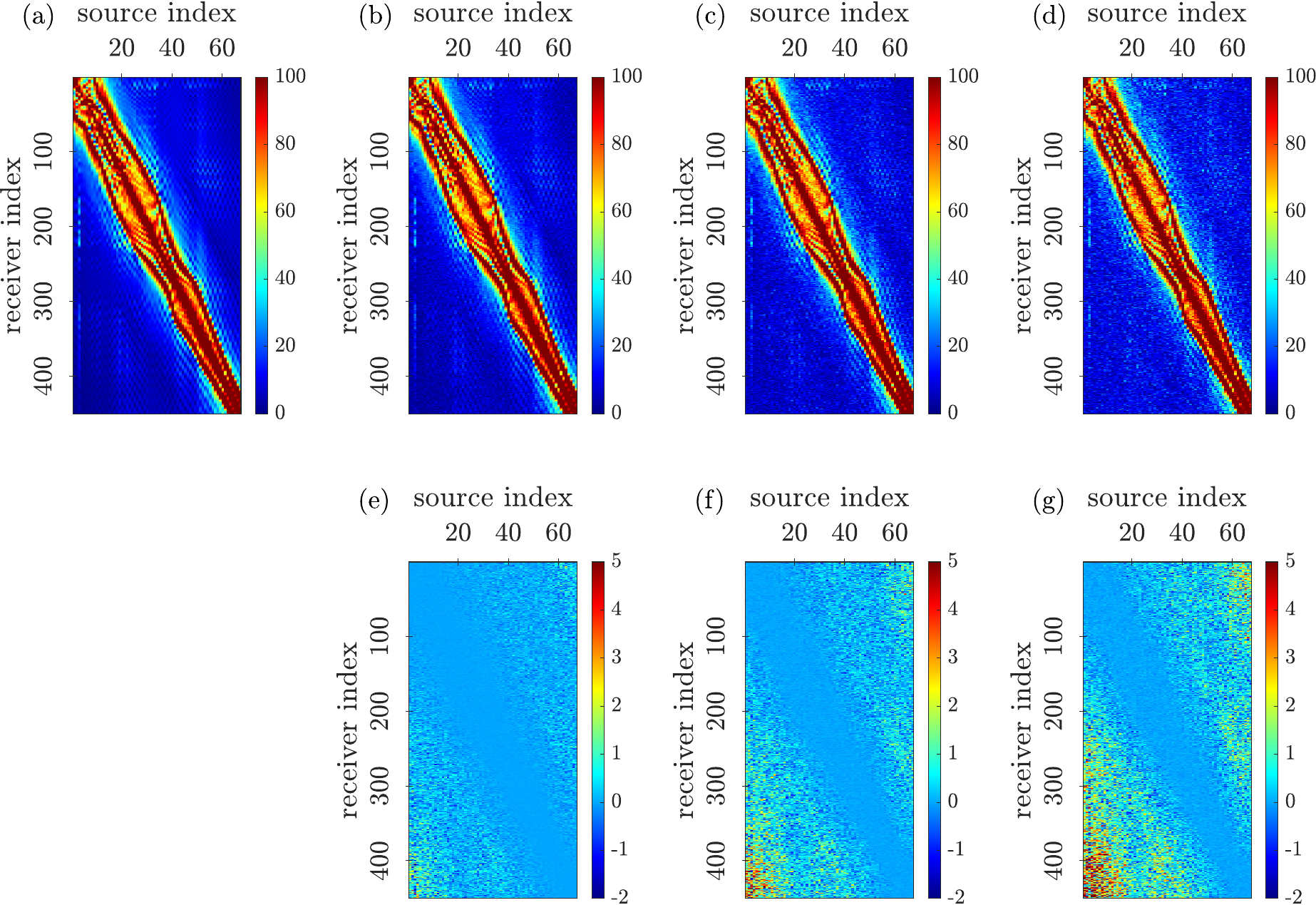}
   \caption{(a) The clean data for the 2004 BP salt model. (b-d) Noisy data for (b) 10\%, (c) 20\%, and (d) 30\% noise levels (computed as a percentage of the mean absolute value of the data). (e)-(g) Relative magnitude of the added noise.}
   \label{BP_data}
\end{figure}

\begin{figure}[h]
   \centering
   \includegraphics[width=1\linewidth,trim={0cm 0cm 0 0},clip]{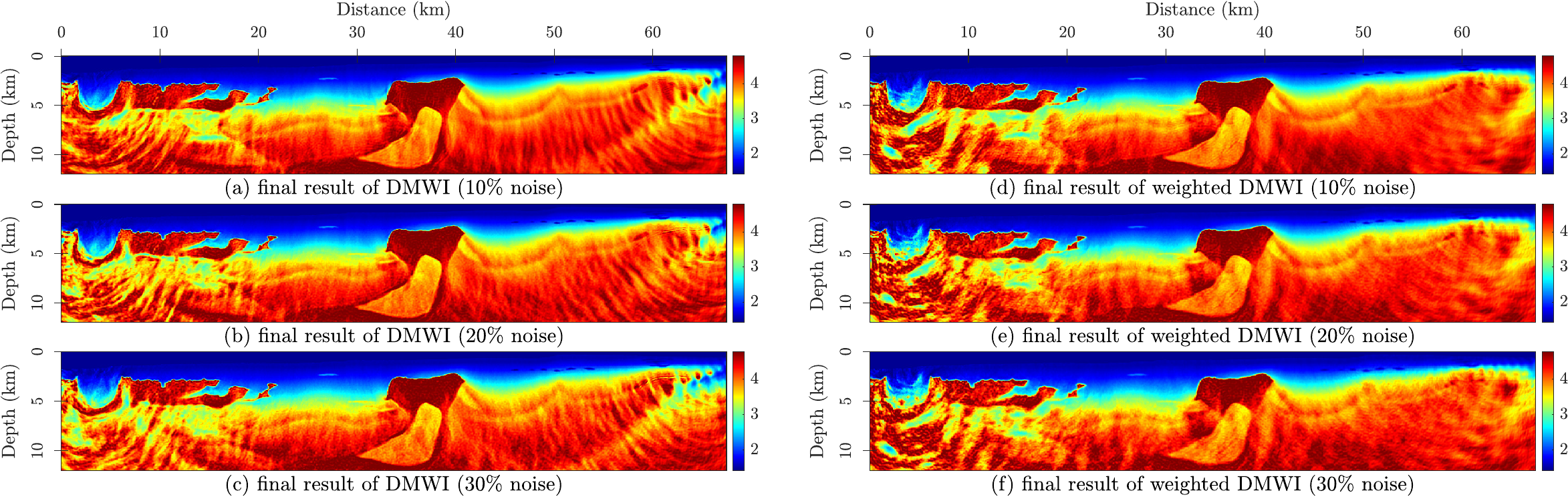}
   \caption{Inverted velocity models obtained in the presence of 10\%, 20\%, and 30\% noise from (a-c) DMWI and (d-f) weighted DMWI. }
   \label{BP_results_noisy}
\end{figure}
\begin{figure}[h]
   \centering
   \includegraphics[width=0.5\linewidth,trim={0cm 0cm 0 0},clip]{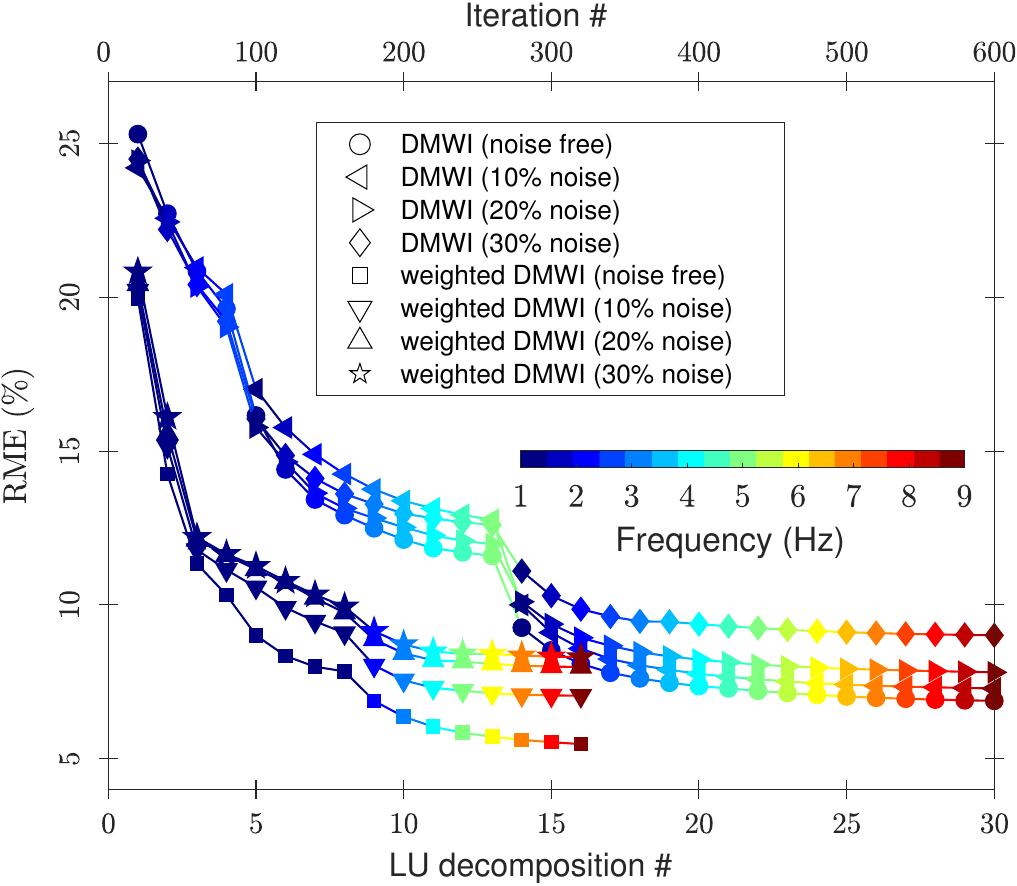}
   \caption{Relative model error as a function of the iteration and the number of LU factorizations during the inversion of the 2004 BP salt model. The color coding indicates the frequency of the input data used for each inversion step.}
   \label{BP_errors}
\end{figure}
\begin{figure}[!h]
    \centering
    \includegraphics[width=0.5\linewidth,trim={0 0cm 0 0},clip]{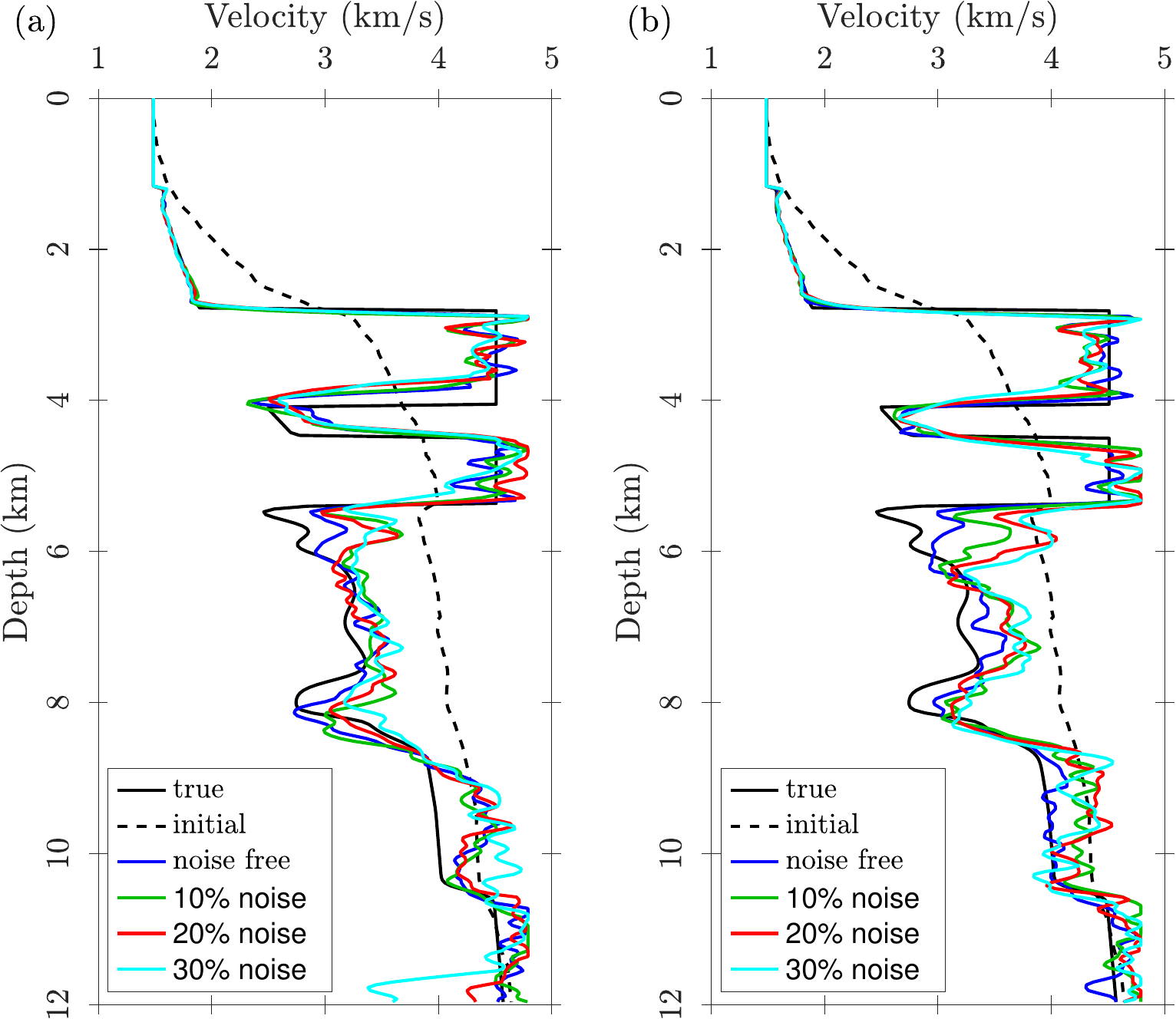}
    \caption{Direct comparison of velocity profiles at a horizontal distance of 18 km. The true model is shown in solid black, and the initial model in dashed black. Inverted models are shown for (a) DMWI and (b) weighted DMWI, using noise-free data (blue) and noisy data with 10\% (green), 20\% (red), and 30\% (cyan) noise levels.
    }
    \label{BP_logs}
\end{figure}
\subsection{BP salt model.}
We further evaluated the weighted method using the challenging 2004 BP salt model (Figure~\ref{BP_model}a), which spans 67.5 km in width and 12 km in depth. The subsurface model was discretized with a 37.5 m grid interval, resulting in a mesh of 320 × 1800 points. The ultra-long-offset stationary acquisition included 67 hydrophones spaced approximately 1 km apart on the seabed and 450 pressure sources positioned about 150 m apart at a depth of 25 m. Notably, for the weighted DMWI, neither the sources nor the receivers were aligned with grid points. Single-frequency data from 1 Hz to 9 Hz were used for inversion, starting from a simple 1D initial velocity model (Figure~\ref{BP_model}b). 

Figures \ref{BP_wavefields}(a) and (b) display the simulated wavefields in the true and initial models at $f = 1$ Hz, respectively. The corresponding magnitude and phase differences are shown in \fref{BP_wavefields}(c) and (d). For this model also, the phase plot reveals strong phase wrapping on both sides of the source which makes the inversion highly susceptible to local minima. 
To address this, we construct the $\bphis(\bx)$ using $\sigma=1.5$ km. Applying the weighting function to the wavefields in \fref{BP_wavefields}(a–b) produces the weighted results shown in \fref{BP_wavefields}(e–f). After weighting, the amplitudes of the wavefields become more balanced. Finally, Figures \ref{BP_wavefields}(g) and (h) show the weighting function and its inverse square, respectively. In the latter case, the color axis is clipped at the value corresponding to $\tfrac{1}{4}\lambda_{\mathrm{w}}$ for better visualization.

The inversion was performed sequentially, beginning with the 1 Hz data and progressively incorporating higher frequencies up to 9 Hz. Low frequencies are particularly crucial for building the velocity model. Therefore, we conducted 160 iterations for the 1 Hz inversion, updating the model (and performing an LU factorizations) every 20 iterations ($\texttt{maxit}=20$ in Algorithms \ref{alg_dual} and \ref{alg_SI_dual}), resulting in 8 LU factorizations in total. However, the DMWI method failed to converge at 1 Hz—the RME increased during the iterations—so we retained the solution after the first 20 iterations. The corresponding velocity models are shown in \fref{BP_results}(a) for DMWI and \fref{BP_results}(b) for weighted DMWI. The weighted method successfully reconstructed key structural features—such as the low-wavenumber background and the top salt bodies in the left and central regions—using only the 1 Hz data. By contrast, the standard DMWI failed to recover these critical features.

The inversion was then extended by progressively incorporating higher frequencies from 2 Hz to 9 Hz, in steps of 1 Hz. Each frequency inversion consisted of 20 iterations, reusing the LU factorizations from the model estimated at the previous frequency. The final velocity model obtained using weighted DMWI is shown in \fref{BP_results}d.
In contrast, DMWI did not produced satisfactory results (no shown here) thus we carried out an additional inversion cycle over the frequency range 1–9 Hz. The final DMWI result after this second cycle is shown in \fref{BP_results}c. A comparison between \fref{BP_results}(c) and \fref{BP_results}(d) clearly demonstrates that the weighted DMWI yields a much more accurate reconstruction while requiring only a single frequency sweep. 

\subsection{BP salt model: Noisy data.}
To further evaluate the robustness of the weighted method, we repeated the inversion using noisy data by adding random Gaussian noise to the observed data.
\fref{BP_data}(a) displays the magnitude of the noise-free data in the source–receiver plane, while \fref{BP_data}(c)–(d) show the noisy data for 10\%, 20\%, and 30\% noise levels (computed as a percentage of the mean absolute value of the data). The relative magnitude of the added noise is presented in \fref{BP_data}(e)–(g).

We applied the same sequential inversion strategy as in the noise-free case, starting from 1 Hz and incrementally increasing to 9 Hz in 1 Hz steps. Figures~\ref{BP_results_noisy}(a)–(c) show the final velocity models recovered by DMWI with 10\%, 20\%, and 30\% noise.
Figures~\ref{BP_results_noisy}(d)–(f) present the corresponding results for the weighted DMWI method.
Despite the added noise, the main structural features—including the salt geometry and background velocity variations—are successfully recovered. These results demonstrate that both methods remain stable and effective under noisy conditions.

The convergence behavior for all cases is shown in \fref{BP_errors}, where we compare the error curves for noise-free and noisy inversions.
A direct comparison of the true model, initial model, and inverted models along two vertical profiles at 18 km is provided in \fref{BP_logs}.
The weighted method consistently improves the conditioning of the inverse problem, reducing noise sensitivity and accelerating convergence. Although noise slightly increases the model error, the overall convergence trend remains stable and reliable.


\section{Discussions.} \label{Discussions}
We have introduced a general weighted proximal Lagrangian objective function for FWI and developed several algorithms for its solution. The performance of the proposed approach critically depends on the proper design of the weighting functions $\Wd, \Wm, \Ws$, which weight the data residual, the model parameters, and the Lagrange multipliers, respectively. Beyond the impact of these weighting functions, the proposed dual-space ADMM implementation significantly enhances computational efficiency. This is achieved by avoiding repeated LU factorizations of the forward operator during inner iterations, relying instead on efficient forward–backward substitutions. Further acceleration of these inner iterations was facilitated by Anderson acceleration \cite{Anderson_1965_IPN}.

In this work, we did not attempt a detailed design of $\Wd$, the data weighting matrix, leaving it for future investigations. 
However, as outlined in Subsection \ref{subsec_Wd}, various strategies exist, such as emphasizing specific data portions based on source–receiver offset, which could be explored in future work. The regularization parameter $\mu$, associated with $\Wd$, was adaptively updated using the Residual Whiteness Principle (RWP) method \cite{Almeida_2013_PEB,Lanza_2013_VID}. Similarly, we set $\Wm$ to be a scalar multiple of the identity matrix, thereby applying a homogeneous regularization over the entire model space. However, $\Wm$ can be designed much more flexibly. For example, a space-variant weighting function could be employed to concentrate model updates in specific regions of interest. Increasing $\Wm$ with depth would emphasize near-surface updates, potentially improving the conditioning of the inversion. Alternatively, $\Wm$ could be replaced by a discrete finite-difference operator to promote smoothness, or more advanced regularizations such as total-variation (TV) or hybrid Tikhonov–TV regularization could be incorporated to preserve sharp contrasts while maintaining stability \cite{Gholami_2013_BCT}.

Among the three weighting functions, we focused in particular on $\Ws$, introducing a distance-dependent weighting function that plays a key role in eliminating the explicit source signature from the optimization problem \cite{Huang_2018_SEW}. This property significantly simplifies practical implementation by relaxing the need for source signature estimation or source alignment with grid points. 
This is achieved by compensating for the natural decay of seismic energy and promoting a focused multiplier estimates, thereby improving the convexity of the objective function and leading to more robust and focused model updates. This mechanism causes model updates to be largest near the sources and receivers, resembling a layer-stripping approach that improves stability by sequentially building the model from shallow to deep regions. The proposed weighting function contains two free parameters, $(\sigma, \gamma)$. In all presented numerical examples, we set $\gamma = 10$ and chose $\sigma$ as a scale factor in the range $[1,3]$ of the dominant wavelength of the propagating wavefield. Numerical experiments consistently demonstrate the method's enhanced resilience to cycle skipping, its ability to handle off-grid sources and receivers, and its robustness against significant levels of random Gaussian noise.

\begin{figure}[h]
    \centering
\begin{tabular}{cccc}
\hspace{0cm}\footnotesize{(a)} & 
\hspace{0cm}\footnotesize{(b)} & 
\hspace{0cm}\footnotesize{(c)} & 
\hspace{0cm}\footnotesize{(d)}\\ 
\hspace{0cm}\includegraphics[scale=0.325]{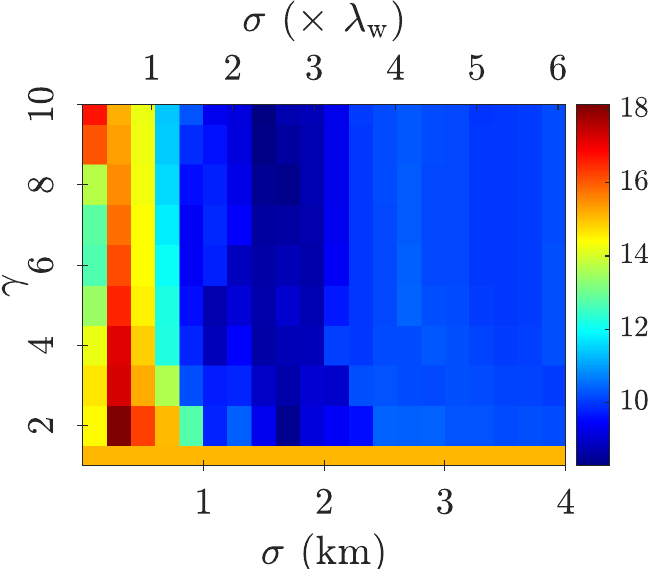} &
\hspace{0cm}\includegraphics[scale=0.325]{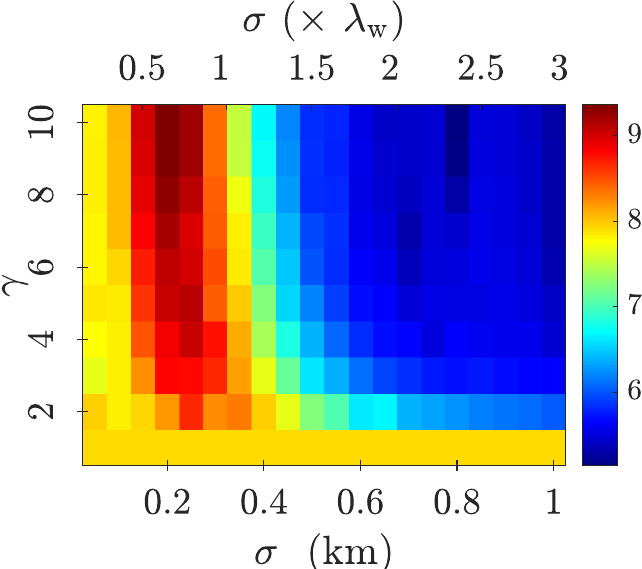} &
\hspace{0cm}\includegraphics[scale=0.325]{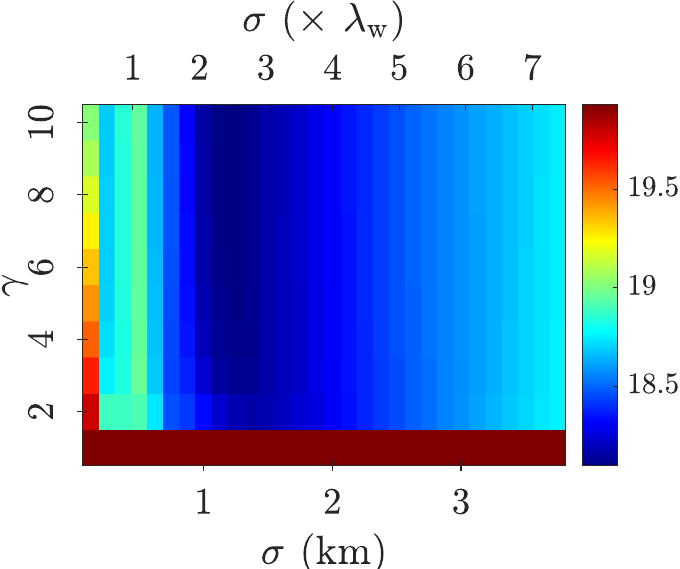} &
\hspace{0cm}\includegraphics[scale=0.325]{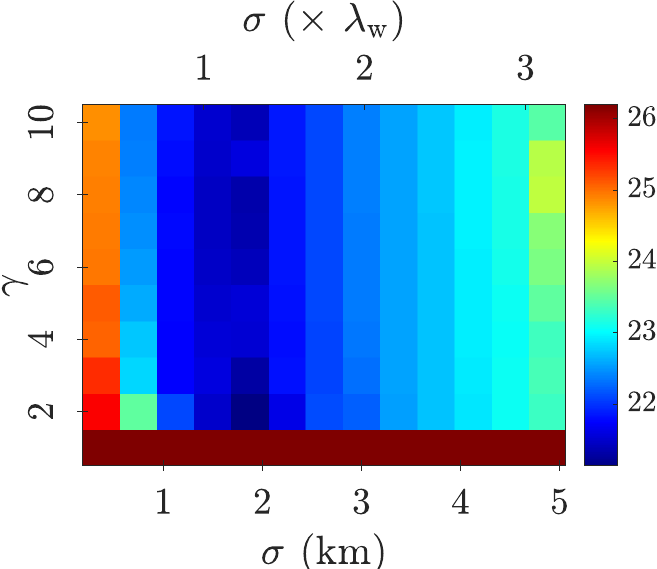}
\end{tabular} 
\caption{Relative model error map as a function of $\sigma$ and $\gamma$ for (a) Camembert model, (b) Gaussian model, (c) Marmousi model, and (d) BP model.}
\label{sigma_gamma_map}
\end{figure}

To further explore the effect of $(\sigma, \gamma)$, we performed a sensitivity study by constructing weighting functions over a grid of parameter values and running 20 iterations of the weighted DMWI algorithm. Figure~\ref{sigma_gamma_map}(a)–(d) shows the mean-squared error (MSE) across the $(\sigma, \gamma)$-parameter space for the four numerical examples considered. Two main observations can be drawn: (i) The inversion results are significantly more sensitive to $\sigma$ than to $\gamma$. For $\sigma > \lambda_{\mathrm{w}}$, the inversion remains stable and yields satisfactory reconstructions over a broad range of $\gamma$, suggesting that $\gamma$ acts primarily as a secondary tuning parameter.
(ii) The MSE shows a clear trend with respect to $\sigma$: small $\sigma$ values lead to higher MSE, indicating poor model recovery due to an overly localized weighting function, whereas larger $\sigma$ values improve model reconstruction and stability.

The distance-dependent dual method proposed in this paper shares similarities with matched source waveform inversion (MSWI) \cite{Huang_2018_SEW, Symes_2025_NIM}, but there are three key differences:
\begin{itemize}
    \item[(i)] MSWI relies on a penalty-based objective function, whereas our method employs a proximal-point Lagrangian objective function. This distinction provides two major advantages: (1) the AL approach enables more accurate solutions to the inverse problem through the incorporation of source multipliers, and (2) it reduces the sensitivity of the optimization process to fine-tuning of the regularization parameter $\mu$, as demonstrated in \cite{Gholami_2024_FWI}, unlike the penalty formulation in which the penalty parameter has a profound influence on the optimization problem \cite{Symes_2020_FWI}.  
    \item[(ii)]  MSWI depends on accurately inverting the normal operator associated with the extended source, expressed in our notation as $\bold{S}^{\top}\bold{S} + \mu \Ws$ (\cite{Huang_2018_SEW}, their Equation 11). 
The forward operator $\bold{S}$ can be computed once and reused for all sources. However, this operator resides in the source space, is dense, and is inherently ill-conditioned. In MSWI, the authors addressed this by employing iterative Krylov subspace methods to solve the resulting system. Nevertheless, the combination of the operator's ill-conditioned nature and its source-dependent characteristics poses significant challenges for large-scale problems (see \cite{Huang_2018_SEW}, their DISCUSSION section). In contrast, our formulation solves the system in the data space \cite{Gholami_2022_EFW}, requiring the inversion of the data-space Hessian matrix $\bold{S}\Ws^{-1}\bold{S}^{\top} + \mu \bI$. This approach allows the inversion of the normal operator by direct methods even for large-scale implementations and efficient selection of the regularization parameter $\mu$ by using standard methods like GCV and RWP.
\item[(iii)] In MSWI, the optimization is performed over both the model parameters and the source, requiring operator updates at each iteration. In contrast, the proposed method adopts a dual formulation, where the source multipliers are estimated for each frequency while keeping the background model fixed. This approach requires only a single LU factorization per extended source inversion, significantly improving computational efficiency and reducing the overall cost of inversion, making it highly suitable for large-scale problems.
\end{itemize}

For a given challenging initial model, the proposed weighted method can successfully converge to accurate results by inverting data at the lowest available frequency, a scenario where standard multiplier-based methods typically fail to converge to the true solution and become trapped in local minima due to cycle skipping. However, while the proposed weighting approach significantly broadens the basin of attraction and extends the frequency range for accurate inversion, numerical results also indicate that it cannot entirely eliminate non-convexity at very high frequencies, where both weighted and unweighted methods may eventually fail. This suggests areas for further methodological refinement. 

\section{Conclusions.}\label{Conclusions}
We introduced a weighted multiplier method for full-waveform inversion (FWI), incorporating distance-dependent penalty weights to improve the conditioning of traditional augmented Lagrangian (AL) approaches. Our work specifically focused on the design and impact of the distance-dependent penalty weights, demonstrating their critical role in the method's performance. By spatially varying the penalty parameter, the method compensates for the natural scaling of wave equation residuals, ensuring more effective constraint enforcement across the model domain. Compared to traditional methods that use a uniform penalty parameter, the proposed variable weighting approach eliminates the need for explicit source signature estimation, enhances resilience to cycle skipping, and accelerates convergence. Furthermore, the algorithm relaxes the requirement for sources and receivers to be positioned on finite-difference grid points, increasing its flexibility in practical applications. While this study primarily focused on the design of $\Ws$, with $\Wm$ set as a simple scalar regularization and $\Wd$ largely kept uniform for experimental focus, the flexible framework allows for more sophisticated designs of these weighting functions to further enhance control over the inversion.
In this framework, FWI is reformulated as a source estimation problem within a fixed-point iteration scheme implemented via dual-space Alternating Direction Method of Multipliers (ADMM). This approach ensures that model parameters, and consequently the required LU factorizations of the forward operator, remain fixed during the inner loops of each source multiplier inversion, significantly reducing computational costs to only a few factorizations per frequency. Further acceleration of the inner iterations was achieved through the application of Anderson acceleration. Numerical experiments on challenging models demonstrated the method’s ability to accurately reconstruct complex subsurface features without relying on source signature estimation, low-frequency data, or a highly accurate initial model. These results highlight the method’s potential as a robust and efficient approach for FWI. Future work will explore more advanced designs for $\Wd$ and $\Wm$, and extend the method's application to even more complex geological structures and 3D scenarios.



\bibliographystyle{siamplain.bst}
\newcommand{\SortNoop}[1]{}

\end{document}